\RequirePackage{fix-cm}
\documentclass[twocolumn]{svjour3}          
\smartqed  
\usepackage{graphicx}
\usepackage{subfig}

\usepackage{color}
\usepackage{amsfonts}
\usepackage{amsmath}
\usepackage{comment}
\usepackage{soul}

\usepackage{multirow}
\usepackage{rotating}
\usepackage{pdflscape}

\usepackage{tikz, pgfplots}
\usetikzlibrary{shapes,shadows,arrows,backgrounds}
\usetikzlibrary[arrows,decorations.pathmorphing,backgrounds,positioning,fit,petri]
\usepgfmodule{plot}

\journalname{The VLDB Journal}
\begin{document}

\title{Secret Sharing for Cloud Data Security}
\subtitle{A Survey}


\author{Varunya Attasena		\and
        J\'{e}r\^{o}me Darmont	\and
        Nouria Harbi
}


\institute{
	Varunya Attasena \at
    	The computer engineering department, Engineering faculty at Kamphaeng Saen, Kasetsart University Kamphaeng Saen Campus, Nakhon Pathom, 73140, Thailand \\
		Tel.: +66-34-281-074 \#7528\\
		Fax: +66-99-695-415 \\
		\email{fengvry@ku.ac.th}   
           \and
	J\'{e}r\^{o}me Darmont \at
    	Universit\'{e} de Lyon, Lyon 2, ERIC EA 3083, 5~avenue Pierre Mend\`{e}s France, 69676~Bron Cedex, France \\
		Tel.: +33-478-774-403 \\
		Fax: +33-478-772-375 \\
		\email{jerome.darmont@univ-lyon2.fr}         
           \and
	Nouria Harbi \at
    	Universit\'{e} de Lyon, Lyon 2, ERIC EA 3083, 5~avenue Pierre Mend\`{e}s France, 69676~Bron Cedex, France \\
		Tel.: +33-478-774-492 \\
		Fax: +33-478-772-375 \\
		\email{nouria.harbi@univ-lyon2.fr}           
}

\date{Received: date / Accepted: date}

\maketitle

\begin{abstract}
Cloud computing helps reduce costs, increase business agility and deploy solutions with a high return on investment for many types of applications. However, data security is of premium importance to many users and often restrains their adoption of cloud technologies. Various approaches, i.e., data encryption, anonymization, replication and verification, help enforce different facets of data security. Secret sharing is a particularly interesting cryptographic technique. Its most advanced variants indeed simultaneously enforce data privacy, availability and integrity, while allowing computation on encrypted data. The aim of this paper is thus to wholly survey secret sharing schemes with respect to data security, data access and costs in the pay-as-you-go paradigm.

\end{abstract}

\section{Introduction}
\label{sec:introduction}

Cloud computing is currently booming, with companies of all sizes adopting associated technologies to benefit from resource and cost elasticity. However, data security remains one of the top concerns for cloud users and would-be users. Security issues, both inherited from classical distributed architectures and specific to the new framework of the cloud, are indeed numerous, especially at the data storage level of public clouds \cite{C-security-issues}. 

\begin{figure*}[htb]
	\centering
\resizebox{\textwidth}{!} {
\begin{tikzpicture}	

\def\y {0}

\node at (-2.25,\y) {\tiny{\textbf{Cloud Computing}}};	
\node at (6.2,\y) {\tiny{\textbf{Intruders}}};

\draw [semithick] (-2.25,\y-0.15) -- (-5.5,\y-0.5); 
\draw [semithick] (-2.25,\y-0.15) -- (1,\y-0.5);
\draw [semithick] (6.2,\y-0.15) -- (5.2,\y-0.5); 
\draw [semithick] (6.2,\y-0.15) -- (7.2,\y-0.5);

\edef\y {\y-0.7}
\node at (-5.5,-0.7) {\tiny{\textbf{Service provider policies}}};
\node at (1,-0.7) {\tiny{\textbf{Characteristics of cloud architectures}}};
\node at (5.2,-0.7) {\tiny{\textbf{Inside intruders}}};
\node at (5.2,-0.95) {\tiny{Service provider staffs}};
\node at (5.2,-1.15) {\tiny{\& other customers}};
\node at (7.2,-0.7) {\tiny{\textbf{Outside}}};
\node at (7.2,-0.95) {\tiny{\textbf{intruders}}};

\draw [semithick] (-5.5,-0.85) -- (-7,-1.2);
\draw [semithick] (-5.5,-0.85) -- (-4,-1.2);
\draw [semithick] (1,-0.85) -- (-1.1,-1.35);
\draw [semithick] (1,-0.85) -- (0.3,-1.2);
\draw [semithick] (1,-0.85) -- (1.7,-1.2);
\draw [semithick] (1,-0.85) -- (3.1,-1.35);


\node at (-7,-1.4) {\tiny{\textbf{Policies for}}};
\node at (-7,-1.65) {\tiny{\textbf{taking benefits}}};
\node at (-7,-1.9) {\tiny{e.g., rollback feature,}};
\node at (-7,-2.15) {\tiny{super-user access}};
\node at (-4,-1.4) {\tiny{\textbf{Control \& modification policies}}};
\node at (-4,-1.65) {\tiny{e.g., Data \& system migration}};
\node at (-1.1,-1.55) {\tiny{\textbf{Grid}}};
\node at (-1.1,-1.8) {\tiny{\textbf{technology}}};
\node at (0.3,-1.4) {\tiny{\textbf{Virtual}}};
\node at (0.3,-1.65) {\tiny{\textbf{machine}}};
\node at (0.3,-1.9) {\tiny{\textbf{technology}}};
\node at (1.7,-1.4) {\tiny{\textbf{Virtual}}};
\node at (1.7,-1.65) {\tiny{\textbf{network}}};
\node at (1.7,-1.9) {\tiny{\textbf{technology}}};
\node at (3.1,-1.55) {\tiny{\textbf{Network}}};
\node at (3.1,-1.8) {\tiny{\textbf{presence}}};

\draw [decorate,decoration={brace,amplitude=4pt}](2.4,-2)--(-1.8,-2);


\draw [semithick,->] (-7.0,-2.3) -- (-5.6,-2.7);
\draw [semithick,->] (0.3,-2.2) -- (-5.4,-2.7);

\draw [semithick,->] (-4.0,-1.9) -- (-1.1,-2.7);
\draw [semithick,->] (0.3,-2.2) -- (-0.9,-2.7);

\draw [semithick,->] (0.3,-2.2) -- (3.4,-2.7);
\draw [semithick,->] (5.2,-1.3) -- (3.6,-2.7);

\draw [semithick,->] (3.1,-1.9) -- (4.4,-2.7);
\draw [semithick,->] (7.2,-1.1) -- (4.6,-2.7);

\node at (-5.5,-2.8) {\tiny{\textbf{+}}};
\node at (-1,-2.8) {\tiny{\textbf{+}}};
\node at (3.5,-2.8) {\tiny{\textbf{+}}};
\node at (4.5,-2.8) {\tiny{\textbf{+}}};

\draw [semithick,->] (-5.5,-2.9) -- (-5.5,-3.1);
\draw [semithick,->] (-1,-2.9) -- (-1,-3.1);
\draw [semithick,->] (3.5,-2.9) -- (3.5,-3.1);
\draw [semithick,->] (4.5,-2.9) -- (4.5,-3.1);

\node at (-5.5,-3.20) {\tiny{\textbf{Data loss, damage,}}};
\node at (-5.5,-3.45) {\tiny{\textbf{alteration and pilfering}}};

\node at (-1,-3.20) {\tiny{\textbf{Data loss and damage ;}}};
\node at (-1,-3.45) {\tiny{\textbf{Service down ;}}};
\node at (-1,-3.70) {\tiny{\textbf{Data transfer bottlenecks}}};

\node at (4,-3.20) {\tiny{\textbf{Data loss, damage, alteration and pilfering ;}}};
\node at (4,-3.45) {\tiny{\textbf{Service down ;}}};
\node at (4,-3.70) {\tiny{\textbf{Data transfer bottlenecks}}};

\draw [semithick] (-5.5,-3.6) -- (-4,-4.4);
\draw [semithick] (-1,-3.8) -- (-4,-4.4);
\draw [semithick] ( 4,-3.8) -- (-4,-4.4);

\draw [semithick] (-5.5,-3.6) -- (-1,-4.4);
\draw [semithick] (-1,-3.8) -- (-1,-4.4);
\draw [semithick] ( 4,-3.8) -- (-1,-4.4);

\draw [semithick] (-1,-3.8) -- (2,-4.4);
\draw [semithick] ( 4,-3.8) -- (2,-4.4);

\node [rectangle,fill=black!100,minimum size=3mm,rounded corners=2mm]  at (-4,-4.65) {\color{white} \tiny{\textbf{Data privacy}}};
\node [rectangle,fill=black!100,minimum size=3mm,rounded corners=2mm]  at (-1,-4.65) {\color{white} \tiny{\textbf{Data integrity}}};
\node [rectangle,fill=black!100,minimum size=3mm,rounded corners=2mm]  at (2,-4.65) {\color{white} \tiny{\textbf{Data availability}}};

\end{tikzpicture}
}
	\caption{Data security issues in the cloud}
	\label{fig:intro+Cloud-data-security}
\end{figure*}
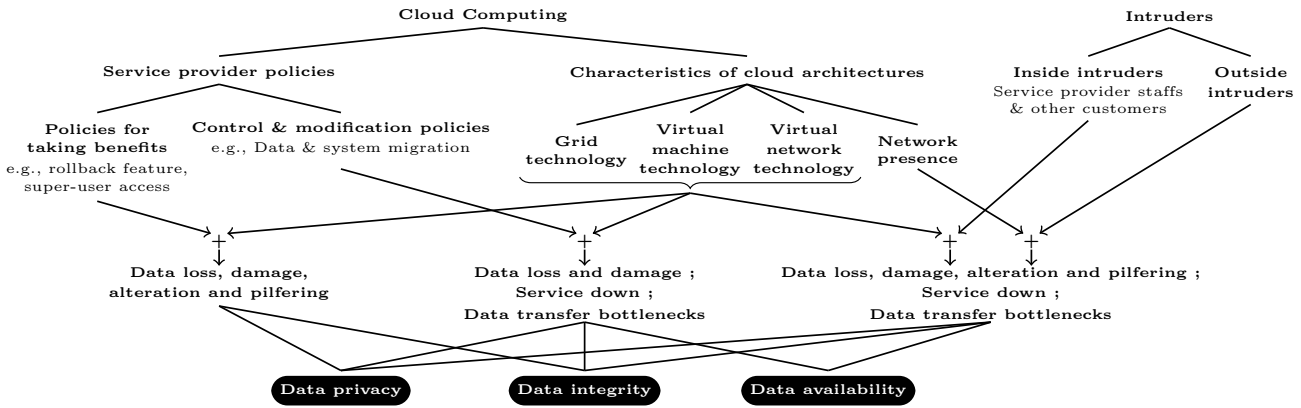

Critical security concerns in cloud storage are depicted in Figure~\ref{fig:intro+Cloud-data-security}, which highlights the major issues in cloud data security, i.e., data privacy, availability and integrity. 
In particular, cloud architectures might not be sufficiently safeguarded from inside attacks. In virtual environments, a malicious user might be able to break into "neighboring" virtual machines located on the same hardware, and then steal, modify or delete the other users' data \cite{MAK-2016,MA-SK-AV-2015,PD-SD-EG-SS-2016,JJ-HT-GA-2010,DZ-DL-2012,MZ-RZ-WX-WQ-AZ-2010,KH-DR-EF-EF-2013}. In such environments, users are indeed usually granted with superuser access for managing their virtual machines. A malicious superuser can access real network components and thus launch attacks \cite{MA-SK-AV-2015,KB-SM-SK-AZ-2014}.
Moreover, virtualization allows the rollback of a virtual machine to some previous state if necessary. Although this rollback feature provides flexibility to the users, it can also revert the virtual machine to previous security policies and configuration control \cite{MA-SK-AV-2015,KH-DR-EF-EF-2013}. Eventually, virtual machine migration is run to improve quality of service. During such migration processes, which typically do not shut down services, virtual machine contents are exposed to the network, and problems such as network transfer bottlenecks and data damage may occur \cite{MA-SK-AV-2015,KH-DR-EF-EF-2013,FZ-HC-2013}.  


\begin{figure}[hbt]
\resizebox{0.48\textwidth}{!}{
\begin{tikzpicture}
		
			\draw [thick] ( 0.0, 0) circle (1cm);
			\draw [thick] ( 0.6,-1) circle (1cm);
			\draw [thick] (-0.6,-1) circle (1cm);
			
			\node [circle,draw,thick,fill=red,opacity=.2] at ( 0.0, 0) {~~~~~~~~~~~~~~~~~~};
			\node [circle,draw,thick,fill=blue,opacity=.2] at ( 0.6,-1) {~~~~~~~~~~~~~~~~~~};
			\node [circle,draw,thick,fill=green,opacity=.2] at (-0.6,-1) {~~~~~~~~~~~~~~~~~~};
			
			\node at ( 0.0,1.60) {\textbf{Data}};
			\node at ( 0.0,1.30) {\textbf{privacy}};
			\node at (-1.9,-2.3) {\textbf{Data availability}};
			\node at ( 1.8,-2.3) {\textbf{Data integrity}};

			\draw [decorate,decoration={brace,amplitude=4pt}](1.5,-0.4)--(1.5,1.9) node[midway, left, font=\footnotesize, xshift=2pt] {~};			
			
			\draw[-](0.0,0.3) -- (1.35,0.75);	
			\node at (2.7,1.8) {\small{Data encryption}};
			\node at (3.0,1.5) {\textit{\scriptsize{e.g. homomorphic and}}};
			\node at (3.0,1.25) {\textit{\scriptsize{incremental encryption}}};
			
			\node at (3.0,0.9) {\small{Data anonymization}}; 
			\node at (3.2,0.6) {\textit{\scriptsize{e.g. k-anonymity}}};
			\node at (3.2,0.35) {\textit{\scriptsize{and l-diversity}}};
			
			\node at (3.0,0.0) {\small{Differential privacy}};
			\node at (2.78,-0.3) {\small{Data separation}};
			
			\draw[-](-0.7,-0.3) -- (-1.8,0.6);			
			\node at (-2.9,0.9) {\small{Secret sharing}};
			
			\draw[-](0,-0.7) -- (-2.1,-0.3);			
			\node at (-3.0,-0.0) {\small{Verifiable}};
			\node at (-3.1,-0.3) {\small{secret sharing}};

			\draw[-](0.9,-1.1) -- (1.75,-1.3);
			\node at (3.0,-1.3) {\small{Data verification}};

			\draw[-](-0.9,-1.1) -- (-1.75,-1.3);
			\node at (-3.0,-1.3) {\small{Data replication}};

		\end{tikzpicture}
}
\caption{Features of data security approaches}
\label{fig:security-features}       
\end{figure}
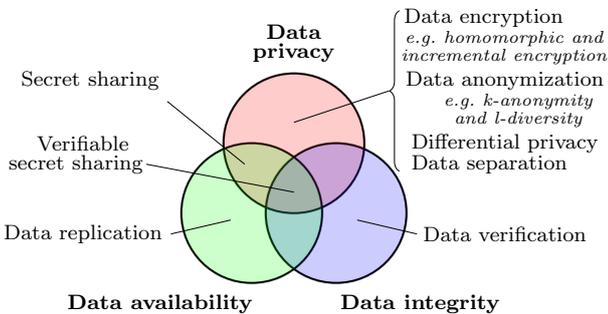

Classical data security approaches, i.e., data encryption \cite{incremental-encryption-I,homomorphic-encryption-II}, data anonymization \cite{data-anonymization-I}, replication \cite{data-replication}, data verification \cite{data-verification-II}, data separation \cite{CZ-EC-RY-2014,KY-SM-VK-MK-2015,ZZ-HZ-XD-PL-XY-2013} and differential privacy \cite{CD-2006}, can solve most data security issues within cloud computing environments (Figure~\ref{fig:security-features}), but usually one at a time.
Many data-centric cloud applications do not only require data to be secure, but also efficiently accessed, sometimes through complex, analytical queries akin to on-line analysis processing (OLAP) operations. With users seeking to reduce costs in the cloud's pay-as-you-go pricing model, achieving the best tradeoff between data security and access power and efficiency is a great challenge \cite{C-security-issues,survey-DB-5}.

Existing surveys about distributed data security list security services in distributed storage: authentication and authorization, availability, confidentiality and integrity, key sharing and management, auditing and intrusion detection, and finally useability, manageability and performance \cite{survey-storage-1,PD-SD-EG-SS-2016}. Then, network file systems, cryptographic file systems and storage intrusion detection systems are discussed and compared. This pre-cloud review is complemented by a thorough comparison of storage-centric data protection (i.e., network storage devices) in user-centric data protection systems (i.e., cryptographic storage systems and cloud-based storage) \cite{survey-storage-2,PD-SD-EG-SS-2016}. Finally, \cite{C-security-issues-3,PD-SD-EG-SS-2016} provide a short overview of what should be done in terms of data auditing and encryption in the cloud.

Although these surveys do mention secret sharing, they provide few details about this particular cryptographic technique, which was simultaneously introduced by Shamir \cite{Shamir-1979} and Blakley \cite{Blakley-1979} in 1979 and can be particularly useful nowadays in the context of cloud computing, e.g., to safely manage and analyze big data. Threshold secret sharing schemes indeed transform sensitive data into individually meaningless data pieces (called shares) that are distributed to $n$ participants akin to CSPs. Computations can then be performed onto shares, but yield meaningless individual results. The global result can only be reconstructed knowing individual results from several participants (more than threshold $t \leq n$). Moreover, some secret sharing variants simultaneously enforce data privacy, availability and integrity, which no other security scheme achieves. Eventually, secret sharing can be used by both CSPs, with data being shared within their cloud infrastructure, and users, who can dispatch sensitive data over several providers. Since some secret sharing schemes also support homomorphism, they allow data analysis on shares, thus allowing data access cost optimization.

To the best of our knowledge, secret sharing schemes (SSSs) up to 2008 have only been surveyed with respect to bounds on share size and global data volume \cite{Survey-SS}. In this paper, we also include the most recent SSSs and complement \cite{Survey-SS} by analyzing the objectives of each SSS, the security and data analysis features a user can expect, and the costs implied in a cloud computing environment.


\begin{figure*}[thb]
	\centering 		
	\resizebox{0.8\textwidth}{!} {\begin{tikzpicture}
		
		\node [rectangle,draw,thick,rounded corners=2mm,minimum size=5mm,minimum width=45mm] at (0,5) (S1) {\scriptsize{\textbf{\ref{sec:introduction} Introduction}}};
		
		\node [rectangle,draw,thick,rounded corners=2mm,minimum size=5mm,minimum width=45mm] at (0,4) (S2) {\textbf{\scriptsize{\ref{sec:Secret-sharing-schemes} Secret Sharing Schemes}}};
		\node [draw,thick,minimum width=40mm] at (0.5,3.0) (S2x1) {\textbf{\scriptsize{\begin{tabular}{l}
			Describes SSS types: \\
			classic SSSs, MSSSs, \\ 
			VSSSs, VMSSSs, PSSSs, \\
			WSSSs, SSSSs, RSSSs
			\end{tabular}
		}}};
		\node [rectangle,draw,thick,rounded corners=2mm,minimum size=4mm,minimum width=40mm] at (0.5,2.0) (S2x2) {\textbf{\scriptsize{2.1-2.11 SSS groups}}};
		
		\draw [->,line width=1mm] (S1) -- (S2);
		\draw [->,line width=0.5mm] (-2,3.0)--(S2x1);
		\draw [->,line width=0.5mm] (-2,3.75) -- (-2,2.0)--(S2x2);

		\node [rectangle,draw,thick,rounded corners=2mm,minimum size=5mm,minimum width=45mm] at (6,5) (S3) {\textbf{\scriptsize{\ref{sec:Discussion} Discussion}}};
		\node [rectangle,draw,thick,rounded corners=2mm,minimum size=4mm,minimum width=45mm] at (6.8,4.45) (S3x1) {\textbf{\scriptsize{\ref{sec:Evolution-sss} Evolution of SSSs}}};
		\node [rectangle,draw,thick,rounded corners=2mm,minimum size=4mm,minimum width=45mm] at (6.8,3.95) (S3x2) {\textbf{\scriptsize{\ref{sec:Encryption-sss} Sharing and Reconstruction}}};
		\node [rectangle,draw,thick,rounded corners=2mm,minimum size=4mm,minimum width=45mm] at (6.8,3.45) (S3x3) {\textbf{\scriptsize{\ref{sec:Properties-sss} Features of SSSs}}};
		\node [draw,thick,minimum width=70mm] at (8.6,2.5) (S3x3x1) {\textbf{\scriptsize{
			\begin{tabular}{ll}
			\ref{sec:dPdA} & Data Privacy and Availability \\
			\ref{sec:dI} & Data Integrity \\
			\ref{sec:dA} & Data Access \\
			\ref{sec:dO} & Other Features
			\end{tabular} }}};
			
		\node [rectangle,draw,thick,rounded corners=2mm,minimum size=4mm,minimum width=45mm] at (6.8,1.55) (S3x4) {\textbf{\scriptsize{\ref{sec:cost-sss} Costs}}};
		\node [draw,thick,minimum width=70mm] at (8.6,1.05) (S3x4x1) {\textbf{\scriptsize{\ref{sec:cT} Time Complexity \ref{sec:cS} Storage Volume }}};
				
		\draw [->,line width=1mm] (S2) -- (2.6,4)--(3.3,5)-- (S3);
		\draw [->,line width=0.5mm] (4,4.45) -- (S3x1);
		\draw [->,line width=0.5mm] (4,3.95) -- (S3x2);
		\draw [->,line width=0.5mm] (4,3.45) -- (S3x3);
		\draw [->,line width=0.5mm] (4,4.75) -- (4,1.55)--(S3x4);
		\draw [->,line width=0.5mm] (4.7,3.25) -- (4.7,2.50)--(S3x3x1);
		\draw [->,line width=0.5mm] (4.7,1.35) -- (4.7,1.05)--(S3x4x1);
		
		\node [rectangle,draw,thick,rounded corners=2mm,minimum size=5mm,minimum width=55mm] at (12.5,4.9) (S4) {\textbf{\scriptsize{
			\begin{tabular}{l}
			\ref{sec:Frameworks} Frameworks and Architectures \\
			for Sharing Secrets in the Cloud \\
			\end{tabular}
		}}};
		\draw [->,line width=1mm] (S3) -- (9.7,5);
		
		\node [rectangle,draw,thick,rounded corners=2mm,minimum size=5mm,minimum width=55mm] at (12.5,3.8) (S5) {\textbf{\scriptsize{\ref{sec:Conclusions} Conclusion}}};
		\draw [->,line width=1mm] (S4) -- (S5);
		
	\end{tikzpicture} }

 	\caption{Schematic map of the paper}
	\label{fig:paper-flows}
\end{figure*}
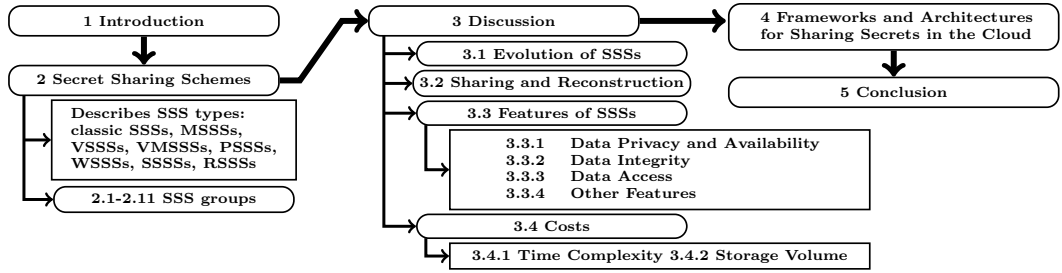  

The remainder of this paper is organized as follows (Figure~\ref{fig:paper-flows}).
Section~\ref{sec:Secret-sharing-schemes} describes the principles of secret sharing and classifies SSSs into eleven groups, whose properties 
are thoroughly detailed. SSSs in a given group are also positioned with respect to one another. In Section~\ref{sec:Discussion}, we compare all surveyed SSSs with respect to data security, queries over shares, and storage and computing costs. 
Moreover, we present SSS-based frameworks that provide secure storage, e.g., databases or data warehouses, in the cloud in Section~\ref{sec:Frameworks}. Finally, Section~\ref{sec:Conclusions} concludes this paper, recaps open research issues and describes sample applications in the cloud.

\section{Secret Sharing Schemes}
\label{sec:Secret-sharing-schemes}

\begin{figure}[hbt]
	\centering
	\subfloat[Sharing process]{\resizebox{0.3\textwidth}{!} {			
	\begin{tikzpicture}
		
			\node [rectangle, thick, draw=black, fill=black!20,minimum size=5mm,rounded corners=2mm,minimum width=45mm] at (0,0) 
			(process) {~};
			\node at (0,0) {\textbf{\small{$\left( t,n\right) $}\scriptsize{ SSS}}};
			
			\node [rectangle,draw,thick,rounded corners=2mm,minimum size=5mm,minimum width=7mm] at (0,0.75) (data) {\textbf{\small{$d$}}};
			\node at (1.3,0.75) {\tiny{Data owner}};
			\draw [-stealth,thick] (data) -- (process);
			
			\node [rectangle,draw,thick,rounded corners=2mm,minimum size=5mm,minimum width=7mm] at (-1.5,-0.75) (E1){\textbf{\small{$e_1$}}};
			\node [rectangle,draw,thick,rounded corners=2mm,minimum size=5mm,minimum width=7mm] at (-0.5,-0.75) (E2){\textbf{\small{$e_2$}}};
			\node [rectangle,draw,thick,rounded corners=2mm,minimum size=5mm,minimum width=7mm] at (1.5,-0.75) (En){\textbf{\small{$e_n$}}};
			\node at (0.5,-0.75) {\textbf{\Large{...}}};
			\draw [-stealth,thick] (-1.5,-0.24) -- (E1);
			\draw [-stealth,thick] (-0.5,-0.24) -- (E2);
			\draw [-stealth,thick] ( 1.5,-0.24) -- (En);
			\node at (-1.5,-1.2) {\textbf{\scriptsize{$PT_1$}}};
			\node at (-0.5,-1.2) {\textbf{\scriptsize{$PT_2$}}};
			\node at ( 1.5,-1.2) {\textbf{\scriptsize{$PT_n$}}};
			
		\end{tikzpicture}				
	}} \\
	\subfloat[Reconstruction process]{\resizebox{0.3\textwidth}{!} {		
	\begin{tikzpicture}
			\node [rectangle, thick, draw=black,fill=black!20,minimum size=5mm,rounded corners=2mm,minimum width=45mm] at (0,0) 
			(process) {~};
			\node at (0,0) {\textbf{\small{$\left( t,n\right) $}\scriptsize{ SSS}}};
			
			\node [rectangle,draw,thick,rounded corners=2mm,minimum size=5mm,minimum width=7mm] at (0,-0.75) (data) {\textbf{\small{$d$}}};
			\node at (1.3,-0.75) {\tiny{Data owner}};
			\draw [-stealth,thick] (process) -- (data);
			
			\node [rectangle,draw,thick,rounded corners=2mm,minimum size=5mm,minimum width=7mm] at (-1.5,1.1) (E1){\textbf{\small{$e_1$}}};
			\node [rectangle,draw,thick,rounded corners=2mm,minimum size=5mm,minimum width=7mm] at (-0.5,1.1) (E2){\textbf{\small{$e_2$}}};
			\node [rectangle,draw,thick,rounded corners=2mm,minimum size=5mm,minimum width=7mm] at (1.5,1.1) (En){\textbf{\small{$e_n$}}};
			\node at (0.5,1.05) {\textbf{\Large{...}}};
			\draw [-stealth,thick] (E1) -- (-1.5,0.6);
			\draw [-stealth,thick] (E2) -- (-0.5,0.6);
			\draw [-stealth,thick] (En) -- ( 1.5,0.6);
			\draw [ultra thick] (-1.6,0.6) -- ( 1.6,0.6);
			\draw [-stealth,ultra thick] (0,0.6) -- (process);
			\node at (-1.5,1.5) {\textbf{\scriptsize{$PT_1$}}};
			\node at (-0.5,1.5) {\textbf{\scriptsize{$PT_2$}}};
			\node at ( 1.5,1.5) {\textbf{\scriptsize{$PT_n$}}};
			\node at (0.95,0.45) {\tiny{select }\scriptsize{$t$}\tiny{ from }\scriptsize{$n$}};
			
		\end{tikzpicture}
	}}
 	\caption{Classical secret sharing}
	\label{fig:SSS}
\end{figure}
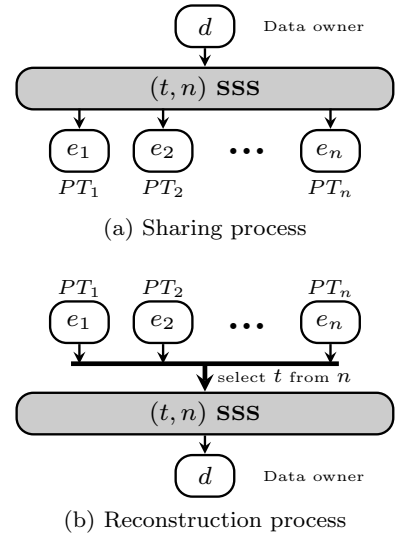

The threshold SSSs we survey in this paper are primarily aimed at enforcing privacy. Individual secret $d$ is divided into $n$ so-called shares $\lbrace e_i\rbrace _{i=0,\cdots,n}$, each share $e_i$ being stored by a different participant (PT) $PT_i$ (Figure~\ref{fig:SSS}(a)). Each share $e_i$ is meaningless to $PT_i$. A subset of $t \leq n$ PTs is required to reconstruct the secret (Figure~\ref{fig:SSS}(b)). Thence, a convenient side effect of SSSs is data availability, since up to $n-t$ PTs may disappear without preventing secret reconstruction. Classical SSSs \cite{Shamir-1979,Blakley-1979,Asmuth-Bloom-1983,He-Dawson-1994,Iftene-2007,Parakh-Kak-2009,Harn-Lin-2010,Parakh-Kak-2011,Liu-et-al-2012} mainly differ in sharing methods, which bear different security properties with different data storage and CPU requirements.

A major drawback of initial SSSs is the multiplication of the initial data volume by the number of PTs. Multi secret sharing schemes (MSSSs) thus aim to reduce computation, storage and data transfer costs by sharing and reconstructing more than one secret at once. Some MSSSs achieve an overall shared data volume (i.e., at all PTs') that is close to that of original secret data. We categorize MSSSs into two types. 

\begin{figure}[hbt]
	\centering
	\subfloat[Sharing process]{\resizebox{0.3\textwidth}{!} {
	\begin{tikzpicture}
		
			\node [rectangle, thick, draw=black, fill=black!20,minimum size=5mm,rounded corners=2mm,minimum width=45mm] at (0,0) 
			(process) {~};
			\node at (0,0) {\textbf{\small{$\left( m,t,n\right) $}\scriptsize{ MSSS type I}}};
			
			\node [rectangle,draw,thick,rounded corners=2mm,minimum size=5mm,minimum width=7mm] at (-1.65,0.75) (D1) {\textbf{\small{$d_1$}}};
			\node [rectangle,draw,thick,rounded corners=2mm,minimum size=5mm,minimum width=7mm] at (-0.8 ,0.75) (D2) {\textbf{\small{$d_2$}}};
			\node [rectangle,draw,thick,rounded corners=2mm,minimum size=5mm,minimum width=7mm] at ( 0.4,0.75) (Dm) {\textbf{\small{$d_m$}}};
			\node at (-0.2,0.75) {\textbf{\Large{...}}};
			\node at (1.5,0.75) {\tiny{Data owner}};
			\draw [-stealth,thick] (D1) -- (-1.65,0.25);
			\draw [-stealth,thick] (D2) -- (-0.8,0.25);
			\draw [-stealth,thick] (Dm) -- ( 0.4,0.25);
			
			\node [rectangle,draw,thick,rounded corners=2mm,minimum size=5mm,minimum width=7mm] at (-1.85,-0.75) (K1){\textbf{\small{$k_1$}}};
			\node at (-1.2,-0.75) {\textbf{\Large{...}}};
			\node [rectangle,draw,thick,rounded corners=2mm,minimum size=5mm,minimum width=7mm] at (-0.6,-0.75) (Kn){\textbf{\small{$k_n$}}};
			\node [rectangle,draw,thick,rounded corners=2mm,minimum size=5mm,minimum width=7mm] at (0.6,-0.75) (E1){\textbf{\small{$c_1$}}};
			\node at (1.2,-0.75) {\textbf{\Large{...}}};
			\node [rectangle,draw,thick,rounded corners=2mm,minimum size=5mm,minimum width=7mm] at (1.85,-0.75) (Ex){\textbf{\small{$c_x$}}};
			\draw [-stealth,thick] (-1.85,-0.25) -- (K1);
			\draw [-stealth,thick] (-0.6 ,-0.25) -- (Kn);
			\draw [-stealth,thick] (0.6 ,-0.25) -- (E1);
			\draw [-stealth,thick] (1.85,-0.25) -- (Ex);
			\node at (-1.85,-1.2) {\textbf{\scriptsize{$PT_1$}}};
			\node at (-0.6,-1.2) {\textbf{\scriptsize{$PT_2$}}};
			\node at ( 1.2,-1.2) {\scriptsize{NB}};
			
		\end{tikzpicture}
	}} \\
	\subfloat[Reconstruction process]{ \resizebox{0.3\textwidth}{!} {
	\begin{tikzpicture}
			\node [rectangle, thick, draw=black ,fill=black!20,minimum size=5mm,rounded corners=2mm,minimum width=45mm] at (0,0) 
			(process) {~};
			\node at (0,0) {\textbf{\small{$\left( m,t,n\right) $}\scriptsize{ MSSS type I}}};
			
			\node [rectangle,draw,thick,rounded corners=2mm,minimum size=5mm,minimum width=7mm] at (-1.65,-0.75) (D1) {\textbf{\small{$d_1$}}};
			\node [rectangle,draw,thick,rounded corners=2mm,minimum size=5mm,minimum width=7mm] at (-0.8 ,-0.75) (D2) {\textbf{\small{$d_2$}}};
			\node [rectangle,draw,thick,rounded corners=2mm,minimum size=5mm,minimum width=7mm] at ( 0.4,-0.75) (Dm) {\textbf{\small{$d_m$}}};
			\node at (-0.2,-0.75) {\textbf{\Large{...}}};
			\node at (1.5,-0.75) {\tiny{Data owner}};
			\draw [-stealth,thick] (-1.65,-0.25) -- (D1);
			\draw [-stealth,thick] (-0.8 ,-0.25) -- (D2);
			\draw [-stealth,thick] ( 0.4 ,-0.25) -- (Dm);
			
			\node [rectangle,draw,thick,rounded corners=2mm,minimum size=5mm,minimum width=7mm] at (-1.85,1.1) (K1){\textbf{\small{$k_1$}}};
			\node at (-1.2,1.05) {\textbf{\Large{...}}};
			\node [rectangle,draw,thick,rounded corners=2mm,minimum size=5mm,minimum width=7mm] at (-0.6,1.1) (Kn){\textbf{\small{$k_n$}}};
			\node [rectangle,draw,thick,rounded corners=2mm,minimum size=5mm,minimum width=7mm] at (0.6,1.1) (E1){\textbf{\small{$c_1$}}};
			\node at (1.2,1.05) {\textbf{\Large{...}}};
			\node [rectangle,draw,thick,rounded corners=2mm,minimum size=5mm,minimum width=7mm] at (1.85,1.1) (Ex){\textbf{\small{$c_x$}}};
			\draw [-stealth,thick] (K1) -- (-1.85,0.6);
			\draw [-stealth,thick] (Kn) -- (-0.6 ,0.6);
			\draw [ultra thick] (-2.2,0.6) -- ( -0.25,0.6);
			\draw [-stealth,ultra thick] (-0.6,0.6) -- (-0.6,0.25);
			\draw [-stealth,thick] (E1) -- (0.6 ,0.25);
			\draw [-stealth,thick] (Ex) -- (1.85,0.25);
			\node at (-1.85,1.5) {\textbf{\scriptsize{$PT_1$}}};
			\node at (-0.6,1.5) {\textbf{\scriptsize{$PT_2$}}};
			\node at ( 1.2,1.5) {\scriptsize{NB}};
			\node at (-1.5,0.45) {\tiny{select~}\scriptsize{$t$}\tiny{~from~}\scriptsize{$n$}};
			
		\end{tikzpicture}
	}}
 	\caption{Multiple secret sharing type~I}
	\label{fig:MSSSI}
\end{figure}
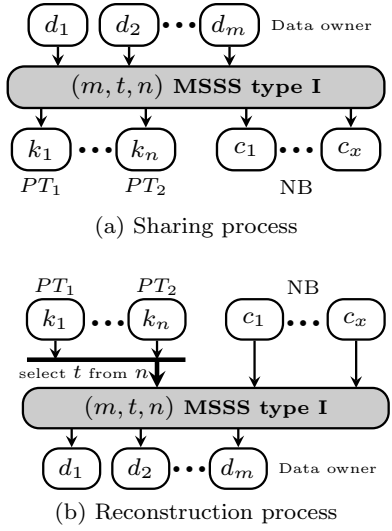

In MSSSs type~I \cite{Yang-et-al-2004,Waseda-Soshi-2012}, data are shared with the help of keys. 
$m$ secrets $\lbrace d_j\rbrace _{j=1,\cdots,m}$ and $n$ keys \\$\lbrace k_i\rbrace _{i=1, \cdots, n}$ are used to construct $x$ shares  $\lbrace c_h\rbrace _{h=1,\cdots,x}$, where $m \leq x$. Shares are stored in a news bulletin board (NB), whereas each key $k_i$ is stored at $PT_i$ (Figure~\ref{fig:MSSSI}(a)). To reconstruct the $m$ secrets, all or some (depending on the MSSS) shares and $t$ keys are used (Figure~\ref{fig:MSSSI}(b)).  

\begin{figure}[hbt]
	\centering
	\subfloat[Sharing process]{\resizebox{0.3\textwidth}{!} {
	\begin{tikzpicture}
			\node [rectangle, thick, draw=black ,fill=black!20,minimum size=5mm,rounded corners=2mm,minimum width=45mm] at (0,0) 
			(process) {~};
			\node at (0,0) {\textbf{\small{$\left( m,t,n\right) $}\scriptsize{ MSSS type II}}};
			
			\node [rectangle,draw,thick,rounded corners=2mm,minimum size=5mm,minimum width=7mm] at (-1.65,0.75) (D1) {\textbf{\small{$d_1$}}};
			\node [rectangle,draw,thick,rounded corners=2mm,minimum size=5mm,minimum width=7mm] at (-0.8 ,0.75) (D2) {\textbf{\small{$d_2$}}};
			\node [rectangle,draw,thick,rounded corners=2mm,minimum size=5mm,minimum width=7mm] at ( 0.4,0.75) (Dm) {\textbf{\small{$d_m$}}};
			\node at (-0.2,0.75) {\textbf{\Large{...}}};
			\node at (1.5,0.75) {\tiny{Data owner}};
			\draw [-stealth,thick] (D1) -- (-1.65,0.25);
			\draw [-stealth,thick] (D2) -- (-0.8,0.25);
			\draw [-stealth,thick] (Dm) -- ( 0.4,0.25);
			
			\node [rectangle,draw,thick,rounded corners=2mm,minimum size=5mm,minimum width=7mm] at (-1.5,-0.75) (E1){\textbf{\small{$e_1$}}};
			\node [rectangle,draw,thick,rounded corners=2mm,minimum size=5mm,minimum width=7mm] at (-0.5,-0.75) (E2){\textbf{\small{$e_2$}}};
			\node [rectangle,draw,thick,rounded corners=2mm,minimum size=5mm,minimum width=7mm] at (1.5,-0.75) (En){\textbf{\small{$e_n$}}};
			\node at (0.5,-0.75) {\textbf{\Large{...}}};
			\draw [-stealth,thick] (-1.5,-0.24) -- (E1);
			\draw [-stealth,thick] (-0.5,-0.24) -- (E2);
			\draw [-stealth,thick] ( 1.5,-0.24) -- (En);
			\node at (-1.5,-1.2) {\textbf{\scriptsize{$PT_1$}}};
			\node at (-0.5,-1.2) {\textbf{\scriptsize{$PT_2$}}};
			\node at ( 1.5,-1.2) {\textbf{\scriptsize{$PT_n$}}};
			
		\end{tikzpicture}
	}} \\
	\subfloat[Reconstruction process]{\resizebox{0.3\textwidth}{!} {		
	\begin{tikzpicture}
			\node [rectangle, thick, draw=black ,fill=black!20,minimum size=5mm,rounded corners=2mm,minimum width=45mm] at (0,0) 
			(process) {~};
			\node at (0,0) { \textbf{\small{$\left( m,t,n\right) $}\scriptsize{ MSSS type II}}};
			
			\node [rectangle,draw,thick,rounded corners=2mm,minimum size=5mm,minimum width=7mm] at (-1.65,-0.75) (D1) {\textbf{\small{$d_1$}}};
			\node [rectangle,draw,thick,rounded corners=2mm,minimum size=5mm,minimum width=7mm] at (-0.8 ,-0.75) (D2) {\textbf{\small{$d_2$}}};
			\node [rectangle,draw,thick,rounded corners=2mm,minimum size=5mm,minimum width=7mm] at ( 0.4,-0.75) (Dm) {\textbf{\small{$d_m$}}};
			\node at (-0.2,-0.75) {\textbf{\Large{...}}};
			\node at (1.5,-0.75) {\tiny{Data owner}};
			\draw [-stealth,thick] (-1.65,-0.25) -- (D1);
			\draw [-stealth,thick] (-0.8 ,-0.25) -- (D2);
			\draw [-stealth,thick] ( 0.4 ,-0.25) -- (Dm);
			
			\node [rectangle,draw,thick,rounded corners=2mm,minimum size=5mm,minimum width=7mm] at (-1.5,1.1) (E1){\textbf{\small{$e_1$}}};
			\node [rectangle,draw,thick,rounded corners=2mm,minimum size=5mm,minimum width=7mm] at (-0.5,1.1) (E2){\textbf{\small{$e_2$}}};
			\node [rectangle,draw,thick,rounded corners=2mm,minimum size=5mm,minimum width=7mm] at (1.5,1.1) (En){\textbf{\small{$e_n$}}};
			\node at (0.5,1.05) {\textbf{\Large{...}}};
			\draw [-stealth,thick] (E1) -- (-1.5,0.6);
			\draw [-stealth,thick] (E2) -- (-0.5,0.6);
			\draw [-stealth,thick] (En) -- ( 1.5,0.6);
			\draw [ultra thick] (-1.6,0.6) -- ( 1.6,0.6);
			\draw [-stealth,ultra thick] (0,0.6) -- (process);
			\node at (-1.5,1.5) {\textbf{\scriptsize{$PT_1$}}};
			\node at (-0.5,1.5) {\textbf{\scriptsize{$PT_2$}}};
			\node at ( 1.5,1.5) {\textbf{\scriptsize{$PT_n$}}};
			\node at (0.9,0.45) {\tiny{select }\scriptsize{$t$}\tiny{ from }\scriptsize{$n$}};
			
		\end{tikzpicture}
	}}
 	\caption{Multiple secret sharing type~II}
	\label{fig:MSSSII}
\end{figure}
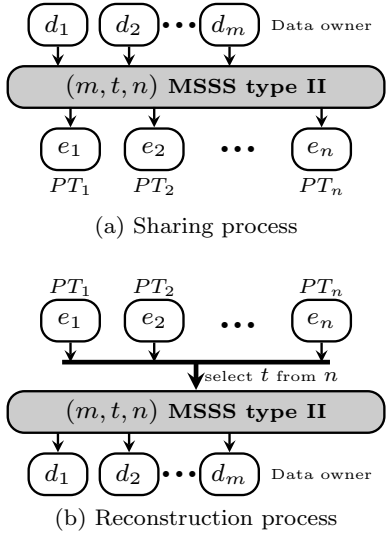

In MSSSs type~II \cite{Chan-Chang-2005,Runhua-et-al-2008,Liu-et-al-2012,Takahashi-Iwamura-2013}, $m$ secrets \\$\lbrace d_j\rbrace _{j=1,\cdots,m}$ are divided into $n$ shares $\lbrace e_i\rbrace _{i=1,\cdots,n}$, where $m\leq t\leq n$. In case $m>t$,  secrets are first organized into blocks that are fewer than $t$. Then, each block is  divided into $n$ shares at once. Finally each share $e_i$ is stored by $PT_i$ (Figure~\ref{fig:MSSSII}(a)). As in SSSs, reconstructing the secrets requires $t$ PTs (Figure~\ref{fig:MSSSII}(b)).  

SSSs and MSSSs assume that all players, i.e., PTs and NB, are honest and always provide valid information (data and keys). However, in reality, they might not, intentionally or not. Thus, verifiable secret sharing schemes (VSSSs) \cite{Pedersen-1991,Tang-Yao-2008,Yue-Hong-2009,Hwang-Chang-1998,Zhao-et-al-2012} and verifiable multi secret sharing schemes (VMSSSs) \cite{Eslami-Ahmadabadi-2010,Zhao-et-al-2007,Dehkordi-Mashhadi-2008a,Dehkordi-Mashhadi-2008b,Wang-et-al-2011,Eslami-Rad-2012,Chen-et-al-2012,Li-et-al-2012,Hu-et-al-2012,Lin-We-1999,Chang-et-al-2005,Shao-Cao-2005,Wei-et-al-2007,Das-Adhikari-2010,Bu-Yang-2012,DB-Attasena-et-al-2014-J,DB-Attasena-et-al-2014-P,Shi-et-al-2007} verify the correctness of data and/or keys before or after reconstruction. Therefore, VSSSs and VMSSSs enforce data integrity in addition to privacy and availability. 

Eventually, some SSSs aim at speciﬁc goals. Proactive secret sharing schemes (PSSSs) are based on classical SSSs or VSSSs but, in addition, periodically refresh shares \cite{PSSS1,PSSS02,PSSS02-2,PSSS05,PSSS15,PSSS16,PSSS17}. Refreshing consists in generating a random number at each PT's and sharing it at all other PTs' to modify existing shares. ln most PSSSs \cite{PSSS1,PSSS02-2,PSSS15,PSSS16,PSSS17,PSSS17}, refreshing is synchronous, i.e., shares cannot be reconstructed during the process, but there are also asynchronous refreshing protocols \cite{PSSS02,PSSS05} that allow share reconstruction at all times. All PSSSs include a share verification process that verifies whether shares are up-to-date when refreshing. If shares are obsolete, they may be safely deleted or recovered from other shares. Since shares are periodically refreshed even if 
secrets have not been updated, an intruder has little time to compromise them.

However, the refreshing process in PSSSs induces extra costs, i.e., computing costs for periodically sharing random numbers among PTs and modifying shares (time complexity is $O(mn^2)$ \cite{PSSS1}); and high communication costs for commuting PTs with each other, whose cost is at least $n$ times that of sharing secrets. Because of these costs, and since PSSSs reuse the data sharing and reconstruction processes of the classical SSSs that are detailed in this section, we do not survey PSSSs further.

Weighted secret sharing schemes (WSSSs) extend classical SSSs by introducing a priority among PTs by assigning to each PT a weight, i.e., the number of shares it stores \cite{WSSS1,WSSS99,WSSS14,WSSS16}. More precisely, in these schemes, any secret $d$ is divided into $w$ shares such that $w\geq n$. Each PT$_i$ holds $w_i$ shares such that $w = \sum_{i=1}^n w_i$. If $n = w$ or $w_i=1$ $\forall i$, we fall back to a classical SSS. Secret reconstruction is only possible by a group of PTs holding at least $t$ shares, with $w_i<t\leq w$ $\forall i$. One single PT cannot reconstruct the secret, since $w_i<t$ $\forall i$. 

Social secret sharing schemes (SSSSs) extend from WSSSs by allowing weights to be adjusted depending on the situation, e.g., if some PTs are found insincere \cite{SSSS-Nojoumian-et-al-2010,SSSS-Nojoumian-et-al-2012,SSSS-Zheng-et-al-2012,SSSS-Nojoumian-et-al-2012-2} .
Even though WSSSs and SSSSs bring in a more flexible PT management, they induce a higher share volume, i.e., at least $n$ times the original  data volume \emph{vs.} at most $n$ times for previous SSSs, supposing that individual shares use up the same volume as secrets. Thus, we do not survey them further.


Finally, 
function secret sharing schemes (FSSSs) \cite{FSS15,FSS15D,FSS16} aim at protecting data transfers over networks when keyword search is performed on outsourced, replicated data. A  function $f$ is shared into $n$ functions $f_1,\cdots,f_n$ such that $f = \sum_{i=1}^n f_i$. Each function $f_i$ is associated with  a data  node akin to a participant PT$_i$ in classical secret sharing. When the user issues a search query with some keyword $k$, $f_i(k)$ is sent to PT$_i$ $\forall i=1 \cdots n$. Then data at each PT$_i$ are matched with $f_i(k)$. The local result $R_i$ is shared as $f_i(R_i)$ and sent back to the user, who can finally reconstruct a global result with $t \leq n$ values of $f_i(R_i)$.
However, FSSSs do not fit in our data outsourcing scenario since data are replicated in clear form.  Thus, we do not survey them further. Yet, FSSSs are quite recent and hybridizing them with other SSSs surveyed in this section could help solve this issue.


We categorize SSSs into eleven groups (Table~\ref{tab:Groups-of-Secret-sharing}) with respect to their basic type, i.e., SSSs and MSSSs types I and II, as well as eventual data or key verification. We survey all groups in the following subsections. Moreover, we introduce the parameters and notations used throughout this section in Table~\ref{tab:Parameters-of-all-scheme}.

\begin{table}[hbt]
	\renewcommand{\arraystretch}{1.3}
	\caption{Classification of secret sharing schemes}
	\label{tab:Groups-of-Secret-sharing}
	\centering
		\resizebox{0.5\textwidth}{!} {
    		\begin{tabular}{  |c|c|c|c|c|c|}
    		
    			\cline{3-5}
    			\multicolumn{2}{c|}{} &
    			\textbf{SSSs} &
    			\textbf{MSSSs type~I} &
    			\textbf{MSSSs type~II} &
    			\multicolumn{1}{c}{} \rule{0pt}{10pt} \\ [1ex] \cline{1-6}
    			
				\multirow{14}{*}{\textbf{\begin{turn}{90} Verification \end{turn}}} & 
				\multirow{5}{*}{\textbf{None}} &
				\textbf{Group 1} &
				\textbf{Group 2} &
				\textbf{Group 3} &
				\multirow{5}{*}{\textbf{\begin{turn}{270} \hspace{-0.2cm}\scriptsize{SSSs/MSSSs} \end{turn}}} \\
    			
    			& & 
    			\cite{Shamir-1979,Blakley-1979,Asmuth-Bloom-1983} &
    			\cite{Yang-et-al-2004,Waseda-Soshi-2012} &
    			\cite{Chan-Chang-2005,Runhua-et-al-2008,Takahashi-Iwamura-2013}   
    			& \\
    			
    			& & \cite{He-Dawson-1994,Iftene-2007,Harn-Lin-2010} & & \cite{Liu-et-al-2012} MSSS  & \\
    			
    			& & \cite{Parakh-Kak-2009,Parakh-Kak-2011}& & & \\
    			& & \cite{Liu-et-al-2012} SSS & & & \\
    			\cline{2-6}
				
    			& \multirow{2}{*}{\textbf{Data}} &
    			\textbf{Group 4} &
    			\textbf{Group 7} & 
    			\textbf{Group 10} & 
    			\multirow{9}{*}{\textbf{\begin{turn}{270} VSSSs/VMSSSs \end{turn}}} \\ 
    			
    			& &
    			\cite{Pedersen-1991,Tang-Yao-2008,Yue-Hong-2009} & 
    			\cite{Eslami-Ahmadabadi-2010} & 
    			\cite{DB-Attasena-et-al-2014-J,DB-Attasena-et-al-2014-P} & \\ 
    			\cline{2-5} 
    				
    			& \multirow{4}{*}{\textbf{Keys}} & 
    			\textbf{Group 5} & 
    			\textbf{Group 8} & & \\ 
    			
    			& & \cite{Hwang-Chang-1998} & 
    			\cite{Zhao-et-al-2007,Dehkordi-Mashhadi-2008a,Dehkordi-Mashhadi-2008b} & & \\
    			
    			& & & \cite{Wang-et-al-2011,Eslami-Rad-2012,Chen-et-al-2012} & & \\
    			
    			& & & \cite{Li-et-al-2012},\cite{Hu-et-al-2012}-I\&II & & \\
    			\cline{2-5}
    			
    			& \textbf{Data} & 
    			\textbf{Group 6} &
    			\textbf{Group 9} &
    			\textbf{Group 11} & \\ 
    			
    			& \textbf{\&} & 
    			\cite{Zhao-et-al-2012} & 
    			\cite{Lin-We-1999,Chang-et-al-2005,Shao-Cao-2005} & 
    			\cite{Shi-et-al-2007} & \\
    			
    			& \textbf{keys}& & \cite{Wei-et-al-2007,Das-Adhikari-2010,Bu-Yang-2012} & & \\
    			\cline{1-6}
    			 
    		\end{tabular}
    	} 
\end{table}

\begin{table}[!hbtp]
	\renewcommand{\arraystretch}{1.3}
	\caption{Secret sharing schemes' parameters}
	\label{tab:Parameters-of-all-scheme}
	\centering
		\resizebox{0.49\textwidth}{!} {    		
    		
    		\begin{tabular}{ | c | p{55mm} |}
    			\hline 
    			\textbf{Parameter} & \textbf{Definition} \rule{0pt}{3mm}\\ 
    			\hline  
    			$m$ & Number of secrets \\ \hline
    			$D$ & Secret data such that $D=\{ d_1,\cdots,d_m\}$ and $D=\{ b_1,\cdots,b_o\}$ \\ \hline
    			$d$ & Secret in integer format \\ \hline	
				$\Vert d\Vert $ & Storage size of $d$ \\ \hline							
				$d_j$ & $j^{th}$ element of $D$ in integer format \\ \hline	    			
    			$n$ & Number of PTs \\ \hline
    			$t$ & Number of shares necessary for reconstructing the secret \\ \hline				
    			$\gamma$ & Number of PTs of the first group in \cite{Takahashi-Iwamura-2013} \\ \hline
    			$PT_i$ & PT number $i$ \\ \hline
				$ID_i$ & Identifier of $PT_i$ \\ \hline				
    			$g$ & Number of groups of PTs \\ \hline
    			$G_r$ & $r^{th}$ group of PTs such that $G_r\subseteq\{ PT_i\} _{i=1,\cdots,n}$ and $G_r=\{ PT_{r,1},\cdots,PT_{r,g}\} $ \\ \hline		
    			$PT_{r,v}$ & PT number $v$ of $G_r$ \\ \hline				
				$ID_{r,v}$ & Identifier of $PT_{r,v}$ of $G_r$ \\ \hline					
    			$o$ & Number of data blocks \\ \hline				
    			$b_l$ & $l^{th}$ block of $D$ such that $b_l=\{ d_{l,1},\cdots,d_{l,t}\}$ with fixed-sized blocks and $b_l=\{ d_{l,1},\cdots,d_{l,t_{l}}\}$ with variable-sized blocks \\ \hline
    			$t_l$ & Number of shares necessary for reconstructing the secret in $b_l$ (in case of variable-sized blocks) \\ \hline				
				$d_{l,q}$ & $q^{th}$ element of $b_{l}$ in integer format \\ \hline
				$e_i$ & Share stored at $PT_i$ \\ \hline
				$e_{j,i}$ & $j^{th}$ share stored at $PT_i$ \\ \hline
				$e_{l,i}$ & Share of $b_l$ stored at $PT_i$ \\ \hline
				$c_h$ & $h^{th}$ share stored at the NB \\ \hline
				$c_{j,h}$ & $h^{th}$ share of $d_j$ stored at the NB \\ \hline
				$c_{l,h}$ & $h^{th}$ share of $b_l$ stored at the NB \\ \hline
				$c_{l,q,h}$ & $h^{th}$ share of $d_{l,q}$ in $b_l$ stored at the NB \\ \hline
				$c_{r,l,h}$ & $h^{th}$ share of $b_l$ from  $G_r$ stored at the NB \\ \hline
				$k_i$ & Key stored at $PT_i$ \\ \hline
				$k_{i,q}$ & Key number $q$ stored at $PT_i$ \\ \hline
				$k_{r,i}$ & Key stored at $PT_i$ of $G_r$ \\ \hline
				$\Vert k\Vert $ & Storage size of keys \\ \hline				
				$s\_d_i$ & Signature stored at $PT_i$ \\ \hline 
				$s\_d_l$ & Signature of $d_l$ \\ \hline
				$s\_d_{l,i}$ & Signature of $b_l$ stored at $PT_i$ \\ \hline
				$s\_d_{l,q}$ & Signature number $q$ of $b_l$ \\ \hline
				$s\_k_i$ & Signature of $PT_i$'s key \\ \hline
				$s\_k_{r,v}$ & Signature of $PT_{r,v}$'s key of $G_r$ \\ \hline
				$\Vert s\Vert $ & Storage size of signatures \\ \hline				
				$p, p_1, p_2\ldots$ & Big prime numbers \\ \hline
				$A, A_1, A_2\ldots$ & Matrices \\ \hline
				$f, f_1, f_2\ldots$ & Functions \\ \hline
				$H, H_1, H_2\ldots$ & Hash functions \\ \hline
    		\end{tabular}
    	}
\end{table}

\subsection{Group 1: Classical Secret Sharing Schemes}
\label{classic-sss}



\begin{figure}[hbt]
	\centering
	\resizebox{0.4\textwidth}{!} {

		\begin{tikzpicture}[domain=0:6]
			\draw[-stealth,thick] (-0.1,0) -- (6.4,0) node[right] {\textbf{\scriptsize{$i$}}};
			\draw[-stealth,thick] (0,-0.7) -- (0,3.4) node[above] {\textbf{\scriptsize{$f(i)$}}};
			
			\node at (1,-0.2) {\scriptsize{1}};
			\node at (2,-0.2) {\scriptsize{2}};
			\node at (3,-0.2) {\scriptsize{3}};
			\node at (4,-0.2) {\scriptsize{4}};
			\node at (5,-0.2) {\scriptsize{5}};
			\node at (6,-0.2) {\scriptsize{6}};
			\draw [thick] (1,0) -- (1,-0.05);
			\draw [thick] (2,0) -- (2,-0.05);
			\draw [thick] (3,0) -- (3,-0.05);
			\draw [thick] (4,0) -- (4,-0.05);
			\draw [thick] (5,0) -- (5,-0.05);
			\draw [thick] (6,0) -- (6,-0.05);
			\node at (-0.25,-0.5) {\scriptsize{-10}};
			\node at (-0.2,0.0) {\scriptsize{0}};
			\node at (-0.2,0.5) {\scriptsize{10}};
			\node at (-0.2,1.0) {\scriptsize{20}};
			\node at (-0.2,1.5) {\scriptsize{30}};
			\node at (-0.2,2.0) {\scriptsize{40}};
			\node at (-0.2,2.5) {\scriptsize{50}};
			\node at (-0.2,3.0) {\scriptsize{60}};
			\draw [thick] (0,-0.5) -- (-0.05,-0.5);
			\draw [thick] (0,0.5) -- (-0.05,0.5);
			\draw [thick] (0,1.0) -- (-0.05,1.0);
			\draw [thick] (0,1.5) -- (-0.05,1.5);
			\draw [thick] (0,2.0) -- (-0.05,2.0);
			\draw [thick] (0,2.5) -- (-0.05,2.5);
			\draw [thick] (0,3.0) -- (-0.05,3.0);

			\draw[color=red,ultra thick] plot (\x,\x*\x*\x*0.05-5*\x*\x*0.05+2*\x*0.05+4*0.05) ;
			\node at (3,2) {\color{red} \textbf{\scriptsize{$y=f\left(x\right)=x^{3}-5x^{2}+2x+4$}}};
			
			\node at (-0.5,0.2) {\color{red} \textbf{\scriptsize{$d=4$}}};
			\node at (0,0.2) {\color{red} \textbf{\Huge{.}}};
			
			\node at (1, 2*0.05) {\color{blue} \textbf{\Huge{.}}};
			\node at (2,-4*0.05) {\color{blue} \textbf{\Huge{.}}};
			\node at (3,-8*0.05) {\color{blue} \textbf{\Huge{.}}};
			\node at (4,-4*0.05) {\color{blue} \textbf{\Huge{.}}};
			\node at (5,14*0.05) {\color{blue} \textbf{\Huge{.}}};
			\node at (6,52*0.05) {\color{blue} \textbf{\Huge{.}}};
		\end{tikzpicture}}
	\caption{Secret sharing by polynomial interpolation}
	\label{fig:SSS+Sharmir}
\end{figure}
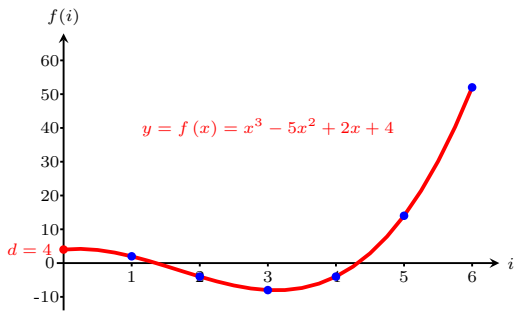

The very first $(t,n)$ SSS \cite{Shamir-1979} enforces data security by using a random polynomial (Equation~\ref{eq:Shamir-1979-Eq1}). This polynomial is generated over a finite field such that coefficient $c_0$ is the secret and other coefficients $c_{u=1,\cdots,t-1}$ are random integers. 
Then, each share $e_i$ is created by Equation~\ref{eq:Shamir-1979-Eq2} and stored at $PT_i$. A number $t \leq n$ of PTs can reconstruct the original polynomial by Lagrange interpolation over a finite field, which enforces data availability even if $n - t$ PTs fail. A sample application of this scheme is given in Figure~\ref{fig:SSS+Sharmir}, where $t=4$ and $n=6$. The random polynomial of degree $t-1=3$ is $e_i = f(i) = i^3-5i^2+2i+4$, where 4 is the secret. The six shares $\lbrace (i,e_i)\rbrace_{i=1,\cdots,6}$ (plotted in blue) are (1,2), (2,-4), (3,-8), (4,-4), (5,14) and (6,52).

\begin{eqnarray}
	f(i) &=& \sum_{u=0}^{t-1}c_u \times i^u 
	\label{eq:Shamir-1979-Eq1} \\
	e_i &=& f(i) 
	\label{eq:Shamir-1979-Eq2}
\end{eqnarray}

\begin{figure}
	\begin{center}
	\includegraphics[width=0.12\textwidth]{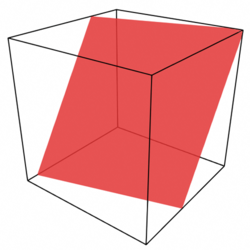}
	\includegraphics[width=0.12\textwidth]{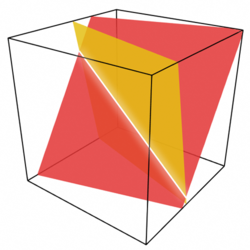}
	\includegraphics[width=0.12\textwidth]{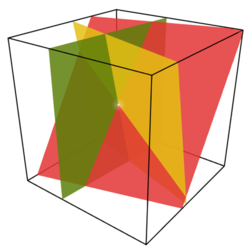}
	\end{center}
	
	\centering\textit{\scriptsize{from: http://en.wikipedia.org/wiki/Secret\_sharing}} 
	\caption{Secret sharing through hyperplan intersection}	
	\label{fig:SSS+Blakley}
\end{figure}

In Blakley's SSS \cite{Blakley-1979}, each PT is associated with an hyperplane in a $t$-dimensional space over a finite field. Hyperplanes, i.e., shares, intersect in a point that is the secret, which can be reconstructed by solving the hyperplanes' equation system. A sample application of this scheme is given in Figure~\ref{fig:SSS+Blakley}, where $t=2$ and $n=3$ (there are thus three hyperplans). 

\cite{Asmuth-Bloom-1983} exploits the Chinese remainder theorem \cite{crt}. 
First, $n+1$ uniquely relatively primes\footnote{Uniquely relatively primes are random prime numbers that are related to each other by some conditions.} $\{ p_i\} _{i=0,\cdots,n}$ are determined such that $p_0<p_1<\cdots<p_n$ and $\prod_{i=1}^{t} p_i>p_0 \prod_{i=1}^{t-1}p_{n-i+1}$. Then, $n$ shares $\lbrace e_{l}\rbrace_{l=1,\cdots,n}$ are created by Equations~\ref{eq:Asmuth-and-Bloom-1983-Eq1} and \ref{eq:Asmuth-and-Bloom-1983-Eq2}, where $u$ is a random positive integer. Finally, secret $d$ is reconstructed from $t$ shares by Equations~\ref{eq:Asmuth-and-Bloom-1983-Eq3} and \ref{eq:Asmuth-and-Bloom-1983-Eq4}. 

\begin{eqnarray}
	e_i &=& y\bmod p_i
	\label{eq:Asmuth-and-Bloom-1983-Eq1} \\
	y &=& d+u\times p_0
	\label{eq:Asmuth-and-Bloom-1983-Eq2} \\
	d &=& y\bmod p_0
	\label{eq:Asmuth-and-Bloom-1983-Eq3} \\
	y &\equiv & e_i\bmod m_i
	\label{eq:Asmuth-and-Bloom-1983-Eq4}
\end{eqnarray}

All subsequent SSSs extend the three foundation schemes above.
\cite{Iftene-2007} extends from \cite{Asmuth-Bloom-1983} to reduce the size of shares. Moreover, this SSS can reconstruct  a secret from $t$ or more shares, whereas previous schemes exploit exactly $t$ shares. In the sharing process, the secret is split in $t$. Share creation from the $t$ splits and secret reconstruction proceed as in \cite{Asmuth-Bloom-1983}.
All other SSSs seek to improve polynomial interpolation.

\cite{Parakh-Kak-2009} proceeds in two steps. 
First, secret $d$ is divided into $t$ intermediate shares $\{ u_v\} _{v=1,\cdots,t}$ by mapping $d$ to the x-axis of a random polynomial. Second, these $t$ shares are divided  again into $n$ actual shares $\{ e_i\} _{i=1,\cdots,t}$ by Equation~\ref{eq:Parakh-and-Kak-2009.a-Eq1}, where $A_1$ is an $n \times t$ random matrix. 
Secret $d$ is reconstructed from a polynomial of degree $t$ created by Equation~\ref{eq:Parakh-and-Kak-2009.a-Eq2}. $\{ u_v\} _{v=1,\cdots,t}$ are reconstructed by Equation~\ref{eq:Parakh-and-Kak-2009.a-Eq3}, where $A_2$ is a $t \times t$ inverse matrix seeded from $t$ rows of matrix $A_1$.

\begin{eqnarray} 
	\left[ e_1,\cdots, e_n \right]^T & = & A_1 \times \left[ u_1,\cdots,u_t\right]^T
	\label{eq:Parakh-and-Kak-2009.a-Eq1} \\
	\prod_{a=1}^{t} (x-u_a) & \equiv & 0 \bmod p 
	\label{eq:Parakh-and-Kak-2009.a-Eq2}  \\
	\left[ u_1,\cdots,u_t\right]^T & = & A_2 \times \left[e_1,\cdots,e_t\right]^T
	\label{eq:Parakh-and-Kak-2009.a-Eq3}
\end{eqnarray}

\begin{figure} [hbt]
	\centering
	\resizebox{0.48\textwidth}{!} {

		\begin{tikzpicture}[domain=0:6]
			\draw[-stealth,thick] (-0.05,0) -- (6.4,0) node[right] {\textbf{\scriptsize{$x$}}};
			\draw[-stealth,thick] (0,-0.05) -- (0,3.4) node[above] {\textbf{\scriptsize{$f(x)$}}};
			\draw[thick,blue] (-0.5,31/10) -- (6,31/10)node[right] {\textbf{\scriptsize{$p$}}};
			\draw[color=black,ultra thick,red] plot [smooth] coordinates { 
			(0/5,21/10) (1/5,2/10) (2/5,9/10) (3/5,17/10) (4/5,1/10)
			(5/5,29/10) (6/5,14/10) (7/5,24/10) (8/5,3/10) (9/5,19/10)
			(10/5,16/10) (11/5,0/10) (12/5,8/10) (13/5,15/10) (14/5,27/10)
			(15/5,19/10) (16/5,28/10) (17/5,29/10) (18/5,28/10) (19/5,0/10)
			(20/5,13/10) (21/5,11/10) (22/5,0/10) (23/5,17/10) (24/5,6/10)
			(25/5,4/10) (26/5,17/10) (27/5,20/10) (28/5,19/10) (29/5,20/10)
			(30/5,29/10) (31/5,21/10) };
			
			\node at (-0.45,26/10) {\color{blue} \textbf{\scriptsize{$d$}}};
			\draw[thick,blue] (-0.5,21/10) -- (0,21/10);
			\draw[-stealth,thick,blue] (-0.45,27/10) -- (-0.45,31/10);
			\draw[-stealth,thick,blue] (-0.45,25/10) -- (-0.45,21/10);
			
			\node at (18/5,0.15) {\color{blue} \textbf{\scriptsize{$u_1$}}};
			\node at (19/5,0) {\color{blue} \textbf{\Huge{.}}};
			
			\node at (23/5,0.15) {\color{blue} \textbf{\scriptsize{$u_2$}}};
			\node at (22/5,0) {\color{blue} \textbf{\Huge{.}}};
			
			\node at (12.3/5,0.15) {\color{blue} \textbf{\scriptsize{$u_3$}}};
			\node at (11/5,0) {\color{blue} \textbf{\Huge{.}}};
			
			\node at (1,-0.2) {\scriptsize{5}};
			\node at (2,-0.2) {\scriptsize{10}};
			\node at (3,-0.2) {\scriptsize{15}};
			\node at (4,-0.2) {\scriptsize{20}};
			\node at (5,-0.2) {\scriptsize{25}};
			\node at (6,-0.2) {\scriptsize{30}};
			\draw [thick] (1,0) -- (1,-0.05);
			\draw [thick] (2,0) -- (2,-0.05);
			\draw [thick] (3,0) -- (3,-0.05);
			\draw [thick] (4,0) -- (4,-0.05);
			\draw [thick] (5,0) -- (5,-0.05);
			\draw [thick] (6,0) -- (6,-0.05);
			\node at (-0.2,0.0) {\scriptsize{~0}};
			\node at (-0.2,0.5) {\scriptsize{~5}};
			\node at (-0.2,1.0) {\scriptsize{10}};
			\node at (-0.2,1.5) {\scriptsize{15}};
			\node at (-0.2,2.0) {\scriptsize{20}};
			\node at (-0.2,2.5) {\scriptsize{25}};
			\node at (-0.2,3.0) {\scriptsize{30}};
			\draw [thick] (0,0.5) -- (-0.05,0.5);
			\draw [thick] (0,1.0) -- (-0.05,1.0);
			\draw [thick] (0,1.5) -- (-0.05,1.5);
			\draw [thick] (0,2.0) -- (-0.05,2.0);
			\draw [thick] (0,2.5) -- (-0.05,2.5);
			\draw [thick] (0,3.0) -- (-0.05,3.0);
		\end{tikzpicture}}
	\caption{\cite{Parakh-Kak-2009}'s secret mapping step}
	\label{fig:SSS-Online-data-storage+ex-graph}
\end{figure}
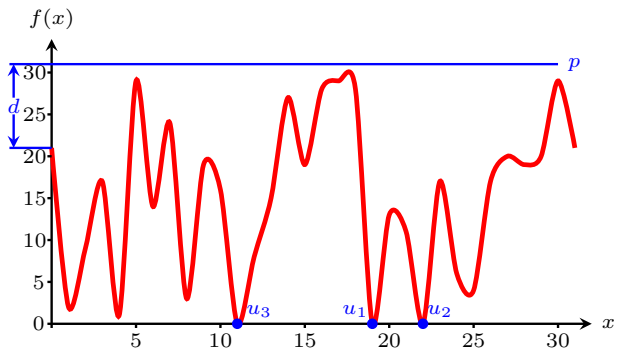

The second step enforces availability and is optional.
A sample application of the first step is given in Figure~\ref{fig:SSS-Online-data-storage+ex-graph}, where $d=10$ and $t=3$. The polynomial equation of degree 3 $(x-u_1)(x-u_2)(x-u_3)\equiv x^3-21x^2+x-10 \equiv 0 \bmod 31$ is created with the help of prime $p=31$ and random positive integers $u_1=19$, $u_2=22$ and $u_3=11$, where $u_1,u_2,u_3$ match with condition $u_3 \equiv d \times (u_1 \times u_2 )\bmod p$.

In \cite{Parakh-Kak-2011}, a secret $d$ is split into $t-1$ smaller data units $\lbrace u_v\rbrace_{v=1,\cdots,t-1}$ to reduce global share volume. Then, a polynomial equation of degree $t-1$ is created by running recursive functions $t-1$ times  (Equation~\ref{eq:Parakh-and-Kak-2011-Eq1}, where $y$ is a random integer) to improve security. Next, $n$ shares $\lbrace e_i\rbrace_{i=1,\cdots,n}$ are created by Equation~\ref{eq:Parakh-and-Kak-2011-Eq2}. Finally, data are reconstructed through $t-1$ steps by Lagrange interpolation.

\begin{eqnarray}
	f_v(x) &=& \begin{cases}
		\begin{array}{ll}
			u_v+ y\times x & \text{if~} v=1 \\
			u_v+\sum_{w=1}^{v}f_{v-1}(w)\times x^w~ & \text{otherwise}
		\end{array}
	\end{cases}
	\label{eq:Parakh-and-Kak-2011-Eq1} \\
	e_i &=& f_{t-1}(i)
	\label{eq:Parakh-and-Kak-2011-Eq2}
\end{eqnarray}

\cite{He-Dawson-1994} extends from \cite{Shamir-1979} to guarantee the $t$-consistency of shares, i.e., any subset of $t$ shares or more always reconstruct the same  secret. A random polynomial function $f(x)$ is created as in \cite{Shamir-1979} ($f(0)=d$). However, $k_{i,1}$ and $k_{i,2}$ are random keys stored at $PT_i$ and $k_{i,2}$ is do not need to be distinct from each other. Next, $n$ shares $\lbrace c_i\rbrace _{i=1,\cdots,n}$ are created by Equation~\ref{eq:He-Dawson-1994} and stored on the NB. Secret $d$ can be reconstructed by Lagrange interpolation from $t$ pairs $\lbrace k_{i,1},c_i+k_{i,2}\rbrace$.

\begin{equation}
	c_i = f(k_{i,1})-k_{i,2}
	\label{eq:He-Dawson-1994}
\end{equation}

\cite{Harn-Lin-2010,Liu-et-al-2012} extend from \cite{Pedersen-1991} (Section~\ref{G4}). 
However, none verifies the correctness of shares. In addition, both approaches verify a strong $t$-consistency property. The verification processes guarantee that any subset of $t$ shares or more (created by summing $n$ random polynomial functions of degree $t-1$ in \cite{Pedersen-1991}) always reconstruct the same secret, but that any subset of $t$ shares or fewer cannot. Verification time is slower in \cite{Harn-Lin-2010} than in \cite{Liu-et-al-2012}.

Eventually, in $(t, L, n)$ threshold ramp SSSs (RSSSs) introduced by Blackley \cite{DBLP:conf/crypto/BlakleyM84}, the secret cannot be reconstructed from $t-L$ or less shares (vs. $t-1$ or less in above SSSs), with $1 \leq \ell \leq L-1$ shares being allowed to leak information about the secret. Thus, RSSSs propose a tradeoff between security and efficiency (measured by entropy) \cite{DBLP:journals/ipl/IwamotoY06}. Let $H(d)$ and $H(e_i)_{i=1,\cdots,n}$ be the entropy of the secret and its shares, respectively. In SSSs, $H(e_i) \geq H(d)$, while in RSSSs, $H(e_i) = H(d) \div L$. \cite{DBLP:journals/ipl/IwamotoY06} also introduces the notion of strong and weak RSSSs, and shows that Shamir-based SSSs may be weak. Yet, most of the following RSSSs still extend Shamir's SSS.

\subsection{Group 2: Multi Secret Sharing Schemes Type~I}

The first $(m,t,n)$ MSSS type I \cite{Yang-et-al-2004} 
extends from \cite{Shamir-1979} to reduce share volume and execution time. All secrets are shared at once, with share volume being controlled to remain close to that of secrets. To share $m$ secrets $\lbrace d_j \rbrace _{j=1,\cdots,m}$ among $n$ PTs, $n$ keys $\lbrace k_i\rbrace_{i=1,\cdots,n}$ are created with a two-variables one-way function. Then, a polynomial (Equation~\ref{eq:Shamir-1979-Eq1}) is created over a finite field \cite{Shamir-1979}, with a degree  $w=\max (m,t)-1$. 

Moreover, coefficients $\lbrace u_j\rbrace_{j=1,\cdots,m}$ are secrets \\$\lbrace d_j\rbrace_{j=1,\cdots,m}$ and other coefficients $\lbrace u_j\rbrace_{j=(m+1),\cdots,t}$ are random integers. Next, $m+n-t$ shares $\lbrace c_h\rbrace_{h=1,\cdots,(m+n-t)}$ are generated by Equation~\ref{eq:Yang-et-al-2004-Eq2} and are published on a NB. Finally, secrets are reconstructed by Lagrange interpolation from $t$ or more keys and $w$ shares. 

\begin{eqnarray}
	c_{h} &=& \begin{cases}
		\begin{array}{ll}
			f\left(k_{h}\right) & \text{if~} 1\leq h\leq n \\
			f\left(H(h)\right)~ & \text{if~} n+1\leq h\leq n+m-t
		\end{array}
	\end{cases}
	\label{eq:Yang-et-al-2004-Eq2}
\end{eqnarray}

\cite{Waseda-Soshi-2012} 
extends from \cite{Shi-et-al-2007} (Section~\ref{G11}) by reducing execution time and dynamically adjusting data block size. In the sharing process, secrets are organized into $o$ unfixed size blocks $\lbrace b_l \rbrace _{l=1,\cdots,o}$. Data block $b_l$ stores $t_l$ secrets $\lbrace d_{l,q} \rbrace _{q=1,\cdots,t_l}$. Next, keys $k_i$ are randomly selected and matrix $A=[a_{x,y}]_{n \times \max (t_1,\cdots,t_o)}$ is created by Equation~\ref{eq:Waseda-and-Soshi-2012-Eq1}, where $l=1,\cdots,o$, $u_{l,q}$ is a random integer and $A_{l}=[a_{x,y}]_{n \times t_l}$ such that $A_{l}$ is made of the first $t_l$ columns of $A$. Next, $o\times t_l$ shares $\{c_{l,h}\}_{l=1,\cdots,o;h=1,\cdots,t_l}$ are created by Equation~\ref{eq:Waseda-and-Soshi-2012-Eq2}, where $v$ is a random integer. Finally, each key $k_i$ is shared at $PT_i$ and $\{c_{l,h}\}_{l=1,\cdots,o;}$ $_{h=1,\cdots,t_l}$, $A$ and $t_l$ are published on the NB. In the reconstruction process, $\lbrace u_{l,q} \rbrace _{q=1,\cdots,t_l}$ is created by solving Equation~\ref{eq:Waseda-and-Soshi-2012-Eq1}. Then, secrets are reconstructed by solving Equation~\ref{eq:Waseda-and-Soshi-2012-Eq2}. 

\begin{equation}
	\left[ f_l(k_1),\cdots,f_l(k_n) \right]^T = A_l \times \left[ u_{l,1},\cdots,u_{l,t_l} \right]^T
	\label{eq:Waseda-and-Soshi-2012-Eq1}
\end{equation}
\vspace{-5mm}
\begin{equation}
	\begin{array}{lll}
		c_{l,h}&=& \sum_{q=1}^{t_l} d_{l,q} \times v^{(q-1)( \sum_{l=1}^{h-1}t_l+h-1) } \\
			   &+& \sum_{q=1}^{t_l} u_{l,q} \times v^{ (t_l+q-1)( \sum_{l=1}^{h-1}t_l+h-1 t) }
	\end{array}
	\label{eq:Waseda-and-Soshi-2012-Eq2}
\end{equation}

\subsection{Group 3: Multi Secret Sharing Schemes Type~II}

The first $(m,t_m,n)$ MSSS type II \cite{Chan-Chang-2005} extends \cite{Shamir-1979} to share $m$ secrets with different threshold access structures. In the sharing process, PT identifiers $\lbrace ID_i\rbrace_{i=1,\cdots,n}$ are randomly chosen from distinct integers. With respect to secret $d_j$, $t_j$ and a prime $p_j$ are selected such that $t_1\leq t_2\leq \cdots \leq t_m$, $p_1<p_2< \cdots <p_m$, $P=\prod _{j=1}^m p_j$ and $d_j<p_j$. Next, a polynomial of degree $t_m-1$ (Equation~\ref{eq:Chan-and-Chang-2005-Eq1}) is created with coefficients $\lbrace u_v\rbrace_{v=1,\cdots,t-1}$ being integers chosen by the Chinese remainder theorem \cite{crt} and the uniqueness theorem of interpolating polynomial.
$\forall_v \in [0,t-1], u_v \equiv u_{j,v}\bmod p_j \forall j=1,\cdots,m$ where $u_{j,v}$ is a coefficient of a random polynomial function of degree $t_j-1$ ($f_j(x)=\sum _{w=0} ^{t_j-1} u_{j,v}\times x^w$  \cite{Shamir-1979}) and $u_{j,0}=d_j$. 
Shares $\lbrace e_i\rbrace_{i=1,\cdots,n}$ are generated by equation~\ref{eq:Chan-and-Chang-2005-Eq2}. Finally, $ID_i$ and $e_i$ are stored at $PT_i$, whereas $\lbrace t_j\rbrace_{j=1,\cdots,m}$ and $\lbrace p_j\rbrace_{j=1,\cdots,m}$ are retained at the user's. Secret $d_j$ is reconstructed from $p_j$ and $t_j$ pairs $(ID_i,e_i)$ by equations~\ref{eq:Chan-and-Chang-2005-Eq3} and \ref{eq:Chan-and-Chang-2005-Eq4}.

\begin{eqnarray}
	f(x) &=&\sum_{v=0}^{t_m-1}u_v \times x^v
	\label{eq:Chan-and-Chang-2005-Eq1} \\
	e_i &=& f\left(ID_i\right) \bmod P
	\label{eq:Chan-and-Chang-2005-Eq2} \\
	f_j (0) & \equiv & d_j \bmod p_j
	\label{eq:Chan-and-Chang-2005-Eq3} \\
	f_j (x) & \equiv & f (x) \bmod p_j
	\label{eq:Chan-and-Chang-2005-Eq4} 
\end{eqnarray}

\cite{Runhua-et-al-2008} shares unfixed sized data blocks with a linear equation.
There are $t_l$ secrets $\lbrace d_{l,q} \rbrace _{q=1,\cdots,t_l}$ in block $b_l$ ($t_{l_1}<t_{l_2}$ if $l_1<l_2$). Then, $o\times n$ shares $\{e_{l,i}\}_{l=1,\cdots,o;}$ $_{i=1,\cdots,n}$ are created by multiplying $b_l$ with random matrix $A= [a_{x,y}] _{n \times \max (t_1,\cdots,t_o)}$ by Equation~\ref{eq:Runhua-et-al-2004-Eq1}, where $A_l=[a_{x,y}]_{n \times t_l}$ and $A_{l}$ is built from the first $t_l$ columns of $A$. Next, $o$ shares $\lbrace e_{l,i} \rbrace_{l=1,\cdots,o}$ are stored at $PT_{i}$ and matrix $A$ is published on the NB. Finally, secrets from block $b_{l}$ are reconstructed from matrix $A_l$ and $t_{l}$ shares $\lbrace e_{l,i}\rbrace_{i=1,\cdots,t_l}$ by solving linear Equation~\ref{eq:Runhua-et-al-2004-Eq1}. 

\begin{equation}
	\left[ e_{l,1},\cdots,e_{l,n} \right]^T = A_l \times \left[ b_l \right]^T
	\label{eq:Runhua-et-al-2004-Eq1}
\end{equation}

\cite{Liu-et-al-2012}'s MSSS 
extends from \cite{Liu-et-al-2012}'s SSS (Section~\ref{classic-sss}) with a new sharing process. At $PT_i$, shares $\lbrace u_{i,a,j} \rbrace_{a=1,\cdots,n}$ of secrets are computed and distributed to other PTs \cite{Pedersen-1991} (Section~\ref{G4}). However, $PT_i$'s actual share $e_{i,j}$ is computed by weighting the sum of other PTs' shares (Equation~\ref{eq:Liu-et-al-2012.II-Eq1}), where $w_{a}$ is a random integer (weight).

\begin{equation}
	e_{i,j}  = \sum_{a=1}^{n}w_{a} \times u_{i,a,j}
	\label{eq:Liu-et-al-2012.II-Eq1} 
\end{equation}

In \cite{Takahashi-Iwamura-2013},  PTs are categorized into two groups: $G_1=\lbrace PT_i\rbrace_{i=1,\cdots,\gamma}$ and $G_2=\lbrace PT_i\rbrace_{i=\gamma+1,\cdots,n}$, with the objective of reducing share volume. PTs of $G_1$ store only one key and one share.
To share $m$ secrets \\$\lbrace d_j\rbrace_{j=1,\cdots,m}$, a key $k_i$ and an identifier $ID_i$ are defined for each $PT_i$. Next, a first polynomial $f_1\left(x\right)$ is defined by Equation~\ref{eq:Takahashi-Iwamura-2013-Eq1}, where coefficients $\lbrace u_{1,v} \rbrace _{v=1,\cdots,t-1}$ are random integers. Then, $n$ shares $\lbrace e_{1,i}\rbrace _{i=1,\cdots,n}$ are created by Equation~\ref{eq:Takahashi-Iwamura-2013-Eq2}. Moreover, $\left(m-1\right)\times \gamma$ pseudo shares  $\lbrace e_{j,i}\rbrace_{j=2,\cdots,m;i=1,\cdots,\gamma}$ are generated with a pseudo-random number generator, keys $\lbrace k_i\rbrace_{i=1,\cdots,\gamma}$ and shares $\lbrace e_{1,i}\rbrace_{i=1,\cdots,\gamma}$. Next,  $m-1$ polynomials $f_2(x),\cdots,f_m(x)$ (Equation~\ref{eq:Takahashi-Iwamura-2013-Eq1}) are solved from $m\times\gamma$ pseudo shares \\$\lbrace e_{j,i}\rbrace_{j=2,\cdots,m;i=1,\cdots,\gamma}$ and $m$ secrets $\lbrace d_j\rbrace_{j=2,\cdots,m}$  to construct the other $\left(m-1\right)\times \left(n-\gamma\right)$ shares \\$\lbrace e_{j,i}\rbrace_{j=2,\cdots,m;i=\gamma+1,\cdots,n}$ (Equations~\ref{eq:Takahashi-Iwamura-2013-Eq1} and \ref{eq:Takahashi-Iwamura-2013-Eq2}). Eventually, each $PT_i\in G_1$ stores $k_i$ and one share $e_{1,i}$; and each $PT_i\in G_2$ stores shares $\lbrace e_{j,i}\rbrace _{j=1,\cdots,m}$. 

\begin{eqnarray}
	f_j\left(x\right) &=& d_j+\sum_{v=1}^{t-1}u_{j,v} \times x^{v}
	\label{eq:Takahashi-Iwamura-2013-Eq1} \\
	e_{j,i} &=& f_j\left(ID_i\right)
	\label{eq:Takahashi-Iwamura-2013-Eq2}
\end{eqnarray}

To reconstruct the secrets, $t$ of $n$ PTs in both $G_1$ and $G_2$ are selected. If $PT_i \in G_1$, pseudo shares $\lbrace e_{j,i}\rbrace _{j=2,\cdots,m}$ are generated as above. Then, secret data are reconstructed by Lagrange interpolation from their shares,  $m \times t$ pseudo shares and $t$ IDs.

\subsection{Group 4: Data-Verifiable Secret Sharing Schemes}
\label{G4}

There are only three $(t,n)$ VSSSs in this group. 
\cite{Pedersen-1991} helps each PT verify other PTs' shares with the help of an RSA cryptosystem \cite{PerlnerC09}. To share secret $d$ at $PT_i$, a random polynomial function $f_i$ (Equation~\ref{eq:Harn-and-Lin-2010-Eq1}) is created such that $d = \sum_{i=1}^n w_{i,0}$. Then, $t$ signatures \\$\lbrace s\_d_{i,v}\rbrace _{v=0,\cdots,t-1}$ are created (Equation~\ref{eq:Harn-and-Lin-2010-Eq2}, where $p$ is a prime and $d=\log _p \prod _{i=1}^n y_i$) and shared on the NB. Then, shares $\lbrace u_{i,a} \rbrace _{a=1,\cdots,n}$ are created by Equation~\ref{eq:Harn-and-Lin-2010-Eq3} and distributed to other PTs. $PT_i$'s actual share $e_i$ is created by summing  other PTs' shares (Equation~\ref{eq:Harn-and-Lin-2010-Eq4}) if they are correct (Equation~\ref{eq:Harn-and-Lin-2010-Eq5}). Secrets are reconstructed by Lagrange interpolation.

\begin{eqnarray}
	f_i(x) &=& \sum_{v=0}^{t-1} w_{i,v} \times x^v
	\label{eq:Harn-and-Lin-2010-Eq1} \\
	s\_d_{i,v} &=& \begin{cases} 
		\begin{array}{ll}
			y_i & \text{if~} v=0 \\
			p^{w_{i,v}}~~ & \text{otherwise}
		\end{array}
	\end{cases}
	\label{eq:Harn-and-Lin-2010-Eq2} \\	
	u_{i,a} &=& f_i(a)
	\label{eq:Harn-and-Lin-2010-Eq3} \\	
	e_i &=& \sum_{a=1}^n u_{a,i}
	\label{eq:Harn-and-Lin-2010-Eq4} \\	
	p^{u_{a,i}} &=& \prod _{v=0}^{t-1} (s\_d_{a,v})^{i^v}
	\label{eq:Harn-and-Lin-2010-Eq5} 	
\end{eqnarray}

\cite{Tang-Yao-2008} 
extends from \cite{Shamir-1979} by verifying  the correctness of reconstructed secrets. To this aim, in the sharing process, a signature $s\_d$ is created for each secret $d$ (Equation~\ref{eq:Tang-Yao-2008-Eq1}, where $u$ is a random integer). Then, $s\_d$ is published on the NB. 

\begin{equation}
	s\_d = u^d \bmod p
	\label{eq:Tang-Yao-2008-Eq1} 
\end{equation}

In the reconstruction process, secret $d$ is reconstructed from $t$ shares by secure multi-party computation (SMC) \cite{SMC} (Equation~\ref{eq:Tang-Yao-2008-Eq2}). Next, a multi-prover zero-knowledge argument \cite{MPZKA} helps verify correctness. Secret $d$ is correct if $u^{v''_1+\cdots+v''_n} \times s\_d^{v_0}=\prod_{i=1}^{n}v'_i \bmod p$, where \\$\lbrace v'_i\rbrace_{i=1,\cdots,n}$ and $\lbrace v''_i\rbrace_{i=1,\cdots,n}$ are generated by Equations~\ref{eq:Tang-Yao-2008-Eq3} and \ref{eq:Tang-Yao-2008-Eq4}, respectively, and $\lbrace v_i\rbrace_{i=0,\cdots,n}$ and \\$\lbrace w_i\rbrace_{i=0,\cdots,n}$ are random integers such that $d=\sum _{i=1}^n w_i$.

\begin{eqnarray}
	d &=& \sum_{i\in G} \left( e_i \times \prod_{j\in G, j\neq i} j/\left(j-i\right) \right) 
	\label{eq:Tang-Yao-2008-Eq2} \\
	v'_i &=& u^{v_i} \bmod p
	\label{eq:Tang-Yao-2008-Eq3} \\
	v''_i &=& v_i-v_0 \times w_i \bmod p
	\label{eq:Tang-Yao-2008-Eq4}
\end{eqnarray}

\cite{Yue-Hong-2009} exploits NTRU encryption \cite{PerlnerC09} and a hash function to verify the correctness of shares. First, $n$ pairs of $PT_i$ keys $\left(k_{i,1},k_{i,2}\right)_{i=1,\cdots,n}$ are randomly created with NTRU. Then, shares $e_i$ and signatures $s\_d_i$ are created by Equations~\ref{eq:Yue and Hong's scheme (2009) -Eq1} and \ref{eq:Yue and Hong's scheme (2009) -Eq2}, respectively, where $\lbrace x_i\rbrace_{i=1,\cdots,n}$ are random integers, $w$ is a random polynomial called blinding value and $f$ is a random polynomial \cite{Shamir-1979}. Keys $\left(k_{i,1},k_{i,2}\right)$ and shares $e_i$ are stored at $PT_i$ and $\lbrace x_i\rbrace_{i=1,\cdots,n}$ and signatures $\lbrace s\_d_i\rbrace_{i=1,\cdots,n}$ are published on the NB. Before reconstruction, each share $e_i$ is verified for correctness by Equations~\ref{eq:Yue and Hong's scheme (2009) -Eq3} and \ref{eq:Yue and Hong's scheme (2009) -Eq4}. Finally, $t$ pairs of $\left(e_{i},x_{i}\right)_{i=1,\cdots,n}$ help reconstruct secrets from the polynomial by Lagrange interpolation.

\begin{eqnarray}
	e_{i}& \equiv & \left(w \times k_{i,1}+f(x_{i})\right)\bmod p_1
	\label{eq:Yue and Hong's scheme (2009) -Eq1} \\
	s\_d_{i} & \equiv & \left(w \times k_{i,1}+H(f(x_{i}))\right)\bmod p_1
	\label{eq:Yue and Hong's scheme (2009) -Eq2} \\
	y_{i} & \equiv & k_{i,2} \times e_i\bmod p_1\bmod p_2
	\label{eq:Yue and Hong's scheme (2009) -Eq3} \\
	y_{i} & \equiv & k_{i,2} \times s\_d_i\bmod p_1\bmod p_2
	\label{eq:Yue and Hong's scheme (2009) -Eq4}
\end{eqnarray}

\subsection{Group 5: Key-Verifiable Secret Sharing Schemes}

In \cite{Hwang-Chang-1998}, the only $(t,n)$ VSSS in this group, 
PT keys and signatures are independent. Hence, if some PTs come or go, the keys of other PTs do not change. PT keys $\lbrace k_i\rbrace_{i=1,\cdots,n}$ and identifiers $\lbrace ID_i\rbrace_{i=1..n}$ are randomly chosen. On the other hand, key signatures $\lbrace s\_k_i\rbrace_{i=1,\cdots,n}$ are generated with the help of an RSA cryptosystem (Equation~\ref{eq:Hwang and Chang's scheme (1998) -Eq0}). Then, key $k_{i}$ is stored at $PT_{i}$, while identifiers and key signatures $\left(ID_{i},s\_k_{i}\right)_{i=1,\cdots,n}$ are published on the NB. For sharing secret $d$, several groups of PTs $\lbrace G_r\rbrace_{r=1,\cdots,g}$ are selected, and then shares $\lbrace e_r\rbrace_{r=1,\cdots,g}$ are created by Equations~\ref{eq:Hwang and Chang's scheme (1998) -Eq1}, \ref{eq:Hwang and Chang's scheme (1998) -Eq2}, \ref{eq:Hwang and Chang's scheme (1998) -Eq3} and \ref{eq:Hwang and Chang's scheme (1998) -Eq4}, where $u$ is a random integer, $p_{4}>p_{3}$ and $p_{4}>p_{2}>p_{1}$. Next, $v$, $w$, $\lbrace G_r\rbrace_{r=1,\cdots,g}$ and $\lbrace e_r\rbrace_{r=1,\cdots,g}$ are published on the NB. Before reconstruction, the key signature of $PT_i \in G_r$ is verified to check whether $s\_k_i = v^{k_i}\bmod p_{2}$. If this is true, secrets are reconstructed by Equations~\ref{eq:Hwang and Chang's scheme (1998) -Eq5} and \ref{eq:Hwang and Chang's scheme (1998) -Eq6}.

\setlength{\arraycolsep}{0.0em}
\begin{eqnarray}
	s\_k_i &=& \left(p_1\right)^{k_i}\bmod p_2
	\label{eq:Hwang and Chang's scheme (1998) -Eq0} \\
	v &=& \left(p_1\right)^u \bmod p_2
	\label{eq:Hwang and Chang's scheme (1998) -Eq1} \\
	u \times p_3 &=& a\bmod\phi\left(p_2\right)
	\label{eq:Hwang and Chang's scheme (1998) -Eq2}
\end{eqnarray}
\begin{equation}
	w_r = d\oplus(s\_k_{r,1})^u \bmod p_2 \oplus\cdots\oplus (s\_k_{r,g})^u \bmod p_2 
	\label{eq:Hwang and Chang's scheme (1998) -Eq3} 
\end{equation}
\begin{equation}
	\begin{array}{l}
		e_r = w_r \times \prod_{x=1}^t \frac{1-ID_{r,x}}{-ID_{r,x}}+\\
		\sum_{x=1}^{t} \frac{(s\_k_{r,x})^u\bmod p_2 \times \prod_{y=1,y\neq x}^{t} \frac{1-ID_{r,y}}{ID_{r,x}-ID_{r,y}} }{ID_{r,x}} \bmod p_4 
	\end{array}
	\label{eq:Hwang and Chang's scheme (1998) -Eq4}
\end{equation}
\begin{equation}
	d = w_r\oplus\left(v^{k_{r,1}}\bmod p_2\right)\oplus\cdots\oplus\left(v^{k_{r,g}}\bmod p_2\right)
	\label{eq:Hwang and Chang's scheme (1998) -Eq5}
\end{equation}
\begin{equation}
\begin{array}{l}
	w_r = e_{r}\times\prod_{x=1}^{t}\frac{-ID_{r,x}}{1-ID_{r,x}}+\\
	\sum_{x=1}^{t} \frac{v^{k_{r,x}}\bmod p_2 \times\prod_{y=1,y\neq x}^{t}\frac{-ID_{r,y}}{ID_{r,x}-ID_{r,y}}}{ID_{r,x}-1} \bmod p_4\\
	\end{array}
	\label{eq:Hwang and Chang's scheme (1998) -Eq6}
\end{equation}

\subsection{Group 6: Key \emph{and} data-verifiable secret sharing schemes}

Unlike other schemes, \cite{Zhao-et-al-2012}'s $(t,n)$ VSSS verifies the correctness of both keys and shares. Moreover, it achieves a smaller share size than that of secrets, by splitting secrets before the sharing process. In the sharing process, key $k_0$ and keys $\lbrace k_i\rbrace_{i=1,\cdots,n}$ are randomly selected from a prime and distinct positive integers, respectively. Key signatures $\lbrace s\_k_i\rbrace_{i=0,\cdots,n}$ are constructed by Equation~\ref{eq:Zhao-et-al-2012-Eq1}, where $z$ is a positive integer and $\varphi (p)$ is Euler's totient function \cite{euler}. Next, any secret $d$ is split into $t^2$ smaller pieces stored in Matrix $D = \left[ d_{x,y}\right]_{t\times t}$. Then, two types of shares are created (PTs' shares and NB's shares). PTs' shares $\lbrace E_i=\{e_{i,0},\cdots,e_{i,a}\} \rbrace_{i=1,\cdots,n}$ are sets of randomly distinct positive integers such that $e_{i,0}$ is the sum of all entries in $E_i$ ($e_{i,0}=\sum _{h=1}^{a} e_{i,h}$) and $e_{i,0}<p$.
To construct the NB's shares $\lbrace c_i\rbrace_{i=1,\cdots,n}$, polynomial function $f(x)$ (Equation~\ref{eq:Zhao-et-al-2012-Eq2}) is created from split secrets and PTs' shares by Equations~\ref{eq:Zhao-et-al-2012-Eq3} and \ref{eq:Zhao-et-al-2012-Eq4}, where $A$ is a Jordan normal form of $D$\footnote{$A$ is a Jordan normal form of $D$ if $DY=YA$, where $Y$ is a row matrix and $A$ is a square, upper triangular matrix whose entries are all the same integer values on the diagonal, all 1 on the entries immediately above the diagonal, and 0 elsewhere.
}. 
Finally, NB's shares $\lbrace c_i\rbrace_{i=1,\cdots,n}$ are constructed from Equations~\ref{eq:Zhao-et-al-2012-Eq5} and \ref{eq:Zhao-et-al-2012-Eq6}; and share signatures $\lbrace s\_d_{i,j}\rbrace_{i=1,\cdots,n;j=1,\cdots,m}$ are created from Equation~\ref{eq:Zhao-et-al-2012-Eq7}. Keys $k_i$ and shares $E_i$ are stored at $PT_i$; shares $\lbrace c_i\rbrace_{i=1,\cdots,n}$, share signatures $\lbrace s\_d_{i,j}\rbrace_{j=1,\cdots,n;j=1,\cdots,m}$, key $ k_0$, key signatures \\$\lbrace s\_k_i\rbrace_{i=0,\cdots,n}$, $p$ and $A$ are published on the NB.

\begin{eqnarray}
	s\_k_i &=& \begin{cases} 
		\begin{array}{ll}
			k_0^{-1} \bmod \varphi (p)~ & \text{if~} i=0 \\
			z^{k_i}\bmod p & \text{if~} 1\leq i\leq n
		\end{array}
	\end{cases} \label{eq:Zhao-et-al-2012-Eq1}\\
	f\left(x\right) &=& \sum_{i=1}^{t-1}u_i \times x^{i-1}
	\label{eq:Zhao-et-al-2012-Eq2} \\
	u_i &=& (((z)^{k_0})^{e_{i,0}})^{-1} y_i\bmod p
	\label{eq:Zhao-et-al-2012-Eq3} \\
	D \times \left[ y_1,\cdots,y_t \right]^T &=& \left[ y_1,\cdots,y_t \right]^T \times A
	\label{eq:Zhao-et-al-2012-Eq4} \\
	c_i &=& f(v_i)
	\label{eq:Zhao-et-al-2012-Eq5} \\
	v_i &=& ((z)^{k_0})^{k_i}\bmod p
	\label{eq:Zhao-et-al-2012-Eq6} \\
	s\_d_{i,j} &=& z^{e_{i,j}}\bmod p
	\label{eq:Zhao-et-al-2012-Eq7}
\end{eqnarray}

In the reconstruction process, key $k_i$ is correct if $((z)^{k_i})^{s\_k_{n+1}}=s\_k_i\bmod p$. $PT_i$'s share $e_{i,j}$ is correct if $(((z)^{k_0})^{e_{i,j}})^{s\_k_{n+1}}=s\_d_{i,j}\bmod p$.
Next, polynomial function $f(x)$ is reconstructed from $t$ pairs of key and NB's share $\{k_i,c_i\}$ by Lagrange interpolation and Equation~\ref{eq:Zhao-et-al-2012-Eq6}. Then, $\lbrace y_a\rbrace_{a=1,\cdots,t}$ are created by Equation~\ref{eq:Zhao-et-al-2012-Eq8}. Finally, secret $d$ is reconstructed by solving Equation~\ref{eq:Zhao-et-al-2012-Eq4}.

\begin{equation}
	y_i = u_i\prod_{j=1}^{a} ((z)^{k_0})^{e_{i,j}}  
	\label{eq:Zhao-et-al-2012-Eq8}
\end{equation}

\subsection{Group 7: Data-Verifiable Multi Secret Sharing Schemes Type~I}

The only $(m,t,n)$ VMSSS type I in this group shares and reconstructs all secrets at once with the help of a cellular automaton, to enhance computation performance. Moreover, the correctness of shares is verified before reconstruction \cite{Eslami-Ahmadabadi-2010}. In the sharing process, a set of integers $\left\{ u_1,\cdots,u_{\max(m,t)},\cdots,u_{w+n}\right\} $ is created, where $w$ is a random integer such that $w\geq \max(m,t)$, \\$u_{j}=d_{j}$ if $1\leq j\leq\min(t,m)$ and $u_{j}$ is a random integer when $m<j\leq t$. Others values of $u_j$ are created with the help of the cellular automaton. 
Then, shares $\lbrace c_h\rbrace_{h=1,\cdots,m-t}$ are generated by Equation~\ref{eq:Eslami-and-Ahmadabadi-2012-Eq1}. Shares $\lbrace e_i\rbrace_{i=1,\cdots,n}$ and their signatures $\lbrace s\_d_i\rbrace_{i=1,\cdots,n}$ are created by Equations~\ref{eq:Eslami-and-Ahmadabadi-2012-Eq2} and \ref{eq:Eslami-and-Ahmadabadi-2012-Eq3}, where $v$ is a random integer. Finally, each share $e_i$ is shared at $PT_i$ and shares $\lbrace c_h\rbrace_{h=1,\cdots,m-t}$ and signatures $\lbrace s\_d_h\rbrace_{h=1,\cdots,n}$ are published on the NB.

\begin{eqnarray}
	c_{h} &=& d_{t+h}+u_{t+h}\left(\bmod2\right)
	\label{eq:Eslami-and-Ahmadabadi-2012-Eq1} \\
	e_i &=& u_{m+i}
	\label{eq:Eslami-and-Ahmadabadi-2012-Eq2} \\
	s\_d_i &=& v^{e_i}\bmod p
	\label{eq:Eslami-and-Ahmadabadi-2012-Eq3}
\end{eqnarray}

Before reconstruction, share integrity is verified by Equation~\ref{eq:Eslami-and-Ahmadabadi-2012-Eq3}. Next, $\left\{ u_1,\cdots,u_{\max (m,t)},\cdots,u_{w+n}\right\} $ are reconstructed from $t$ shares with the cellular automaton. Finally, all secrets are regenerated by Equation~\ref{eq:Eslami-and-Ahmadabadi-2012-Eq4}.

\begin{equation}
	d_{j} =
	\begin{cases}
		\begin{array}{ll}
			u_{j} & \text{if~} 1\leq j\leq\min(t,m)\\
			c_{j-t}+u_{j}\left(\bmod p\right)~ & \text{otherwise}
		\end{array} 
	\end{cases}
	\label{eq:Eslami-and-Ahmadabadi-2012-Eq4}
\end{equation}

\subsection{Group 8: Key-Verifiable Multi Secret Sharing Schemes Type~I}

A fair amount of research has been done on $(m,t,n)$ VMSSSs type~I, half of which belong to this group.
\cite{Zhao-et-al-2007} extends from \cite{Yang-et-al-2004} by verifying whether keys shared between PTs are correct. In the sharing process, each key $k_{i}$, its signature $s\_k_{i}$ and public key $v$ are created by Equations~\ref{eq:Zhao-et-al-2007-Eq1}, \ref{eq:Zhao-et-al-2007-Eq2} and \ref{eq:Zhao-et-al-2007-Eq3}, respectively, where prime $p_{1}$ is a multiple of prime $p_{2}$, $\lbrace u_i\rbrace_{i=0,\cdots,n}$ are random integers and $\phi$ is Euler's totient function \cite{euler}. Key $k_{i}$ is stored at $PT_{i}$ and $\lbrace s\_k_i\rbrace_{i=0,\cdots,n}$ and $v$ are published on the NB. 
Before reconstruction, keys are verified. Key $k_{i}$ is correct if $((s\_k_{0})^{k_{i}})^v \equiv u_{i}'\mod p_1$.

\begin{eqnarray}
	k_i &=& (s\_k_i)^{u_0}\bmod p_1
	\label{eq:Zhao-et-al-2007-Eq1} \\	
	s\_k_i&=&(p_2)^{u_i}\bmod p_1
	\label{eq:Zhao-et-al-2007-Eq2} \\	
	v &=& (u_0)^{-1}\bmod \phi(p_1)
	\label{eq:Zhao-et-al-2007-Eq3}
\end{eqnarray}

\cite{Dehkordi-Mashhadi-2008a} also extends from \cite{Yang-et-al-2004} with the same goal. 
Only key and signature generation actually varies. However, the verification process is more efficient. Key $k_{i}$ is created by Equations~\ref{eq:Dehkordi-and-Mashhadi-2008.I-Eq1}, \ref{eq:Dehkordi-and-Mashhadi-2008.I-Eq2} and \ref{eq:Dehkordi-and-Mashhadi-2008.I-Eq4}, where $u_{i=1,2,3}$ are random integers and $f$ is any two-variable one-way function. Signature $s\_k_i$ of key $k_i$  is created by Equation~\ref{eq:Dehkordi-and-Mashhadi-2008.I-Eq5}, where $u_{4}$ is a random integer. Key $k_{i}$ is stored at $PT_{i}$, while $u_{1},\cdots,u_{4}$ and $\lbrace s\_k_i\rbrace_{i=1,\cdots,n}$ are published on the NB. 

\begin{eqnarray}
	k_i &=& f\left(u_{1},w_{i}\right)
	\label{eq:Dehkordi-and-Mashhadi-2008.I-Eq1} \\
	w_i &=& \left(\left(v_{i}\right)^{u_{3}}\right)^{u_{2}}\bmod p
	\label{eq:Dehkordi-and-Mashhadi-2008.I-Eq2} \\
	u_2 \times u_{3} &\equiv& 1\bmod\phi\left(p\right)
	\label{eq:Dehkordi-and-Mashhadi-2008.I-Eq4} \\
	s\_k_i &=& \left(u_{4}\right)^{k_{i}}\bmod p
	\label{eq:Dehkordi-and-Mashhadi-2008.I-Eq5}
\end{eqnarray}

\cite{Dehkordi-Mashhadi-2008b} in turn extends from \cite{Dehkordi-Mashhadi-2008a} by proposing new secret sharing and reconstruction processes to reduce computation costs. After keys and signatures are created, shares $\lbrace c_{j,1}\rbrace_{j=1,\cdots,n}$ and $\lbrace c_{j,2}\rbrace_{j=1,\cdots,m}$ are generated by Equations~\ref{eq:Dehkordi-and-Mashhadi-2008.II-Eq1}, \ref{eq:Dehkordi-and-Mashhadi-2008.II-Eq2}, \ref{eq:Dehkordi-and-Mashhadi-2008.II-Eq3} and \ref{eq:Dehkordi-and-Mashhadi-2008.II-Eq4}, where $u_0$ is a random integer. Next, $\lbrace c_{j,1}\rbrace_{j=1,\cdots,n}$ and $\lbrace c_{j,2}\rbrace_{j=1,\cdots,m}$ are published on the NB. After key verification, secrets are reconstructed by Equations~\ref{eq:Dehkordi-and-Mashhadi-2008.II-Eq5}, \ref{eq:Dehkordi-and-Mashhadi-2008.II-Eq6} and \ref{eq:Dehkordi-and-Mashhadi-2008.II-Eq7}.

\begin{eqnarray}
	c_{j,1} &=& d_{j}-y_{j+n}
	\label{eq:Dehkordi-and-Mashhadi-2008.II-Eq1} \\
	c_{j,2} &=& k_{j}-y_{j-1}
	\label{eq:Dehkordi-and-Mashhadi-2008.II-Eq2} \\
	y_{j} &=&
	\begin{cases}
		\begin{array}{ll}
			k_{j+1} & \text{if~}0\leq j<t\\
			-\sum_{v=1}^t u_v \times y_{j-v} \bmod p~ & \text{otherwise}
		\end{array}
	\end{cases}
	\label{eq:Dehkordi-and-Mashhadi-2008.II-Eq3} \\
	\left(x-u_0\right)^{t} &=& x^{t}+u_1 \times x^{t-1}+\cdots+u_t=0
	\label{eq:Dehkordi-and-Mashhadi-2008.II-Eq4} \\
	d_{j} &=& y_{j+n}+c_{j,2}
	\label{eq:Dehkordi-and-Mashhadi-2008.II-Eq5} \\
	y_{j} &=&
	\begin{cases}
		\begin{array}{ll}
			k_{j+1} 		& \text{if~} 0\leq j<t\\
			k_{j+1}-c_{j+1,1}	& \text{if~} t\leq j<n\\
			f(j)\times\left(u_0 \right)^{j}\bmod p~ & \text{otherwise}
		\end{array}
	\end{cases}
	\label{eq:Dehkordi-and-Mashhadi-2008.II-Eq6} \\
	f(x) &=& \sum_{v=1}^{t}\frac{y_{v-1}}{\left(u_0\right)^{v-1}}\prod_{w=1\& w \neq v}^{t}\frac{x-w+1}{v-w} \bmod p
	\label{eq:Dehkordi-and-Mashhadi-2008.II-Eq7}
\end{eqnarray}

\cite{Wang-et-al-2011}
extends from \cite{Wei-et-al-2007} (Section~\ref{G9}) to improve the efficiency of the sharing and  reconstruction processes. To this aim,  secrets are split into blocks of size $t$ that are each shared and reconstructed all at once. 
Block $b_{l}$ is divided into $n$ shares $\lbrace c_{l,h}\rbrace_{h=1,\cdots,n}$ by Equation~\ref{eq:Wang-et-al-2011-Eq1}, where $A=\left[a_{i,w}\right]_{t\times n}$, $a_{i,w}=H(u_l \times k_i \times v )^{w-1}$, and $\lbrace u_l\rbrace_{l=1,\cdots,o}$ and $v$ are random integers. Key $k_i$ is stored at $PT_{i}$, whereas key signatures $\lbrace s\_k_i\rbrace_{i=1,\cdots,n}$, shares $\lbrace c_{l,h}\rbrace_{l=1,\cdots,o;h=1,\cdots,n}$ and $\lbrace x_l=u_l\times v\rbrace_{l=1,\cdots,o}$ are published on the NB.
To reconstruct secrets, shares and keys are verified for correctness with a bilinear map $f (u_l \times k_i \times v,v )=f (x_l,s\_k_i)$. Then,  secrets are reconstructed by solving Equation~\ref{eq:Wang-et-al-2011-Eq1}. 

\begin{equation}
	\left[c_{l,1},\cdots,c_{l,n}\right]^{T} = A \times [b_l]^T
	\label{eq:Wang-et-al-2011-Eq1}
\end{equation}

\cite{Eslami-Rad-2012} also extends from \cite{Wei-et-al-2007}, pursuing the same goal as \cite{Wang-et-al-2011}. 
The difference is that secrets are divided into $n+m-t$ shares to reduce the number of shares. Shares $\lbrace c_h\rbrace_{h=1,\cdots,(n+m-t)}$ are computed by Equation~\ref{eq:Eslami-and-Rad-2012-Eq1},  where $A=\left[a_{x,y}\right]_{(m+n)\times (m+n-t)}$, $a_{x,y}=\left(w \right)^{x\left(y-1\right)}$, $z_i=H (u \times v \times k_i)$, and $u$, $v$ and $w$ are random integers. Key $k_i$ is stored at $PT_i$, whereas key signatures $\lbrace s\_k_i\rbrace_{i=1,\cdots,n}$, shares $\lbrace c_h\rbrace_{h=1,\cdots,(n+m-t)}$, data signatures $\lbrace s\_d_j\rbrace_{j=1,\cdots,m}$ and $x=u \times v$ are published on the NB. To reconstruct  secrets, shares and keys are verified for correctness with a bilinear map $f (u \times k_i \times v,v )=f (x,s\_k_i)$. Then,  secrets are reconstructed by solving Equation~\ref{eq:Eslami-and-Rad-2012-Eq1}.

\begin{equation}
	\left[ c_1,\cdots,c_{n+m-t} \right]^T =A \times \left[ z_1,\cdots,z_n,d_1,\cdots,d_m \right]^T
	\label{eq:Eslami-and-Rad-2012-Eq1}
\end{equation}

Unlike in other schemes, PTs in \cite{Chen-et-al-2012} can be added or deleted. Moreover, threshold $t$ can vary. To this aim, keys $k_{i}$, key signatures $s\_k_{i}$ and PT identifiers $ID_{i}$ are randomly selected such that they are different from one PT to the other. Then,  secrets are organized into unfixed-sized blocks, where  block $b_{l}$ stores $u_l$ secrets. All secrets $\lbrace d_{l,q}\rbrace_{q=1,\cdots,u_l}$ in block $b_l$ are divided into $n+u_l-t_l$ shares $\lbrace c_{l,h}\rbrace_{h=1,\cdots,(n+u_l-t_l)}$ by Equations~\ref{eq:Chen-et-al-2012-Eq1}, \ref{eq:Chen-et-al-2012-Eq2}, \ref{eq:Chen-et-al-2012-Eq3}, \ref{eq:Chen-et-al-2012-Eq4} and \ref{eq:Chen-et-al-2012-Eq5}, where $z_l$ is a random integer. Each key $k_{i}$ is stored at $PT_{i}$ and identifiers $\lbrace ID_i\rbrace_{i=1,\cdots,n}$, signatures $\lbrace s\_k_i\rbrace_{i=1,\cdots,n}$ and shares $\lbrace y_l\rbrace_{l=1,\cdots,n}$ and $\lbrace c_{l,h}\rbrace _{l=1,\cdots,o;}$ $_{h=1,\cdots,(n+u_l-t_l)}$ are published on the NB. Before reconstruction, keys are verified for correctness with a discrete logarithm modulo and a one-way hash function. Finally, each secret $d_{l,q}$ in  block $b_l$ is reconstructed by Lagrange interpolation.

\begin{eqnarray}
	c_{l,h} &=& f_{l}\left(n+u_l+h\right)
	\label{eq:Chen-et-al-2012-Eq1} \\
	f_l (x) &=& \sum_{v=1}^{u_l}d_{l,v}\times\Delta_1 + \sum_{v=1}^n (s\_k_v)^{z_l}\times\Delta_2 \bmod p_1 
	\label{eq:Chen-et-al-2012-Eq2} \\
	\Delta_1 &=& \prod_{w=1\&w\neq v}^{u_{l}}\frac{x-(n+w)}{v-w}\times\prod_{i=1}^{n}\frac{x-ID_{i}}{(n+v)-ID_{i}}  
	\label{eq:Chen-et-al-2012-Eq3} \\
	\Delta_2 &=& \prod_{i=1\&i\neq v}^{n}\frac{x-ID_i}{ID_v-ID_i}\times\prod_{w=1}^{u_l}\frac{x-\left(n+w\right)}{ID_v-\left(n+w\right)}  
	\label{eq:Chen-et-al-2012-Eq4} \\
	y_l &=& \left(p_2\right)^{z_l}\bmod p_1
	\label{eq:Chen-et-al-2012-Eq5}
\end{eqnarray}

\cite{Li-et-al-2012} 
extends from \cite{Parakh-Kak-2011} to reduce computation cost and verify key correctness. 
Secrets are organized into blocks of size $t-1$. Keys $\lbrace k_i\rbrace_{i=1,\cdots,n}$ are randomly selected and their signatures $\lbrace s\_k_i\rbrace_{i=1,\cdots,n}$ are created by Equation~\ref{eq:Li-et-al-2012-Eq1}, where $H$ is a hash function. In  block $b_{l}$, the first secret $d_{l,1}$ is divided into two shares $c_{l,1,1}$ and $c_{l,1,2}$ by Equation~\ref{eq:Li-et-al-2012-Eq2}, where $u$ is a random integer. Other secrets in  block $b_l$ are shared by Equations~\ref{eq:Li-et-al-2012-Eq3} and \ref{eq:Li-et-al-2012-Eq4}. Key $k_{i}$ is stored at $PT_{i}$ and $\lbrace s\_k_i\rbrace_{i=1,\cdots,n}$ , $\left\{ c_{l,q,h}\right\} _{l=1,\cdots,o;q=1,\cdots,t-2;}$ $_{h=1,\cdots,q+1}$ and $\left\{ c_{l,t-1,h}\right\} _{l=1,\cdots,o;}$ $_{h=1,\cdots,n}$ are published on the NB. Before reconstruction, each key $k_{i}$ is verified for validity by Equation~\ref{eq:Li-et-al-2012-Eq5}. Then, all  secrets in each block are reconstructed by Lagrange interpolation.

\begin{eqnarray}
	s\_k_{i} &=& H\left(H^{t-1}\left(k_{i}\right)\oplus k_{i}\right)
	\label{eq:Li-et-al-2012-Eq1} \\
	c_{l,1,h} &=& u \times h+d_{l,1}-(k_{q}\oplus H\left(k_{i}\right))
	\label{eq:Li-et-al-2012-Eq2} \\
	c_{l,q,h} &=& f_{l,q}\left(h\right)-(k_{q}\oplus H^{q}\left(k_{q}\right))
	\label{eq:Li-et-al-2012-Eq3} \\
	f_{l,q} (x) &=& \begin{cases}
		\begin{array}{ll}
			d_{l,q}+u \times x 		& \text{if~} q=1\\
			d_{l,q}+\sum_{v=1}^{q}x^{v} \times f_{l,q-1}\left(x\right)~ & \text{otherwise}
		\end{array}
		\end{cases}
	\label{eq:Li-et-al-2012-Eq4} \\
	s\_k_{i} &=& H\left(H^{t-1}\left(k_{i}\right)\oplus k_{i}\right)
	\label{eq:Li-et-al-2012-Eq5}
\end{eqnarray}

Finally, \cite{Hu-et-al-2012} 
propose two schemes. They create keys and verify their correctness by using a one-way hash function and a LFSR public key cryptosystem \cite{lfsr1,lfsr2}. The first scheme shares and reconstructs secrets as \cite{Yang-et-al-2004}, while the second scheme does as \cite{Dehkordi-Mashhadi-2008b}, while providing higher security than \cite{Yang-et-al-2004,Dehkordi-Mashhadi-2008b} with keys of same lengths.

\subsection{Group 9: Key \emph{and} Data-Verifiable Multi Secret Sharing Schemes Type~I}
\label{G9}

The other third of $(m,t,n)$ VMSSSs type~I belong to this group. \cite{Lin-We-1999} prevents cheating from malicious PTs by verifying both shares and keys. Keys $\lbrace k_{i}\rbrace_{i=1,\cdots,n}$ and their signatures $\lbrace s\_k_{i}\rbrace_{i=1,\cdots,n}$ are created by Equations~\ref{eq:Lin-We-1999-Eq1}, \ref{eq:Lin-We-1999-Eq2}, \ref{eq:Lin-We-1999-Eq3}, \ref{eq:Lin-We-1999-Eq4} and \ref{eq:Lin-We-1999-Eq5}, where $\lbrace u_v\rbrace_{v=0,\cdots,t-1}$ are random integers and $a_1,\cdots,a_5$ are set as discrete logarithms. Let $p_1$ and $p_2$ be big primes. $a_1$ is a random integer, $a_2=(2\times p_1+1)(2 \times p_2+1)$, $a_3=p_1 \times p_2$ and $a_3\times a_2=\phi (a_5)$, where $\phi$ is Euler's totient function \cite{euler}. Key $k_i$ is stored at $PT_{i}$, while signatures $\{s\_k_i\}_{i=1,\cdots,n}$ and $\lbrace w_v\rbrace_{v=0,\cdots,t-1}$ are published on the NB. Key correctness is checked by Equation~\ref{eq:Lin-We-1999-Eq6}.

\begin{eqnarray}
	f (x) &=& \left(\sum_{v=0}^{t-1} u_v \times x^v \right)\bmod a_3
	\label{eq:Lin-We-1999-Eq1} \\
	w_v &=& \left(p_1\right)^{u_v}\bmod a_2
	\label{eq:Lin-We-1999-Eq2} \\
	y_i &=& \prod_{\forall PT_v,v \neq i}\left(ID_i-ID_v \right)\bmod a_3
	\label{eq:Lin-We-1999-Eq3}\\
	k_i &=& \left(f (ID_i)/y_i\right)\bmod a_3
	\label{eq:Lin-We-1999-Eq4} \\
	s\_k_i &=& (a_1)^{k_i}\bmod a_2
	\label{eq:Lin-We-1999-Eq5} \\
	\left((a_1)^{y_i}\right)^{k_i} &=& \prod_{v=0}^{t-1} (w_v)^{(ID_i)^v}\bmod a_2
	\label{eq:Lin-We-1999-Eq6} 
\end{eqnarray}

A 4-tuple of shares $\left\{ c_{j,1},\cdots,c_{j,4}\right\} $ is created by Equations~\ref{eq:Lin-We-1999-Eq7} and \ref{eq:Lin-We-1999-Eq8}, where $c_{j,1}$ and $c_{j,2}$ are random integers. Shares $\left\{ c_{j,h}\right\} _{j=1,\cdots,m;h=1,\cdots,4}$ are published on the NB. Before reconstruction, each $PT_i$ must verify share and key correctness by Equation~\ref{eq:Lin-We-1999-Eq9}. If verification is positive,  secrets are reconstructed by Equations~\ref{eq:Lin-We-1999-Eq11} and \ref{eq:Lin-We-1999-Eq12}, where $G$ is any group of $t$ PTs.

\setlength{\arraycolsep}{0.0em}
\begin{eqnarray}
	c_{j,3} &=& (a_1)^{-a_5+c_{j,1}}\times (c_{j,2})^{2\times a_5+c_{j,1}+1} \bmod a_2
	\label{eq:Lin-We-1999-Eq7} \\
	c_{j,4} &=& ((c_{j,2})^{u_0}-d_j)(c_{j,3})^{-u_0} \bmod a_2
	\label{eq:Lin-We-1999-Eq8} 
\end{eqnarray}
\begin{equation}
	\begin{array}{ll}
			((c_{j,3})^{k_i})^{a_4} \equiv& (s\_k_i)^{a_4 \times c_{j,1}-1}\times \\
			& ((c_{j,2})^{k_i})^{2+a_4(c_{j,1}+1)} \bmod a_2
	\end{array}
	\label{eq:Lin-We-1999-Eq9} 
\end{equation}
\begin{equation}
	\begin{array}{ll}
			d_j= & \left(\prod _{PT_i\in G}((c_{j,2})^{k_i})^{\bigtriangleup _i} \right)- \\
			& \left(c_{j,4}\prod _{PT_i\in G} ((c_{j_3})^{k_i})^{\bigtriangleup _i} \right) \bmod a_2 
	\end{array}
	\label{eq:Lin-We-1999-Eq11}
\end{equation}
\begin{equation}
	\triangle_i = \prod_{\forall PT_v\in G} -ID_v \times \prod_{\forall PT_v\in G}(ID_i-ID_v)
	\label{eq:Lin-We-1999-Eq12}
\end{equation}

\cite{Chang-et-al-2005} extends from \cite{Lin-We-1999} to improve the efficiency of the sharing and  reconstruction processes. To this aim, $j$ 3-tuples of shares $\left\{ c_{j,1},c_{j,3},c_{j,4}\right\} _{j=1,\cdots,m}$ are created by Equations~\ref{eq:Chang-et-al-2005-Eq7} and \ref{eq:Chang-et-al-2005-Eq8} and published on the NB. Before reconstruction, each PT must verify share and key correctness by Equation~\ref{eq:Chang-et-al-2005-Eq9}. If verification is positive, secrets are reconstructed by Equations~\ref{eq:Chang-et-al-2005-Eq10} and \ref{eq:Lin-We-1999-Eq12}.

\begin{eqnarray}
	c_{j,3} &=& (a_1)^{a_5 \times c_{j,1}}\bmod a_2
	\label{eq:Chang-et-al-2005-Eq7} \\
	c_{j,4} &=& \left( (a_1)^{u_0 \times a_5 \times c_{j,1}}\bmod a_2\right)\oplus d_j
	\label{eq:Chang-et-al-2005-Eq8} \\
	((c_{j,3})^{k_i})^{a_4} &\equiv & (s\_k_i)^{c_{j,1}}\bmod a_2
	\label{eq:Chang-et-al-2005-Eq9} \\
	d_j &=& c_{j,4}\oplus\prod_{\forall PT_i\in G} ((c_{j,3})^{k_i})^{\triangle_i}\bmod a_2
	\label{eq:Chang-et-al-2005-Eq10}
\end{eqnarray}

\cite{Shao-Cao-2005}  
extends from \cite{Yang-et-al-2004} by checking whether keys and shares are valid, with the help of a discrete logarithm. 
Signatures $\lbrace s\_d_j\rbrace_{j=1,\cdots,\max (m,t)}$ are created after secrets are shared by Equation~\ref{eq:Shao-and-Cao-2005-Eq1}, where $\lbrace u_j\rbrace_{j=1,\cdots,m}$ are secrets ($u_j=d_j$) 
and $\lbrace u_j\rbrace_{j=(m+1),\cdots,t}$ are random integers. They are then published on the NB. Before reconstruction, keys are verified first, and then shares are, 
both by Equation~\ref{eq:Shao-and-Cao-2005-Eq2}. Signature $s\_d_j$ is also used to check share integrity.

\begin{eqnarray}
	s\_d_j &=& \left(p_1\right)^{u_j}\bmod p_2
	\label{eq:Shao-and-Cao-2005-Eq1} \\
	(p_1)^{c_i} &=& \prod_{h=1}^{\max(t,m)}(c_{h+n+1})^{f(w,k_i)^h}\bmod p_2
	\label{eq:Shao-and-Cao-2005-Eq2}
\end{eqnarray}

In \cite{Wei-et-al-2007}, each secret $d_j$ is divided independently into vary threshold $t_j$. Keys $\lbrace k_i\rbrace_{i=1,\cdots,n}$ are randomly selected such that their signatures $\lbrace s\_k_i\rbrace_{i=1,\cdots,n}$ (Equation~\ref{eq:Wei-et-al-2007-Eq0}, where $v$ is a random integer) are unique. Each secret $d_{j}$ is divided into $n$ shares $\lbrace c_{j,h}\rbrace_{h=1,\cdots,n}$ by Equations~\ref{eq:Wei-et-al-2007-Eq1} and \ref{eq:Wei-et-al-2007-Eq2}, where $A_{j}=\left[a_{x,y}\right]_{(n\times t_i)}$, $a_{x,y}=(u)^{x(y-1)}$, $Z_{j}=\left[ w_j\times v,d_j \times (k_1)^v,\cdots,d_j \times (k_n)^v\right]$ and $u$ and $w_j$ are random integers. Signature $s\_d_{j}$ of $d_j$ is created by Equation~\ref{eq:Wei-et-al-2007-Eq3}. Keys $k_{i}$ are stored at $PT_{i}$, whereas key signatures $\lbrace s\_k_i\rbrace_{i=1,\cdots,n}$, shares \\$\left\{ w_j,c_{j,1},\cdots,c_{j,n}\right\} _{j=1,\cdots,m}$, signatures $\lbrace s\_d_j\rbrace_{j=1,\cdots,m}$, $u$ and $v$ are published on the NB. Before reconstruction, shares and keys are verified for correctness with a bilinear map $f ((k_i)^{s\_d_j},v )=f (s\_d_j,(k_j)^v)$. Then,  secrets are reconstructed by solving linear Equations~\ref{eq:Wei-et-al-2007-Eq1} and \ref{eq:Wei-et-al-2007-Eq2}.

\begin{eqnarray}
	s\_k_{i} &=& (k_{i})^v
	\label{eq:Wei-et-al-2007-Eq0} \\
	\left[ c_{j,1},\cdots,c_{j,n}\right] ^{T} &=& A_{j} \times [Z_j]^{T}
	\label{eq:Wei-et-al-2007-Eq1} \\
	d_{j} &=& H (w_j \times v)
	\label{eq:Wei-et-al-2007-Eq2} \\
	s\_d_{j} &=& d_{j} \times v
	\label{eq:Wei-et-al-2007-Eq3}
\end{eqnarray}

Unlike other schemes that compute integers over a finite field, \cite{Das-Adhikari-2010} exploits binary strings in all processes to improve the efficiency of both sharing and reconstruction processes. 
In the sharing process, two kinds of keys are randomly created in binary string format: PT keys $\lbrace k_i\rbrace_{i=1,\cdots,n}$ and user keys $\left\{ u_{j,v}\right\} _{j=1,\cdots,m;v=1,\cdots,t_l}$. Then, each share $c_{j,h}$ is created by Equation~\ref{eq:Des-and-Adhikari-2010-Eq1}, where $H$ is a one-way hash function and $\parallel$ is the concatenation operator. Finally, shares $c_{j,h}$, $H(d_j)$, $H\left(H (k_i\parallel j \parallel h )\right)$ with $j=1,\cdots,m$; $h=1,\cdots,t_l$ and $i=1,\cdots,n$, are published on the NB. 

\begin{equation}
	c_{j,h} = d_j\oplus\left\{ \oplus_{i:PT_{i}\in u_{j,v}}H (k_i\parallel j\parallel h )\right\} 
	\label{eq:Des-and-Adhikari-2010-Eq1} 
\end{equation}

Secrets are reconstructed by Equation~\ref{eq:Des-and-Adhikari-2010-Eq2} if all keys pass the verification process, which is split in two steps. Before reconstruction, keys $\lbrace k_i\rbrace_{i=1,\cdots,n}$  are checked for correctness by comparison with signatures $H\left(H(k_i\parallel j \parallel h)\right)$. After reconstruction, secrets $\{d_j\}_{j=1,\cdots,m}$ are checked for correctness by comparison with signatures $H (d_j)$. 

\begin{equation}
	d_{j} = c_{l,h}\oplus\left\{ \oplus_{i:PT_{i}\in u_{j,v}}H (k_i\parallel j\parallel h )\right\} 
	\label{eq:Des-and-Adhikari-2010-Eq2}
\end{equation}

Finally, \cite{Bu-Yang-2012} extends from \cite{Yue-Hong-2009} by sharing multiple  secrets, to improve sharing/reconstruction efficiency and reduce share volume. To this aim, $PT_i$'s identifier $ID_{i}$ is randomly selected and $PT_i$'s key $k_i$ and signatures $\lbrace s\_k_v\rbrace_{v=0,\cdots,(t-1)}$ are created by Equations~\ref{eq:Bu-and-Yang-2012-Eq1} and \ref{eq:Bu-and-Yang-2012-Eq2}, respectively, where $x$ and $y$ are randomly created with NTRU \cite{PerlnerC09} and $w$ is NTRU's blinding value. Each secret $d_j$ is divided into a 3-tuple of shares $\lbrace c_{j,1},c_{j,2},c_{j,3}\rbrace$ by Equations~\ref{eq:Bu-and-Yang-2012-Eq3} and \ref{eq:Bu-and-Yang-2012-Eq4}, where $c_{j,1}$ is a random integer. Key $k_i$ is stored at $PT_i$, whereas identifiers \\$\lbrace ID_i\rbrace_{i=1,\cdots,n}$, signature $\lbrace s\_k_v\rbrace_{v=0,\cdots,(t-1)}$ and shares $\lbrace c_{j,h}\rbrace_{j=1,\cdots,m;h=1,\cdots,3}$ are published on the NB.

\begin{eqnarray}
	k_i &=& \sum_{v=0}^{t-1} u_v \times (ID_i)^v
	\label{eq:Bu-and-Yang-2012-Eq1} \\
	s\_k_v &=& w \times x+u_v \bmod p_1
	\label{eq:Bu-and-Yang-2012-Eq2} \\
	c_{j,2} &=& w \times x+c_{j,1} \bmod p_1
	\label{eq:Bu-and-Yang-2012-Eq3} \\
	c_{j,3} &=& d_{j}\oplus H (u_0 \times c_{j,2})
	\label{eq:Bu-and-Yang-2012-Eq4} 
\end{eqnarray}

Before reconstruction, keys and shares are verified for correctness by Equations~\ref{eq:Bu-and-Yang-2012-Eq5} and \ref{eq:Bu-and-Yang-2012-Eq6}, respectively. Finally, secrets are reconstructed by Equation~\ref{eq:Bu-and-Yang-2012-Eq7}.

\begin{equation}
	k_i = y \sum_{v=0}^{t-1} s\_k_v (ID_i)^v \bmod p_2
	\label{eq:Bu-and-Yang-2012-Eq5}
\end{equation}
\vspace{-4mm} 
\begin{equation}
	y \times k_i \times c_{j,2} = y \sum_{v=0}^{t-1} \left( w_v \times(ID_i)^v  \times c_{j,1}\right) \bmod p_2
	\label{eq:Bu-and-Yang-2012-Eq6}
\end{equation}
\vspace{-4mm} 
\begin{equation}
	d_j=c_{j,3}\oplus H \left( \sum_{i\in G}k_i \times c_{j,2}\times\prod_{v\in G\&v\neq i}\frac{-ID_v}{ID_{i}-ID_v}\right)
	\label{eq:Bu-and-Yang-2012-Eq7}
\end{equation}

\subsection{Group 10: Data-Verifiable Multi Secret Sharing Schemes Type~II}

VMSSSs type~II are recent.
Unlike all previous SSSs, \cite{DB-Attasena-et-al-2014-J} verifies both PT honesty and share correctness with inner and outer signatures, respectively. Inner signatures are  signatures that help verify secret correctness after reconstruction. If one or more shares are erroneous, then reconstructed secrets do not match with their inner signatures. Outer signatures are share signatures. The correctness of shares is checked before reconstructing  secrets.

In the sharing process of \cite{DB-Attasena-et-al-2014-J}, $n$ distinct random  linear equations $\lbrace f_i\rbrace_{i=1,\cdots,n}$ (Equation~\ref{eq:Attasena-et-al-2014-J-Eq1}, where coefficients $u_{i,v}$ are random positive integers) are created. Then, $m$ secrets $\lbrace d_{l,q}\rbrace_{q=1,\cdots,t-1}$ are organized into $o$  blocks $b_l$ of size $t-1$. The inner signature $s\_b_l$ of block $b_l$ is created with the help of an homomorphic function (Equation~\ref{eq:Attasena-et-al-2014-J-Eq2}). Next, $n$ shares $\lbrace e_{l,i}\rbrace_{i=1,\cdots,n}$ are created by Equation~\ref{eq:Attasena-et-al-2014-J-Eq3}. Their outer signatures $\lbrace s\_out_{l,i}\rbrace_{i=1,\cdots,n}$ are created with any hash function. Shares $\lbrace e_{l,i},s\_out_{l,i}\rbrace$ $_{l=1,\cdots,o}$ are stored at $PT_i$. 

\setlength{\arraycolsep}{0.0em}
\begin{eqnarray}
	f_i(x_1,\cdots,x_t) &=& x_t \times u_{i,v}+\sum_{v=1}^{t-1} (x_v+2)\times u_{i,v}
	\label{eq:Attasena-et-al-2014-J-Eq1} \\
	s\_b_l &=& H(b_l)
	\label{eq:Attasena-et-al-2014-J-Eq2} \\
	e_{l,i} &=& f_i(b_l,s\_b_l)
	\label{eq:Attasena-et-al-2014-J-Eq3}
\end{eqnarray}

Before reconstruction, shares from $t$ out of $n$ PTs are verified against their outer signatures. Then,  blocks and their inner signatures are reconstructed by solving the linear equations. Finally, recovered  blocks are verified against their inner signatures. If the test fails, erroneous  blocks can be reconstructed from shares in a new PT group.

\cite{DB-Attasena-et-al-2014-P} 
extends from \cite{Shamir-1979} by sharing each secret at fewer than $n$ PTs'. PT failure is also allowed, more specifically by allowing  data updates at remaining online PTs. Moreover, \cite{DB-Attasena-et-al-2014-P} also protects  from PT \emph{group} cheating by imposing a new constraint: no PT group can hold enough shares to reconstruct the secret when $n<2t-2$. PT honesty and share correctness are checked as in \cite{DB-Attasena-et-al-2014-J}. In addition, this scheme separates outer signature creation and verification from the sharing and reconstruction processes. 

Although \cite{DB-Attasena-et-al-2014-P} is an MSSS, each secret is shared and reconstructed independently. 
Inner signature $s\_d_j$ of secret $d_j$ is created with the help of an homomorphic function. Next, PTs are split into two groups: $n-t+2$ PTs in group $G_1$ and $t+2$ PTs in group $G_2$. 
Then, $t+2$ pseudo shares $\lbrace e_{j,i} \rbrace_{PT_i\in G_2}$ ($G_2$'s shares created to construct polynomial $f_2$ but not stored at $PT_i\in G_2$) are created from $d_j$'s identifier $d\_id_j$ and identifiers $\lbrace ID_i \rbrace_{PT_i\in G_2}$ of PTs in $G_2$ with an homomorphic function (Equation~\ref{eq:Attasena-et-al-2014-P-Eq1}).  

\setlength{\arraycolsep}{0.0em}
\begin{eqnarray}
	e_{j,i} &=& f_1(d\_id_j,ID_i)
	\label{eq:Attasena-et-al-2014-P-Eq1}
\end{eqnarray}

Next, a polynomial $f_2$ of degree $t-1$  is created from $d_j$, inner signature $s\_d_j$, pseudo shares $\lbrace e_{j,i} \rbrace_{PT_i\in G_2}$ and PT identifiers $\lbrace ID_i \rbrace_{PT_i\in G_2}$ by Lagrange interpolation (Equation~\ref{eq:Attasena-et-al-2014-P-Eq2}, where $\lbrace (x_1,y_2),\dots ,(x_t,y_t) \rbrace = \\ \lbrace(H(K_d),d_j),(H(K_s),s\_d_j)\rbrace \cup \lbrace (H(ID_i),e_{j,i})_{PT_i\in G_2} \rbrace$). 

\setlength{\arraycolsep}{0.0em}
\begin{eqnarray}
	f_2(x) &=& \sum_{u = 1}^t \prod _{1\leq v \leq t,u \neq v} \frac{x-x_v}{x_u - x_v}\times y_u
	\label{eq:Attasena-et-al-2014-P-Eq2}
\end{eqnarray}

Shares $\lbrace e_{j,i} \rbrace_{PT_i\in G_1}$ are created by Equation~\ref{eq:Attasena-et-al-2014-P-Eq3} and stored at $PT_i\in G_1$. To reconstruct $d_j$, $t$ out of $n$ PTs from $G_1$ and $G_2$ are selected. Secrets are reconstructed by Lagrange interpolation (Equation~\ref{eq:Attasena-et-al-2014-P-Eq2}) from both shares and pseudo shares (Equation~\ref{eq:Attasena-et-al-2014-P-Eq1}). 

\setlength{\arraycolsep}{0.0em}
\begin{eqnarray}
	e_{j,i} &=& f_2(H(IDi))
	\label{eq:Attasena-et-al-2014-P-Eq3}
\end{eqnarray}

\subsection{Group 11: Key \emph{and} Data-Verifiable Multi Secret Sharing Schemes Type~II}
\label{G11}

\cite{Shi-et-al-2007} is the only $(m,t,n)$ VMSSS type~II. It exploits elliptic curve cryptography to verify the correctness of both shares and keys. In the sharing process, keys $K=\lbrace k_{i,q}\rbrace_{i=1,\cdots,n,q=1,\cdots,t}$ are randomly chosen from small integers. Then, $l \times t$ secrets $\lbrace d_{l,q}\rbrace _{l=1,\cdots,o;q=1,\cdots,t}$ are organized into $o$  blocks $\lbrace b_l\rbrace _{l=1,\cdots,o}$ of size $t$. Each  block $b_{l}$ is divided into $n$ shares $\lbrace e_{l,i}\rbrace _{i=1,\cdots,n}$ by Equation~\ref{eq:Shi-et-al-2007-Eq1}. Signature $s\_d_{l,q}$ of $d_{l,q}$ is created by Equation~\ref{eq:Shi-et-al-2007-Eq2}, where $u$ is an elliptic curve point. Keys $\lbrace k_{i,q}\rbrace_{q=1,\cdots,t}$ and shares $\lbrace e_{l,i}\rbrace _{l=1,\cdots,o}$ are stored at $PT_{i}$, whereas  signatures\\ $\lbrace s\_d_{l,q}\rbrace_{l=1,\cdots,o;q=1,\cdots,t}$ are published on the NB. Before reconstruction, each share $e_{l,i}$ and its keys $\lbrace k_{i,q}\rbrace_{q=1,\cdots,t}$ are verified for correctness by Equation~\ref{eq:Shi-et-al-2007-Eq3}. Finally, each  block is reconstructed by solving $t$ simultaneous linear equations (Equation~\ref{eq:Shi-et-al-2007-Eq1}). 

\begin{eqnarray}
	\left[ e_{l,1},\cdots, e_{l,n} \right]^T & = & K \times \left[ b_l \right]^T \bmod p
	\label{eq:Shi-et-al-2007-Eq1} \\
	s\_d_{l,q} &=& u \times d_{l,q}
	\label{eq:Shi-et-al-2007-Eq2} \\
	u \times \left[ e_{l,1},\cdots, e_{l,n} \right]^T & = & K \times \left[ s\_d_{l,1},\cdots,s\_d_{l,t} \right]^T
	\label{eq:Shi-et-al-2007-Eq3}
\end{eqnarray}

\section{Discussion} 
\label{sec:Discussion}

In this section, we compare the SSSs presented in Section~\ref{sec:Secret-sharing-schemes} along four axes. First, we provide a global view of the evolution of SSSs since their inception (Section~\ref{sec:Evolution-sss}). Second, we synthesize and account for the various sharing and verification techniques used in SSSs to enforce data security (Section~\ref{sec:Encryption-sss}). Third, we compare the features provided by SSSs beyond data privacy and integrity (Section~\ref{sec:Properties-sss}). Finally, we study the factors that influence the cost of cloud SSS-based solutions in the pay-as-you-go paradigm (Section~\ref{sec:cost-sss}).


\subsection{Evolution of Secret Sharing Schemes} 
\label{sec:Evolution-sss}

\begin{sidewaysfigure*} 
\resizebox{\textheight}{!} {
	\begin{tikzpicture}		
		\def\y {-0.25}
		\node  at (0.9,\y+0.2) {\color{gray} \textsf{\textbf{1979-2003}}};
		\draw [ultra thin,gray] (0,\y) -- (24,\y);
			
		\edef\x {3.5}		
		\node [rectangle,fill=black!100,minimum size=4mm,rounded corners=2mm,minimum width=40mm] at (\x,\y+0.9) {~};		
		\node [fill=red!50,minimum size=4mm,minimum width=8mm] at (\x-1.6,\y+0.9) 
				{ \textsf{\textbf{\tiny{\cite{Blakley-1979}}}} };			
		\node [fill=orange!50,minimum size=4mm,minimum width=6mm] at (\x-0.9,\y+0.9) 
				{\textsf{\textbf{\tiny{G1}}} };	
		\node [fill=yellow!50,minimum size=4mm,minimum width=10mm] at (\x-0.1,\y+0.9) 
				{\textsf{\textbf{\tiny{SSS}}} };	
		\node  at (\x,\y+0.5) {\textsf{\tiny{Hyperplane intersection}}};
		
		\edef\x {12}
		\node [rectangle,fill=black!100,minimum size=4mm,rounded corners=2mm,minimum width=40mm] at (\x,\y+0.9) {~};			
		\node [fill=red!50,minimum size=4mm,minimum width=8mm] at (\x-1.6,\y+0.9) 
				{ \textsf{\textbf{\tiny{\cite{Shamir-1979}}}} };			
		\node [fill=orange!50,minimum size=4mm,minimum width=6mm] at (\x-0.9,\y+0.9) 
				{\textsf{\textbf{\tiny{G1}}} };	
		\node [fill=yellow!50,minimum size=4mm,minimum width=10mm] at (\x-0.1,\y+0.9) 
				{\textsf{\textbf{\tiny{SSS}}} };	
		\node  at (\x,\y+0.50) {\textsf{\tiny{Polynomial and Lagrange interpolation}}};	
							
		\edef\x {16.8}		
		\node [rectangle,fill=black!100,minimum size=4mm,rounded corners=2mm,minimum width=40mm] at (\x,\y+0.9) {~};			
		\node [fill=red!50,minimum size=4mm,minimum width=8mm] at (\x-1.6,\y+0.9) 
				{\textbf{\tiny{\cite{Pedersen-1991}}} };
		\node [fill=orange!50,minimum size=4mm,minimum width=6mm] at (\x-0.9,\y+0.9) 
				{\textsf{\textbf{\tiny{G4}}} };	
		\node [fill=yellow!50,minimum size=4mm,minimum width=10mm] at (\x-0.1,\y+0.9) 
				{\textsf{\textbf{\tiny{VSSS}}} };	
		\node [fill=blue!50,minimum size=4mm,minimum width=4mm] at (\x+0.6,\y+0.9) 
				{\textsf{\textbf{\tiny{D}}} };	
		\node  at (\x,\y+0.50) {\textsf{\tiny{Verifies data correctness}}};
		\draw [-stealth,thick] (\x-2,\y+0.9) --(\x-2.8,\y+0.9);
		
		\edef\y {\y+1.15}	
					
		\edef\x {3.5}		
		\node [rectangle,fill=black!100,minimum size=4mm,rounded corners=2mm,minimum width=40mm] at (\x,\y+0.9) {~};			
		\node [fill=red!50,minimum size=4mm,minimum width=8mm] at (\x-1.6,\y+0.9) 
				{ \textsf{\textbf{\tiny{\cite{Lin-We-1999}}}} };		
		\node [fill=orange!50,minimum size=4mm,minimum width=6mm] at (\x-0.9,\y+0.9) 
				{\textsf{\textbf{\tiny{G9}}} };	
		\node [fill=yellow!50,minimum size=4mm,minimum width=10mm] at (\x-0.1,\y+0.9) 
				{\textsf{\textbf{\tiny{VMSSS-I}}} };
		\node [fill=green!50,minimum size=4mm,minimum width=4mm] at (\x+0.6,\y+0.9) 
				{\textsf{\textbf{\tiny{K}}} };	
		\node [fill=blue!50,minimum size=4mm,minimum width=4mm] at (\x+1.0,\y+0.9) 
				{\textsf{\textbf{\tiny{D}}} };	
		\node  at (\x,\y+0.50) {\textsf{\tiny{Factorisation intractability}}};
		\node  at (\x,\y+0.25) {\textsf{\tiny{and discrete logarithm modulo}}};
		
		\edef\x {9.0}		
		\node [rectangle,fill=black!100,minimum size=4mm,rounded corners=2mm,minimum width=40mm] at (\x,\y+0.9) {~};					
		\node [fill=red!50,minimum size=4mm,minimum width=8mm] at (\x-1.6,\y+0.9) 
				{ \textsf{\textbf{\tiny{\cite{He-Dawson-1994}}}} };			
		\node [fill=orange!50,minimum size=4mm,minimum width=6mm] at (\x-0.9,\y+0.9) 
				{\textsf{\textbf{\tiny{G1}}} };	
		\node [fill=yellow!50,minimum size=4mm,minimum width=10mm] at (\x-0.1,\y+0.9) 
				{\textsf{\textbf{\tiny{SSS}}} };	
		\node  at (\x,\y+0.5) {\textsf{\tiny{$t$-consistency property}}};
		\draw [-stealth,thick] (\x+1.9,\y+0.7) --(\x+1.9,0.85);
		
		\def\x {16.3}
		\node [rectangle,fill=black!100,minimum size=4mm,rounded corners=2mm,minimum width=40mm] at (\x,\y+0.9) {~};			
		\node [fill=red!50,minimum size=4mm,minimum width=8mm] at (\x-1.6,\y+0.9) 
				{ \textsf{\textbf{\tiny{\cite{Asmuth-Bloom-1983}}}} };			
		\node [fill=orange!50,minimum size=4mm,minimum width=6mm] at (\x-0.9,\y+0.9) 
				{\textsf{\textbf{\tiny{G1}}} };	
		\node [fill=yellow!50,minimum size=4mm,minimum width=10mm] at (\x-0.1,\y+0.9) 
				{\textsf{\textbf{\tiny{SSS}}} };
		\node  at (\x,\y+0.5) {\textsf{\tiny{Chinese remainder theorem}}};

		\edef\x {22}		
		\node [rectangle,fill=black!100,minimum size=4mm,rounded corners=2mm,minimum width=40mm] at (\x,\y+0.9) {~};			
		\node [fill=red!50,minimum size=4mm,minimum width=8mm] at (\x-1.6,\y+0.9) 
				{ \textsf{\textbf{\tiny{\cite{Hwang-Chang-1998}}}} };			
		\node [fill=orange!50,minimum size=4mm,minimum width=6mm] at (\x-0.9,\y+0.9) 
				{\textsf{\textbf{\tiny{G5}}} };	
		\node [fill=yellow!50,minimum size=4mm,minimum width=10mm] at (\x-0.1,\y+0.9) 
				{\textsf{\textbf{\tiny{VSSS}}} };	
		\node  at (\x,\y+0.50) {\textsf{\tiny{RSA and Lagrange interpolation}}};
		
		\edef\y {\y+1.35}
		\node  at (0.4,\y+0.2) {\color{gray} \textsf{\textbf{2004}}};
		\draw [ultra thin,gray] (0,\y) -- (24,\y);		
	
		\edef\x {9.5}		
		\node [rectangle,fill=black!100,minimum size=4mm,rounded corners=2mm,minimum width=40mm] at (\x,\y+0.9) {~};			
		\node [fill=red!50,minimum size=4mm,minimum width=8mm] at (\x-1.6,\y+0.9) 
				{ \textsf{\textbf{\tiny{\cite{Yang-et-al-2004}}}} };			
		\node [fill=orange!50,minimum size=4mm,minimum width=6mm] at (\x-0.9,\y+0.9) 
				{\textsf{\textbf{\tiny{G2}}} };	
		\node [fill=yellow!50,minimum size=4mm,minimum width=10mm] at (\x-0.1,\y+0.9) 
				{\textsf{\textbf{\tiny{MSSS-I}}} };	
		\node  at (\x,\y+0.5) {\textsf{\tiny{Reduces costs by sharing}}};
		\node  at (\x,\y+0.25) {\textsf{\tiny{and reconstructing all data at once}}};
		\draw [-stealth,thick] (\x+1.9,\y+0.7) --(\x+1.9,0.85);
		
		\edef\y {\y+1.35}
		\node  at (0.4,\y+0.2) {\color{gray} \textsf{\textbf{2005}}};
		\draw [ultra thin,gray] (0,\y) -- (24,\y);	
				
		\edef\x {3.5}		
		\node [rectangle,fill=black!100,minimum size=4mm,rounded corners=2mm,minimum width=40mm] at (\x,\y+0.9) {~};			
		\node [fill=red!50,minimum size=4mm,minimum width=8mm] at (\x-1.6,\y+0.9) 
				{ \textsf{\textbf{\tiny{\cite{Chang-et-al-2005}}}} };			
		\node [fill=orange!50,minimum size=4mm,minimum width=6mm] at (\x-0.9,\y+0.9) 
				{\textsf{\textbf{\tiny{G9}}} };	
		\node [fill=yellow!50,minimum size=4mm,minimum width=10mm] at (\x-0.1,\y+0.9) 
				{\textsf{\textbf{\tiny{VMSSS-I}}} };
		\node [fill=green!50,minimum size=4mm,minimum width=4mm] at (\x+0.6,\y+0.9) 
				{\textsf{\textbf{\tiny{K}}} };	
		\node [fill=blue!50,minimum size=4mm,minimum width=4mm] at (\x+1.0,\y+0.9) 
				{\textsf{\textbf{\tiny{D}}} };	
		\node  at (\x,\y+0.5) {\textsf{\tiny{Improve the efficiency of}}};
		\node  at (\x,\y+0.25) {\textsf{\tiny{the sharing and reconstruction processes}}};
		\draw [-stealth,thick] (\x+1.9,\y+0.7) --(\x+1.9,2.0);
		
		\edef\x {8}		
		\node [rectangle,fill=black!100,minimum size=4mm,rounded corners=2mm,minimum width=40mm] at (\x,\y+0.9) {~};			
		\node [fill=red!50,minimum size=4mm,minimum width=8mm] at (\x-1.6,\y+0.9) 
				{ \textsf{\textbf{\tiny{\cite{Shao-Cao-2005}}}} };			
		\node [fill=orange!50,minimum size=4mm,minimum width=6mm] at (\x-0.9,\y+0.9) 
				{\textsf{\textbf{\tiny{G9}}} };	
		\node [fill=yellow!50,minimum size=4mm,minimum width=10mm] at (\x-0.1,\y+0.9) 
				{\textsf{\textbf{\tiny{VMSSS-I}}} };	
		\node [fill=green!50,minimum size=4mm,minimum width=4mm] at (\x+0.6,\y+0.9) 
				{\textsf{\textbf{\tiny{K}}} };	
		\node [fill=blue!50,minimum size=4mm,minimum width=4mm] at (\x+1.0,\y+0.9) 
				{\textsf{\textbf{\tiny{D}}} };	
		\node  at (\x,\y+0.5) {\textsf{\tiny{New process for verifying}}};
		\node  at (\x,\y+0.25) {\textsf{\tiny{key and data correctness}}};
		\draw [-stealth,thick] (\x+1.9,\y+0.7) --(\x+1.9,3.4);
	
		\edef\x {15.8}		
		\node [rectangle,fill=black!100,minimum size=4mm,rounded corners=2mm,minimum width=40mm] at (\x,\y+0.9) {~};			
		\node [fill=red!50,minimum size=4mm,minimum width=8mm] at (\x-1.6,\y+0.9) 
				{ \textsf{\textbf{\tiny{\cite{Chan-Chang-2005}}}} };			
		\node [fill=orange!50,minimum size=4mm,minimum width=6mm] at (\x-0.9,\y+0.9) 
				{\textsf{\textbf{\tiny{G3}}} };	
		\node [fill=yellow!50,minimum size=4mm,minimum width=10mm] at (\x-0.1,\y+0.9) 
				{\textsf{\textbf{\tiny{MSSS-II}}} };	
		\node  at (\x,\y+0.5) {\textsf{\tiny{Chinese remainder theorem}}};
		\node  at (\x,\y+0.25) {\textsf{\tiny{and polynomial interpolation}}};
		\draw [-stealth,thick] (\x-1.9,\y+0.7) --(\x-1.9,0.85); 
		\draw [-stealth,thick] (\x+1.9,\y+0.7) --(\x+1.9,2.00); 
		
		\edef\y {\y+1.35}
		\node  at (0.4,\y+0.2) {\color{gray} \textsf{\textbf{2007}}};
		\draw [ultra thin,gray] (0,\y) -- (24,\y);
							
		\edef\x {3.5}		
		\node [rectangle,fill=black!100,minimum size=4mm,rounded corners=2mm,minimum width=40mm] at (\x,\y+0.9) {~};			
		\node [fill=red!50,minimum size=4mm,minimum width=8mm] at (\x-1.6,\y+0.9) 
				{ \textsf{\textbf{\tiny{\cite{Shi-et-al-2007}}}} };			
		\node [fill=orange!50,minimum size=4mm,minimum width=6mm] at (\x-0.9,\y+0.9) 
				{\textsf{\textbf{\tiny{G11}}} };	
		\node [fill=yellow!50,minimum size=4mm,minimum width=10mm] at (\x-0.1,\y+0.9) 
				{\textsf{\textbf{\tiny{VMSSS-II}}} };	
		\node [fill=green!50,minimum size=4mm,minimum width=4mm] at (\x+0.6,\y+0.9) 
				{\textsf{\textbf{\tiny{K}}} };	
		\node [fill=blue!50,minimum size=4mm,minimum width=4mm] at (\x+1.0,\y+0.9) 
				{\textsf{\textbf{\tiny{D}}} };	
		\node  at (\x,\y+0.5) {\textsf{\tiny{Linear equations and}}};
		\node  at (\x,\y+0.25) {\textsf{\tiny{elliptic curve cryptography}}};
			
		\edef\x {8.5}		
		\node [rectangle,fill=black!100,minimum size=4mm,rounded corners=2mm,minimum width=40mm] at (\x,\y+0.9) {~};			
		\node [fill=red!50,minimum size=4mm,minimum width=8mm] at (\x-1.6,\y+0.9) 
				{ \textsf{\textbf{\tiny{\cite{Zhao-et-al-2007}}}} };			
		\node [fill=orange!50,minimum size=4mm,minimum width=6mm] at (\x-0.9,\y+0.9) 
				{\textsf{\textbf{\tiny{G8}}} };	
		\node [fill=yellow!50,minimum size=4mm,minimum width=10mm] at (\x-0.1,\y+0.9) 
				{\textsf{\textbf{\tiny{VMSSS-I}}} };	
		\node [fill=green!50,minimum size=4mm,minimum width=4mm] at (\x+0.6,\y+0.9) 
				{\textsf{\textbf{\tiny{K}}} };	
		\node  at (\x,\y+0.5) {\textsf{\tiny{New sharing and reconstruction processes}}};
		\draw [-stealth,thick] (\x+1.9,\y+0.7) --(\x+1.9,3.4);
				
		\edef\x {16.3}		
		\node [rectangle,fill=black!100,minimum size=4mm,rounded corners=2mm,minimum width=40mm] at (\x,\y+0.9) {~};			
		\node [fill=red!50,minimum size=4mm,minimum width=8mm] at (\x-1.6,\y+0.9) 
				{ \textsf{\textbf{\tiny{\cite{Iftene-2007}}}} };			
		\node [fill=orange!50,minimum size=4mm,minimum width=6mm] at (\x-0.9,\y+0.9) 
				{\textsf{\textbf{\tiny{G1}}} };	
		\node [fill=yellow!50,minimum size=4mm,minimum width=10mm] at (\x-0.1,\y+0.9) 
				{\textsf{\textbf{\tiny{SSS}}} };	
		\node  at (\x,\y+0.5) {\textsf{\tiny{Reduces share volume}}};
		\draw [-stealth,thick] (\x+1.9,\y+0.7) --(\x+1.9,2.0);
			
		\edef\x {22}		
		\node [rectangle,fill=black!100,minimum size=4mm,rounded corners=2mm,minimum width=40mm] at (\x,\y+0.9) {~};			
		\node [fill=red!50,minimum size=4mm,minimum width=8mm] at (\x-1.6,\y+0.9) 
				{ \textsf{\textbf{\tiny{\cite{Wei-et-al-2007}}}} };			
		\node [fill=orange!50,minimum size=4mm,minimum width=6mm] at (\x-0.9,\y+0.9) 
				{\textsf{\textbf{\tiny{G9}}} };	
		\node [fill=yellow!50,minimum size=4mm,minimum width=10mm] at (\x-0.1,\y+0.9) 
				{\textsf{\textbf{\tiny{VMSSS-I}}} };	
		\node [fill=green!50,minimum size=4mm,minimum width=4mm] at (\x+0.6,\y+0.9) 
				{\textsf{\textbf{\tiny{K}}} };	
		\node [fill=blue!50,minimum size=4mm,minimum width=4mm] at (\x+1.0,\y+0.9) 
				{\textsf{\textbf{\tiny{D}}} };	
		\node  at (\x,\y+0.5) {\textsf{\tiny{Bilinear map, linear equations}}};
		\node  at (\x,\y+0.25) {\textsf{\tiny{and hash function}}};
		
		\edef\y {\y+1.35}
		\node  at (0.4,\y+0.2) {\color{gray} \textsf{\textbf{2008}}};
		\draw [ultra thin,gray] (0,\y) -- (24,\y);
			
		\edef\x {4.5}		
		\node [rectangle,fill=black!100,minimum size=4mm,rounded corners=2mm,minimum width=40mm] at (\x,\y+0.9) {~};			
		\node [fill=red!50,minimum size=4mm,minimum width=8mm] at (\x-1.6,\y+0.9) 
				{ \textsf{\textbf{\tiny{\cite{Dehkordi-Mashhadi-2008b}}}} };			
		\node [fill=orange!50,minimum size=4mm,minimum width=6mm] at (\x-0.9,\y+0.9) 
				{\textsf{\textbf{\tiny{G8}}} };	
		\node [fill=yellow!50,minimum size=4mm,minimum width=10mm] at (\x-0.1,\y+0.9) 
				{\textsf{\textbf{\tiny{VMSSS-I}}} };	
		\node [fill=green!50,minimum size=4mm,minimum width=4mm] at (\x+0.6,\y+0.9) 
				{\textsf{\textbf{\tiny{K}}} };	
		\node  at (\x,\y+0.5) {\textsf{\tiny{New sharing and reconstruction processes}}};
		\draw [-stealth,thick] (\x+2,\y+0.9)--(7,\y+0.9);
			
		\edef\x {9}		
		\node [rectangle,fill=black!100,minimum size=4mm,rounded corners=2mm,minimum width=40mm] at (\x,\y+0.9) {~};			
		\node [fill=red!50,minimum size=4mm,minimum width=8mm] at (\x-1.6,\y+0.9) 
				{ \textsf{\textbf{\tiny{\cite{Dehkordi-Mashhadi-2008a}}}} };			
		\node [fill=orange!50,minimum size=4mm,minimum width=6mm] at (\x-0.9,\y+0.9) 
				{\textsf{\textbf{\tiny{G8}}} };		
		\node [fill=green!50,minimum size=4mm,minimum width=4mm] at (\x+0.6,\y+0.9) 
				{\textsf{\textbf{\tiny{K}}} };	
		\node [fill=yellow!50,minimum size=4mm,minimum width=10mm] at (\x-0.1,\y+0.9) 
				{\textsf{\textbf{\tiny{VMSSS-I}}} };	
		\node  at (\x,\y+0.5) {\textsf{\tiny{New process for verifying}}};
		\node  at (\x,\y+0.25) {\textsf{\tiny{key correctness}}};
		\draw [-stealth,thick] (\x+1.9,\y+0.7) --(\x+1.9,3.4);
			
		\edef\x {15.3}		
		\node [rectangle,fill=black!100,minimum size=4mm,rounded corners=2mm,minimum width=40mm] at (\x,\y+0.9) {~};			
		\node [fill=red!50,minimum size=4mm,minimum width=8mm] at (\x-1.6,\y+0.9) 
				{ \textsf{\textbf{\tiny{\cite{Tang-Yao-2008}}}} };			
		\node [fill=orange!50,minimum size=4mm,minimum width=6mm] at (\x-0.9,\y+0.9) 
				{\textsf{\textbf{\tiny{G4}}} };	
		\node [fill=yellow!50,minimum size=4mm,minimum width=10mm] at (\x-0.1,\y+0.9) 
				{\textsf{\textbf{\tiny{VSSS}}} };	
		\node [fill=blue!50,minimum size=4mm,minimum width=4mm] at (\x+0.6,\y+0.9) 
				{\textsf{\textbf{\tiny{D}}} };	
		\node  at (\x,\y+0.5) {\textsf{\tiny{New process for verifying}}};
		\node  at (\x,\y+0.25) {\textsf{\tiny{data correctness}}};
		\draw [-stealth,thick] (\x-1.9,\y+0.7) --(\x-1.9,0.85);
		
		\edef\x {21.05}		
		\node [rectangle,fill=black!100,minimum size=4mm,rounded corners=2mm,minimum width=40mm] at (\x,\y+0.9) {~};			
		\node [fill=red!50,minimum size=4mm,minimum width=8mm] at (\x-1.6,\y+0.9) 
				{ \textsf{\textbf{\tiny{\cite{Runhua-et-al-2008}}}} };			
		\node [fill=orange!50,minimum size=4mm,minimum width=6mm] at (\x-0.9,\y+0.9) 
				{\textsf{\textbf{\tiny{G3}}} };	
		\node [fill=yellow!50,minimum size=4mm,minimum width=10mm] at (\x-0.1,\y+0.9) 
				{\textsf{\textbf{\tiny{MSSS-II}}} };	
		\node  at (\x,\y+0.5) {\textsf{\tiny{Linear equations}}};
					
		\edef\y {\y+1.35}
		\node  at (0.4,\y+0.2) {\color{gray} \textsf{\textbf{2009}}};
		\draw [ultra thin,gray] (0,\y) -- (24,\y);	
			
		\edef\x {9}		
		\node [rectangle,fill=black!100,minimum size=4mm,rounded corners=2mm,minimum width=40mm] at (\x,\y+0.9) {~};			
		\node [fill=red!50,minimum size=4mm,minimum width=8mm] at (\x-1.6,\y+0.9) 
				{ \textsf{\textbf{\tiny{\cite{Parakh-Kak-2009}}}} };			
		\node [fill=orange!50,minimum size=4mm,minimum width=6mm] at (\x-0.9,\y+0.9) 
				{\textsf{\textbf{\tiny{G1}}} };	
		\node [fill=yellow!50,minimum size=4mm,minimum width=10mm] at (\x-0.1,\y+0.9) 
				{\textsf{\textbf{\tiny{SSS}}} };	
		\node  at (\x,\y+0.5) {\textsf{\tiny{Polynomial interpolation}}};
			
		\edef\x {14.8}		
		\node [rectangle,fill=black!100,minimum size=4mm,rounded corners=2mm,minimum width=40mm] at (\x,\y+0.9) {~};			
		\node [fill=red!50,minimum size=4mm,minimum width=8mm] at (\x-1.6,\y+0.9) 
				{ \textsf{\textbf{\tiny{\cite{Yue-Hong-2009}}}} };			
		\node [fill=orange!50,minimum size=4mm,minimum width=6mm] at (\x-0.9,\y+0.9) 
				{\textsf{\textbf{\tiny{G4}}} };	
		\node [fill=yellow!50,minimum size=4mm,minimum width=10mm] at (\x-0.1,\y+0.9) 
				{\textsf{\textbf{\tiny{VSSS}}} };	
		\node [fill=blue!50,minimum size=4mm,minimum width=4mm] at (\x+0.6,\y+0.9) 
				{\textsf{\textbf{\tiny{D}}} };		
		\node  at (\x+0.2,\y+0.5) {\textsf{\tiny{NTRU algorithm, Lagrange interpolation}}};
		\node  at (\x,\y+0.25) {\textsf{\tiny{and hash function}}};
		\draw [-stealth,thick] (\x-1.9,\y+0.7) --(\x-1.9,0.85);
		
		\edef\y {\y+1.35}
		\node  at (0.4,\y+0.2) {\color{gray} \textsf{\textbf{2010}}};
		\draw [ultra thin,gray] (0,\y) -- (24,\y);
		
		\edef\x {4}	
		\node [rectangle,fill=black!100,minimum size=4mm,rounded corners=2mm,minimum width=40mm] at (\x,\y+0.9) {~};			
		\node [fill=red!50,minimum size=4mm,minimum width=8mm] at (\x-1.6,\y+0.9) 
				{ \textsf{\textbf{\tiny{\cite{Das-Adhikari-2010}}}} };			
		\node [fill=orange!50,minimum size=4mm,minimum width=6mm] at (\x-0.9,\y+0.9) 
				{\textsf{\textbf{\tiny{G9}}} };	
		\node [fill=yellow!50,minimum size=4mm,minimum width=10mm] at (\x-0.1,\y+0.9) 
				{\textsf{\textbf{\tiny{VMSSS-I}}} };	
		\node [fill=green!50,minimum size=4mm,minimum width=4mm] at (\x+0.6,\y+0.9) 
				{\textsf{\textbf{\tiny{K}}} };	
		\node [fill=blue!50,minimum size=4mm,minimum width=4mm] at (\x+1.0,\y+0.9) 
				{\textsf{\textbf{\tiny{D}}} };	
		\node  at (\x,\y+0.5) {\textsf{\tiny{Shares binary data}}};
			
		\edef\x {9}		
		\node [rectangle,fill=black!100,minimum size=4mm,rounded corners=2mm,minimum width=40mm] at (\x,\y+0.9) {~};			
		\node [fill=red!50,minimum size=4mm,minimum width=8mm] at (\x-1.6,\y+0.9) 
				{ \textsf{\textbf{\tiny{\cite{Eslami-Ahmadabadi-2010}}}} };			
		\node [fill=orange!50,minimum size=4mm,minimum width=6mm] at (\x-0.9,\y+0.9) 
				{\textsf{\textbf{\tiny{G7}}} };	
		\node [fill=yellow!50,minimum size=4mm,minimum width=10mm] at (\x-0.1,\y+0.9) 
				{\textsf{\textbf{\tiny{VMSSS-I}}} };	
		\node [fill=blue!50,minimum size=4mm,minimum width=4mm] at (\x+0.6,\y+0.9) 
				{\textsf{\textbf{\tiny{D}}} };		
		\node  at (\x,\y+0.5) {\textsf{\tiny{Cellular automata}}};
		
		\edef\x {20.6}		
		\node [rectangle,fill=black!100,minimum size=4mm,rounded corners=2mm,minimum width=40mm] at (\x,\y+0.9) {~};			
		\node [fill=red!50,minimum size=4mm,minimum width=8mm] at (\x-1.6,\y+0.9) 
				{ \textsf{\textbf{\tiny{\cite{Harn-Lin-2010}}}} };			
		\node [fill=orange!50,minimum size=4mm,minimum width=6mm] at (\x-0.9,\y+0.9) 
				{\textsf{\textbf{\tiny{G1}}} };	
		\node [fill=yellow!50,minimum size=4mm,minimum width=10mm] at (\x-0.1,\y+0.9) 
				{\textsf{\textbf{\tiny{SSS}}} };	
		\node  at (\x,\y+0.5) {\textsf{\tiny{Verifies a strong consistency property}}};
		\node  at (\x,\y+0.25) {\textsf{\tiny{~~}}};
		\draw [-stealth,thick] (\x-1.9,\y+0.7) --(\x-1.9,0.85);
		
		\edef\y {\y+1.35}
		\node  at (0.4,\y+0.2) {\color{gray} \textsf{\textbf{2011}}};
		\draw [ultra thin,gray] (0,\y) -- (24,\y);
		
		\edef\x {14.3}		
		\node [rectangle,fill=black!100,minimum size=4mm,rounded corners=2mm,minimum width=40mm] at (\x,\y+0.9) {~};			
		\node [fill=red!50,minimum size=4mm,minimum width=8mm] at (\x-1.6,\y+0.9) 
				{ \textsf{\textbf{\tiny{\cite{Parakh-Kak-2011}}}} };			
		\node [fill=orange!50,minimum size=4mm,minimum width=6mm] at (\x-0.9,\y+0.9) 
				{\textsf{\textbf{\tiny{G1}}} };	
		\node [fill=yellow!50,minimum size=4mm,minimum width=10mm] at (\x-0.1,\y+0.9) 
				{\textsf{\textbf{\tiny{SSS}}} };	
		\node  at (\x,\y+0.5) {\textsf{\tiny{Improves security by}}};
		\node  at (\x,\y+0.25) {\textsf{\tiny{sharing \cite{Shamir-1979} $t$ times}}};
		\draw [-stealth,thick] (\x-1.9,\y+0.7) --(\x-1.9,0.85);
			
		\edef\x {21.5}		
		\node [rectangle,fill=black!100,minimum size=4mm,rounded corners=2mm,minimum width=40mm] at (\x,\y+0.9) {~};			
		\node [fill=red!50,minimum size=4mm,minimum width=8mm] at (\x-1.6,\y+0.9) 
				{ \textsf{\textbf{\tiny{\cite{Wang-et-al-2011}}}} };			
		\node [fill=orange!50,minimum size=4mm,minimum width=6mm] at (\x-0.9,\y+0.9) 
				{\textsf{\textbf{\tiny{G9}}} };	
		\node [fill=yellow!50,minimum size=4mm,minimum width=10mm] at (\x-0.1,\y+0.9) 
				{\textsf{\textbf{\tiny{VMSSS-I}}} };	
		\node [fill=green!50,minimum size=4mm,minimum width=4mm] at (\x+0.6,\y+0.9) 
				{\textsf{\textbf{\tiny{K}}} };	
		\node [fill=blue!50,minimum size=4mm,minimum width=4mm] at (\x+1.0,\y+0.9) 
				{\textsf{\textbf{\tiny{D}}} };	
		\node  at (\x,\y+0.5) {\textsf{\tiny{New sharing and reconstruction processes}}};
		\draw [-stealth,thick] (\x+1.9,\y+0.7) --(\x+1.9,6.1);
				
		\edef\y {\y+1.35}
		\node  at (0.8,\y+0.2) {\color{gray} \textsf{\textbf{2012-2016}}};
		\draw [ultra thin,gray] (0,\y) -- (24,\y);
			
		\edef\x {4.5}		
		\node [rectangle,fill=black!100,minimum size=4mm,rounded corners=2mm,minimum width=40mm] at (\x,\y+0.9) {~};			
		\node [fill=red!50,minimum size=4mm,minimum width=8mm] at (\x-1.6,\y+0.9) 
				{ \textsf{\textbf{\tiny{\cite{Hu-et-al-2012}}}} };			
		\node [fill=orange!50,minimum size=4mm,minimum width=6mm] at (\x-0.9,\y+0.9) 
				{\textsf{\textbf{\tiny{G8}}} };	
		\node [fill=yellow!50,minimum size=4mm,minimum width=10mm] at (\x-0.1,\y+0.9) 
				{\textsf{\textbf{\tiny{VMSSS-I}}} };	
		\node [fill=green!50,minimum size=4mm,minimum width=4mm] at (\x+0.6,\y+0.9) 
				{\textsf{\textbf{\tiny{K}}} };	
		\node  at (\x,\y+0.50) {\textsf{\tiny{New process for verifying}}};
		\node  at (\x,\y+0.25) {\textsf{\tiny{key correctness}}};			
		\draw [-stealth,thick] (\x+1.9,\y+0.7) --(\x+1.9,7.4);
		\draw [-stealth,thick] (\x+2,\y+0.9) --(7.5,\y+0.9); 
			
		\edef\x {9.5}		
		\node [rectangle,fill=black!100,minimum size=4mm,rounded corners=2mm,minimum width=40mm] at (\x,\y+0.9) {~};			
		\node [fill=red!50,minimum size=4mm,minimum width=8mm] at (\x-1.6,\y+0.9) 
				{ \textsf{\textbf{\tiny{\cite{Hu-et-al-2012}}}} };			
		\node [fill=orange!50,minimum size=4mm,minimum width=6mm] at (\x-0.9,\y+0.9) 
				{\textsf{\textbf{\tiny{G8}}} };	
		\node [fill=yellow!50,minimum size=4mm,minimum width=10mm] at (\x-0.1,\y+0.9) 
				{\textsf{\textbf{\tiny{VMSSS-I}}} };	
		\node [fill=green!50,minimum size=4mm,minimum width=4mm] at (\x+0.6,\y+0.9) 
				{\textsf{\textbf{\tiny{K}}} };	
		\node  at (\x,\y+0.50) {\textsf{\tiny{New process for verifying}}};
		\node  at (\x,\y+0.25) {\textsf{\tiny{key correctness}}};		
		\draw [-stealth,thick] (\x+1.9,\y+0.7) --(\x+1.9,3.4);
			
		\edef\x {14.3}		
		\node [rectangle,fill=black!100,minimum size=4mm,rounded corners=2mm,minimum width=40mm] at (\x,\y+0.9) {~};			
		\node [fill=red!50,minimum size=4mm,minimum width=8mm] at (\x-1.6,\y+0.9) 
				{ \textsf{\textbf{\tiny{\cite{Li-et-al-2012}}}} };			
		\node [fill=orange!50,minimum size=4mm,minimum width=6mm] at (\x-0.9,\y+0.9) 
				{\textsf{\textbf{\tiny{G8}}} };	
		\node [fill=yellow!50,minimum size=4mm,minimum width=10mm] at (\x-0.1,\y+0.9) 
				{\textsf{\textbf{\tiny{VMSSS-I}}} };	
		\node [fill=green!50,minimum size=4mm,minimum width=4mm] at (\x+0.6,\y+0.9) 
				{\textsf{\textbf{\tiny{K}}} };	
		\node  at (\x,\y+0.50) {\textsf{\tiny{New process for verifying}}};
		\node  at (\x,\y+0.25) {\textsf{\tiny{key correctness}}};		
		\draw [-stealth,thick] (\x-1.9,\y+0.7) --(\x-1.9,11.5);
			
		\edef\x {20.6}		
		\node [rectangle,fill=black!100,minimum size=4mm,rounded corners=2mm,minimum width=40mm] at (\x,\y+0.9) {~};			
		\node [fill=red!50,minimum size=4mm,minimum width=8mm] at (\x-1.6,\y+0.9) 
				{ \textsf{\textbf{\tiny{\cite{Liu-et-al-2012} SSS}}} };			
		\node [fill=orange!50,minimum size=4mm,minimum width=6mm] at (\x-0.9,\y+0.9) 
				{\textsf{\textbf{\tiny{G1}}} };	
		\node [fill=yellow!50,minimum size=4mm,minimum width=10mm] at (\x-0.1,\y+0.9) 
				{\textsf{\textbf{\tiny{SSS}}} };
		\node  at (\x,\y+0.5) {\textsf{\tiny{Stronger $t$-consistency property}}};
		\draw [-stealth,thick] (\x-1.9,\y+0.7) --(\x-1.9,10.15);

	//-------- 2012 line 2 --------------					
		\edef\y {\y+1.15}		
			
		\edef\x {3.5}		
		\node [rectangle,fill=black!100,minimum size=4mm,rounded corners=2mm,minimum width=40mm] at (\x,\y+0.9) {~};			
		\node [fill=red!50,minimum size=4mm,minimum width=8mm] at (\x-1.6,\y+0.9) 
				{ \textsf{\textbf{\tiny{\cite{Waseda-Soshi-2012}}}} };			
		\node [fill=orange!50,minimum size=4mm,minimum width=6mm] at (\x-0.9,\y+0.9) 
				{\textsf{\textbf{\tiny{G2}}} };	
		\node [fill=yellow!50,minimum size=4mm,minimum width=10mm] at (\x-0.1,\y+0.9) 
				{\textsf{\textbf{\tiny{MSSS-I}}} };	
		\node  at (\x,\y+0.5) {\textsf{\tiny{Reduce execution time and}}};
		\node  at (\x,\y+0.25) {\textsf{\tiny{adjusts data block size}}};		
		\draw [-stealth,thick] (\x-1.9,\y+0.7) --(\x-1.9,6.1);
		
		\edef\x {10}		
		\node [rectangle,fill=black!100,minimum size=4mm,rounded corners=2mm,minimum width=40mm] at (\x,\y+0.9) {~};			
		\node [fill=red!50,minimum size=4mm,minimum width=8mm] at (\x-1.6,\y+0.9) 
				{ \textsf{\textbf{\tiny{\cite{DB-Attasena-et-al-2014-P}}}} };			
		\node [fill=orange!50,minimum size=4mm,minimum width=6mm] at (\x-0.9,\y+0.9) 
				{\textsf{\textbf{\tiny{G10}}} };	
		\node [fill=yellow!50,minimum size=4mm,minimum width=10mm] at (\x-0.1,\y+0.9) 
				{\textsf{\textbf{\tiny{VMSSS-II}}} };	
		\node [fill=blue!50,minimum size=4mm,minimum width=4mm] at (\x+0.6,\y+0.9) 
				{\textsf{\textbf{\tiny{D}}} };	
		\node  at (\x,\y+0.5) {\textsf{\tiny{Reduces storage cost and}}};
		\node  at (\x,\y+0.25) {\textsf{\tiny{verifies data correctness}}};		
		\draw [-stealth,thick] (\x+1.9,\y+0.7) --(\x+1.9,0.85); 
		
		\edef\x {14.8}		
		\node [rectangle,fill=black!100,minimum size=4mm,rounded corners=2mm,minimum width=40mm] at (\x,\y+0.9) {~};			
		\node [fill=red!50,minimum size=4mm,minimum width=8mm] at (\x-1.6,\y+0.9) 
				{ \textsf{\textbf{\tiny{\cite{Bu-Yang-2012}}}} };			
		\node [fill=orange!50,minimum size=4mm,minimum width=6mm] at (\x-0.9,\y+0.9) 
				{\textsf{\textbf{\tiny{G9}}} };	
		\node [fill=yellow!50,minimum size=4mm,minimum width=10mm] at (\x-0.1,\y+0.9) 
				{\textsf{\textbf{\tiny{VMSSS-I}}} };	
		\node [fill=green!50,minimum size=4mm,minimum width=4mm] at (\x+0.6,\y+0.9) 
				{\textsf{\textbf{\tiny{K}}} };	
		\node [fill=blue!50,minimum size=4mm,minimum width=4mm] at (\x+1.0,\y+0.9) 
				{\textsf{\textbf{\tiny{D}}} };	
		\node  at (\x,\y+0.5) {\textsf{\tiny{NTRU and Lagrange interpolation}}};
		\draw [-stealth,thick] (\x+1.9,\y+0.7) --(\x+1.9,8.8); 	
			
		\edef\x {20.6}		
		\node [rectangle,fill=black!100,minimum size=4mm,rounded corners=2mm,minimum width=40mm] at (\x,\y+0.9) {~};			
		\node [fill=red!50,minimum size=4mm,minimum width=8mm] at (\x-1.6,\y+0.9) 
				{ \textsf{\textbf{\tiny{\cite{Liu-et-al-2012} MSSS}}} };			
		\node [fill=orange!50,minimum size=4mm,minimum width=6mm] at (\x-0.9,\y+0.9) 
				{\textsf{\textbf{\tiny{G3}}} };	
		\node [fill=yellow!50,minimum size=4mm,minimum width=10mm] at (\x-0.1,\y+0.9) 
				{\textsf{\textbf{\tiny{MSSS-II}}} };	
		\node  at (\x,\y+0.5) {\textsf{\tiny{Stronger $t$-consistency property}}};
		\draw [-stealth,thick] (\x-1.9,\y+0.7) --(\x-1.9,12.8);

	//-------- 2012 line 3 --------------				
		\edef\y {\y+1.1}
				
		\edef\x {3.2}		
		\node [rectangle,fill=black!100,minimum size=4mm,rounded corners=2mm,minimum width=40mm] at (\x,\y+0.9) {~};			
		\node [fill=red!50,minimum size=4mm,minimum width=8mm] at (\x-1.6,\y+0.9) 
				{ \textsf{\textbf{\tiny{\cite{Takahashi-Iwamura-2013}}}} };			
		\node [fill=orange!50,minimum size=4mm,minimum width=6mm] at (\x-0.9,\y+0.9) 
				{\textsf{\textbf{\tiny{G3}}} };	
		\node [fill=yellow!50,minimum size=4mm,minimum width=10mm] at (\x-0.1,\y+0.9) 
				{\textsf{\textbf{\tiny{MSSS-II}}} };	
		\node  at (\x,\y+0.5) {\textsf{\tiny{Does not share at all PTs'}}};
		
		\edef\x {7.9}		
		\node [rectangle,fill=black!100,minimum size=4mm,rounded corners=2mm,minimum width=40mm] at (\x,\y+0.9) {~};			
		\node [fill=red!50,minimum size=4mm,minimum width=8mm] at (\x-1.6,\y+0.9) 
				{ \textsf{\textbf{\tiny{\cite{Zhao-et-al-2012}}}} };			
		\node [fill=orange!50,minimum size=4mm,minimum width=6mm] at (\x-0.9,\y+0.9) 
				{\textsf{\textbf{\tiny{G6}}} };	
		\node [fill=yellow!50,minimum size=4mm,minimum width=10mm] at (\x-0.1,\y+0.9) 
				{\textsf{\textbf{\tiny{VSSS}}} };	
		\node [fill=green!50,minimum size=4mm,minimum width=4mm] at (\x+0.6,\y+0.9) 
				{\textsf{\textbf{\tiny{K}}} };	
		\node [fill=blue!50,minimum size=4mm,minimum width=4mm] at (\x+1.0,\y+0.9) 
				{\textsf{\textbf{\tiny{D}}} };	
		\node  at (\x,\y+0.5) {\textsf{\tiny{Verifies both keys and data}}};
		\node  at (\x,\y+0.25) {\textsf{\tiny{and reduces share volume}}};
		
		\edef\x {12.6}		
		\node [rectangle,fill=black!100,minimum size=4mm,rounded corners=2mm,minimum width=40mm] at (\x,\y+0.9) {~};			
		\node [fill=red!50,minimum size=4mm,minimum width=8mm] at (\x-1.6,\y+0.9) 
				{ \textsf{\textbf{\tiny{\cite{Chen-et-al-2012}}}} };			
		\node [fill=orange!50,minimum size=4mm,minimum width=6mm] at (\x-0.9,\y+0.9) 
				{\textsf{\textbf{\tiny{G8}}} };	
		\node [fill=yellow!50,minimum size=4mm,minimum width=10mm] at (\x-0.1,\y+0.9) 
				{\textsf{\textbf{\tiny{VMSSS-I}}} };	
		\node [fill=green!50,minimum size=4mm,minimum width=4mm] at (\x+0.6,\y+0.9) 
				{\textsf{\textbf{\tiny{K}}} };	
		\node  at (\x,\y+0.5) {\textsf{\tiny{Handles changes in PT configuration}}};
		
		\edef\x {17.3}		
		\node [rectangle,fill=black!100,minimum size=4mm,rounded corners=2mm,minimum width=40mm] at (\x,\y+0.9) {~};			
		\node [fill=red!50,minimum size=4mm,minimum width=8mm] at (\x-1.6,\y+0.9) 
				{ \textsf{\textbf{\tiny{\cite{DB-Attasena-et-al-2014-J}}}} };			
		\node [fill=orange!50,minimum size=4mm,minimum width=6mm] at (\x-0.9,\y+0.9) 
				{\textsf{\textbf{\tiny{G10}}} };	
		\node [fill=yellow!50,minimum size=4mm,minimum width=10mm] at (\x-0.1,\y+0.9) 
				{\textsf{\textbf{\tiny{VMSSS-II}}} };	
		\node [fill=blue!50,minimum size=4mm,minimum width=4mm] at (\x+0.6,\y+0.9) 
				{\textsf{\textbf{\tiny{D}}} };	
		\node  at (\x,\y+0.5) {\textsf{\tiny{Verifies data correctness}}};
		
		\edef\x {22}		
		\node [rectangle,fill=black!100,minimum size=4mm,rounded corners=2mm,minimum width=40mm] at (\x,\y+0.9) {~};			
		\node [fill=red!50,minimum size=4mm,minimum width=8mm] at (\x-1.6,\y+0.9) 
				{ \textsf{\textbf{\tiny{\cite{Eslami-Rad-2012}}}} };			
		\node [fill=orange!50,minimum size=4mm,minimum width=6mm] at (\x-0.9,\y+0.9) 
				{\textsf{\textbf{\tiny{G9}}} };	
		\node [fill=yellow!50,minimum size=4mm,minimum width=10mm] at (\x-0.1,\y+0.9) 
				{\textsf{\textbf{\tiny{VMSSS-I}}} };	
		\node [fill=green!50,minimum size=4mm,minimum width=4mm] at (\x+0.6,\y+0.9) 
				{\textsf{\textbf{\tiny{K}}} };	
		\node [fill=blue!50,minimum size=4mm,minimum width=4mm] at (\x+1.0,\y+0.9) 
				{\textsf{\textbf{\tiny{D}}} };	
		\node  at (\x,\y+0.50) {\textsf{\tiny{New sharing and reconstruction processes}}};
		\draw [-stealth,thick] (\x+1.9,\y+0.7) --(\x+1.9,6.1);

	\end{tikzpicture}    
}
	\caption{Evolution of SSSs}
	\label{fig:evolution-of-sss} 
\end{sidewaysfigure*}
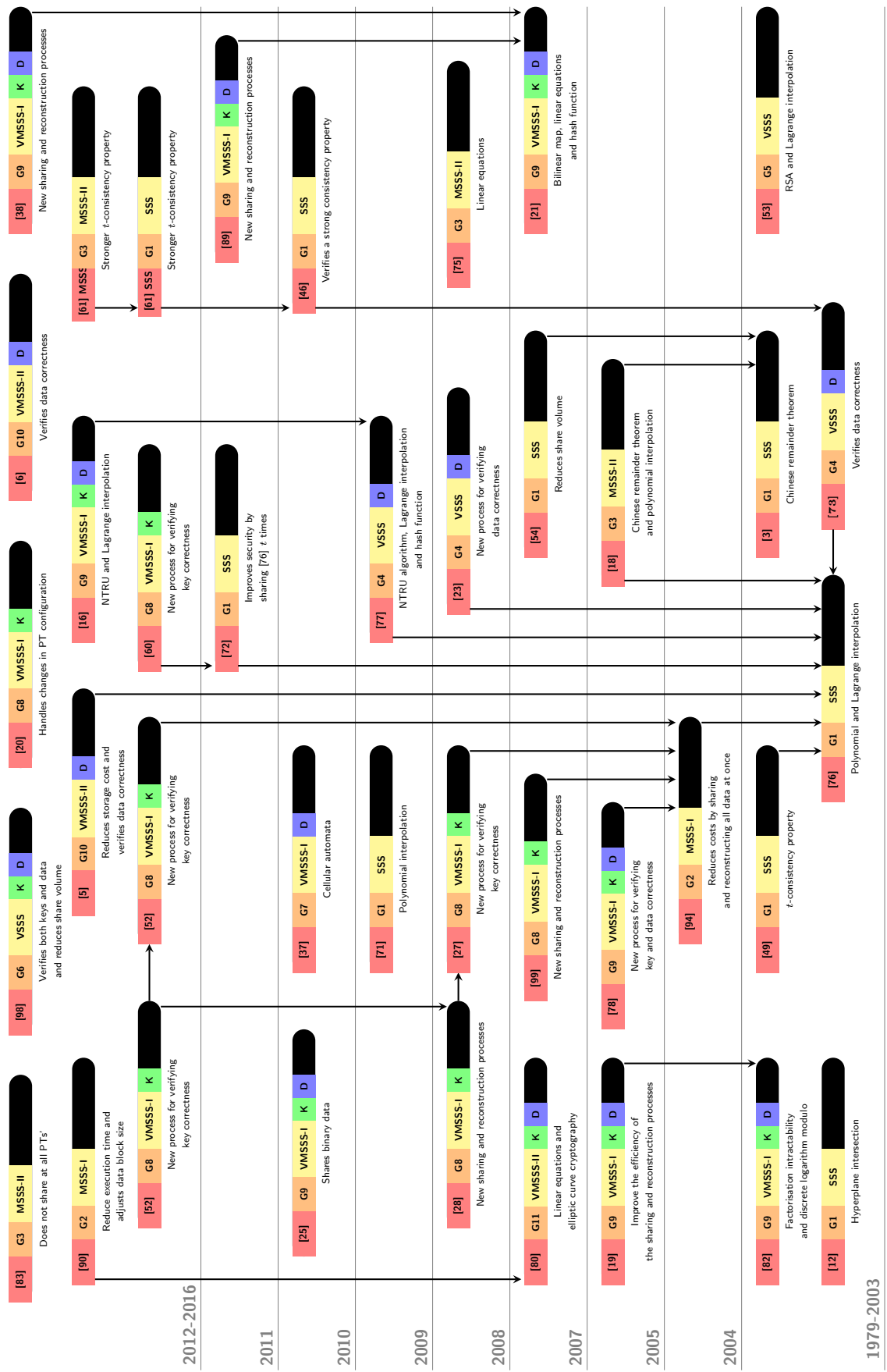

To clarify the historical relationships between the SSSs reviewed in this paper and better visualize the improvements brought to Shamir's \cite{Shamir-1979} and Blakley's \cite{Blakley-1979} schemes since 1979, we refer the reader to Figure \ref{fig:evolution-of-sss}. In this flowchart, each scheme is identified by a bibliographical reference (in red), the group (in orange) and type (in yellow) it belongs to (Section~\ref{sec:Secret-sharing-schemes}), and whether it enforces key (represented by a green K) and/or data  (represented by a blue D) verification. Moreover, a brief text describes the novelty brought by each scheme. Finally, an arrow from scheme $S_1$ to scheme $S_2$ indicates that $S_1$ extends from $S_2$. For example, \cite{Shao-Cao-2005}, proposed in 2005, is a VMSSS type~I belonging to Group~9. This scheme can verify both data and key correctness and extends from \cite{Yang-et-al-2004} to improve sharing and reconstruction efficiency.

\setlength{\rotFPtop}{0pt plus 1fil}

Figure \ref{fig:evolution-of-sss} quite clearly shows that SSSs have been less studied for almost 25 years than since the 2000's, when they attracted new attention in conjunction with the development of new, on-line distributed systems, i.e., clusters, grids and the cloud. Moreover, research about secret sharing seems to have accelerated since 2012, with the wide spread of cloud computing and associated data security concerns.

\subsection{Sharing, Reconstruction and Verification Methods}
\label{sec:Encryption-sss}

SSSs may be subdivided into five subprocesses, i.e., data sharing, data reconstruction, key creation, key verification and data verification. Of course, data sharing and reconstruction are the main processes for all groups of SSSs (Table~\ref{tab:Groups-of-Secret-sharing}). Key creation is always optional. Finally, data verification is the focus of groups 4, 6, 7, 9, 10 and 11; and key verification the focus of groups 5, 6, 8, 9 and 11.
The methods supporting these processes in each studied SSS are summarized in Table~\ref{tab:Encryption-and-verification-methods}. 



\setlength{\rotFPtop}{0pt plus 1fil}
\begin{sidewaystable*}
	\caption{Sharing, Reconstruction and Verification Methods in SSSs}
	\label{tab:Encryption-and-verification-methods}
	\centering
		\resizebox{\textheight}{!} {
    		\begin{tabular}{ |c|c|c|c|c|c|c|c|}
    		
    			\hline
				\multirow{2}{*}{\textbf{Type}} &
    			\multirow{2}{*}{\textbf{Group}} &
    			\multirow{2}{*}{\textbf{Scheme}} &
    			\multicolumn{5}{c|}{\textbf{Method}} \rule{0pt}{4mm}\\ \cline{4-8}
    			& & &
    			\textbf{Data sharing} &
    			\textbf{Data reconstruction} &
    			\textbf{Key creation} &
    			\textbf{Key verification} &
    			\textbf{Data verification} \rule{0pt}{4mm}\\ \hline\hline
    			
    			\multirow{9}{*}{SSS}& 
    			\multirow{9}{*}{Group 1}& 
    			\cite{Shamir-1979} &
    			Polynomial interpolation&
    			Lagrange interpolation&
    			& & \\ \cline{3-8}
    			    			
    			& & 
    			\cite{Blakley-1979} &
    			\multicolumn{2}{c|}{Hyperplane intersection} &
    			& & \\ \cline{3-8}
    			
    			& & 
    			\cite{Asmuth-Bloom-1983,Iftene-2007} &
    			\multicolumn{2}{c|}{Chinese remainder theorem} &
    			Random & & \\ \cline{3-8}
    			
    			& &
    			\cite{He-Dawson-1994} &
    			Polynomial interpolation&
    			Lagrange interpolation&
    			Random & & \\ \cline{3-8}
    			
    			& & 
    			\cite{Parakh-Kak-2009} &
    			\multicolumn{2}{c|}{Polynomial interpolation} &
    			Random & & \\ \cline{3-8}
    			
    			& & 
    			\multirow{2}{*}{\cite{Harn-Lin-2010},\cite{Liu-et-al-2012} SSS}&
    			Homomorphism and &
    			\multirow{2}{*}{Lagrange interpolation}&
    			\multirow{2}{*}{} &
    			\multirow{2}{*}{} &
    			\multirow{2}{*}{} \\ 
    			
    			& & & polynomial interpolation & & & & \\ \cline{3-8}
    			
    			& & 
    			\cite{Parakh-Kak-2011} &
    			\multicolumn{2}{c|}{Polynomial interpolation and recursion} &
    			 & & \\ \hline
    			
    			\multirow{8}{*}{MSSS} & 
    			\multirow{2}{*}{Group 2} & 
    			\cite{Yang-et-al-2004} &
    			Polynomial interpolation &
    			Lagrange interpolation &
    			Two-variable one-way function & & \\ \cline{3-8} 
    			
    			& & 
    			\cite{Waseda-Soshi-2012} &
    			\multicolumn{2}{c|}{Linear equation and matrix multiplication} &
    			Random & & \\ \cline{2-8}
    			
    			& \multirow{6}{*}{Group 3} & 
    			\cite{Chan-Chang-2005} &
    			\multicolumn{2}{c|}{Chinese remainder theorem and polynomial interpolation} &
    			Random & & \\ \cline{3-8}  
    			
				& &   
    			\cite{Runhua-et-al-2008} &
    			\multicolumn{2}{c|}{Linear equation and matrix multiplication} &
    			Random & & \\ \cline{3-8}  
    			
    			& & 
    			\multirow{2}{*}{\cite{Liu-et-al-2012} MSSS} &
    			Polynomial interpolation and &
    			\multirow{2}{*}{Lagrange interpolation} &
    			\multirow{2}{*}{} &
    			\multirow{2}{*}{} &
    			\multirow{2}{*}{} \\
    			
    			& & & homomorphism  & & & & \\ \cline{3-8}
    			
    			& & 
    			\multirow{2}{*}{\cite{Takahashi-Iwamura-2013}} &
    			Polynomial interpolation and &
    			\multirow{2}{*}{Lagrange interpolation} &
    			\multirow{2}{*}{Random} &
    			\multirow{2}{*}{} &
    			\multirow{2}{*}{} \\
    			
    			& & & pseudo-random number generation & & & & \\ \hline
    			
    			\multirow{7}{*}{VSSS} &  
    			\multirow{5}{*}{Group 4} & 
    			\multirow{2}{*}{\cite{Pedersen-1991}}&
    			Homomorphism and &
    			\multirow{2}{*}{Lagrange interpolation}&
    			\multirow{2}{*}{} &
    			\multirow{2}{*}{} &
    			\multirow{2}{*}{RSA cryptosystem} \\ 
    			& & & polynomial interpolation & & & & \\ \cline{3-8}	
    			
    			& & 
    			\cite{Tang-Yao-2008} &
    			Polynomial interpolation &
    			SMC &
    			& & Discrete logarithm modulo \\ \cline{3-8}
    			
    			& & 
    			\multirow{2}{*}{\cite{Yue-Hong-2009}} &
    			NTRU and &
    			\multirow{2}{*}{Lagrange interpolation} &
    			\multirow{2}{*}{NTRU} &
    			&
    			\multirow{2}{*}{NTRU}  \\ 
    			
    			& & & one-way hash function & & & & \\ \cline{2-8}
    			
    			& Group 5 &
    			\cite{Hwang-Chang-1998} &
    			\multicolumn{2}{c|}{RSA cryptosystem and Lagrange interpolation} &
    			\multicolumn{2}{c|}{RSA cryptosystem} &
    			\\ \cline{2-8}
    			
    			& Group 6 & 
    			\cite{Zhao-et-al-2012} &
    			\multicolumn{2}{c|}{Jordan matrix and linear equations} &
    			Random &
    			Discrete logarithm modulo &
    			Discrete logarithm modulo \\ \hline
    			
    			\multirow{20}{*}{VMSSS} &
    			Group 7 &
    			\cite{Eslami-Ahmadabadi-2010} &
    			\multicolumn{2}{c|}{One-dimensional cellular automaton} &
    			& &
    			Discrete logarithm modulo \\ \cline{2-8}
    			
    			& \multirow{13}{*}{Group 8} &
    			\cite{Zhao-et-al-2007} &
    			Polynomial interpolation &
    			Lagrange interpolation &
    			\multicolumn{2}{c|}{Discrete logarithm modulo} & \\ \cline{3-8}
    					
    			& &
    			\multirow{2}{*}{\cite{Dehkordi-Mashhadi-2008a}} &
    			\multirow{2}{*}{Polynomial interpolation} &
    			\multirow{2}{*}{Lagrange interpolation} &
    			Discrete logarithm modulo and &
    			\multirow{2}{*}{Discrete logarithm modulo} &
    			\multirow{2}{*}{} \\
    			
    			& & & & & two-variables one-way function & & \\ \cline{3-8}
    			
    			& &
    			\multirow{2}{*}{\cite{Dehkordi-Mashhadi-2008b}} &
    			\multirow{2}{*}{Homogeneous linear recursion} &
    			\multirow{2}{*}{Lagrange interpolation} &
    			Discrete logarithm modulo and&
    			\multirow{2}{*}{Discrete logarithm modulo} &
    			\multirow{2}{*}{} \\
    			
    			& & & & & two-variables one-way function & & \\ \cline{3-8}
    						
    			& &
    			\cite{Wang-et-al-2011,Eslami-Rad-2012} &
    			\multicolumn{2}{c|}{linear equations and hash function} &
    			Random &
    			\multicolumn{2}{c|}{Bilinear map} \\ \cline{3-8}
    			
    			& &
    			\multirow{2}{*}{\cite{Chen-et-al-2012}} &
    			\multicolumn{2}{c|}{\multirow{2}{*}{Discrete logarithm modulo and Lagrange interpolation}} &
    			\multirow{2}{*}{Random} &
    			Discrete logarithm modulo &
    			\multirow{2}{*}{} \\
    			
    			& & & \multicolumn{2}{c|}{ } & & and one-way hash function & \\ \cline{3-8}
    			
    			&&
    			\multirow{3}{*}{\cite{Li-et-al-2012}} &
    			Polynomial interpolation, &
    			Lagrange interpolation, &
    			\multicolumn{2}{c|}{\multirow{4}{*}{XOR and one-way hash function}} &
    			\multirow{4}{*}{} \\
    			
    			& & & recursion, XOR and & recursion, XOR and & \multicolumn{2}{c|}{}  & \\ 
    			& & & one-way hash function & one-way hash function & \multicolumn{2}{c|}{}  & \\ \cline{3-8}

    			& &
    			\cite{Hu-et-al-2012}-I &
    			Polynomial interpolation &
    			Lagrange interpolation &
    			\multicolumn{2}{c|}{One-way hash function and LFSR public key cryptography} & \\ \cline{3-8}
    			
    			& &
    			\cite{Hu-et-al-2012}-II &
    			Homogeneous linear recursion &
    			Lagrange interpolation &
    			\multicolumn{2}{c|}{One-way hash function and LFSR public key cryptography} & \\ \cline{2-8}
    			
    			& \multirow{5}{*}{Group 9} &
    			\cite{Lin-We-1999,Chang-et-al-2005} &
    			\multicolumn{2}{c|}{Factorisation intractability and discrete logarithm modulo} &
    			Discrete logarithm modulo &
    			\multicolumn{2}{c|}{Factorisation intractability and discrete logarithm modulo} \\ \cline{3-8}
    			
    			& &
    			\cite{Shao-Cao-2005} &
    			Polynomial interpolation &
    			Lagrange interpolation &
    			Two-variable one-way function &
    			\multicolumn{2}{c|}{Discrete logarithm modulo} \\ \cline{3-8}
    						
    			& &
    			\cite{Wei-et-al-2007} &
    			\multicolumn{2}{c|}{linear equations and hash function} &
    			Random &
    			\multicolumn{2}{c|}{Bilinear map} \\ \cline{3-8}
    			
    			& &
    			\cite{Das-Adhikari-2010} &
    			\multicolumn{2}{c|}{One-way hash function and binary operations} &
    			Random binary &
    			\multicolumn{2}{c|}{One-way hash function} \\ \cline{3-8}
    			    			
    			& &
    			\cite{Bu-Yang-2012} &
    			\multicolumn{2}{c|}{NTRU, XOR and one-way hash function} &
    			 NTRU &
    			\multicolumn{2}{c|}{NTRU  and one-way hash function} \\ \cline{2-8}
    			
    			& \multirow{2}{*}{Group 10} &
    			\cite{DB-Attasena-et-al-2014-J} &
    			\multicolumn{2}{c|}{Linear equations and homomorphic function} &
    			Random &
    			One-way hash function & \\ \cline{3-8}
    			
    			& &
    			\cite{DB-Attasena-et-al-2014-P} &
    			\multicolumn{2}{c|}{Lagrange interpolation, homomorphic function and hash function} &
    			Random &
    			Homomorphic functions and tree structure & \\ \cline{2-8}
    			
    			& Group 11 &
    			\cite{Shi-et-al-2007} &
    			\multicolumn{2}{c|}{Linear equations} &
    			Random &
    			\multicolumn{2}{c|}{Elliptic curve cryptography} \\ \hline
    			
    		\end{tabular}
    	}
\end{sidewaystable*}

Approximately half of the surveyed SSSs  share secrets by polynomial interpolation and  reconstruct them by Lagrange interpolation, as Shamir's \cite{Shamir-1979}. Yet, other methods, such as homomorphic encryption, NTRU or RSA enhance security. Similarly, approximately half of the schemes necessitating keys generate them at random,
 while more elaborate methods such as hash functions, LFSR, NTRU or RSA help protect keys. Eventually, the same variety of methods is found in the key and data verification processes, although discrete logarithm modulo and hash functions are by far the most popular. 

Given such variety, it is difficult to crisply rank the security level of all studied schemes. SSSs have indeed been continually addressing different issues over time, and thus adopted ad-hoc methods suited to their objectives. Moreover, the papers describing them typically do not compare to one another. Thence, we push the comparison of SSSs' features and cost in the following subsections.



\subsection{Features of Secret Sharing Schemes} 
\label{sec:Properties-sss}

SSSs mainly aim at enforcing data security (privacy, availability and integrity). However, in the context of cloud data processing, efficient data access (update, search and aggregation operations) must also be made possible by SSSs. Thus, some SSSs allow computation (e.g., sums and averages \cite{DB-Attasena-et-al-2014-P,DB-Attasena-et-al-2014-J,Blakley-1979,He-Dawson-1994,Shamir-1979,Shi-et-al-2007} and exact matches \cite{DB-Attasena-et-al-2014-J,Blakley-1979,Das-Adhikari-2010}) directly over shares, i.e., without reconstructing secrets. To provide a global overview, the features of all studied SSSs are synthesized in Table~\ref{tab:Properties-of-each-scheme}, 
where an X means a particular feature is supported by the corresponding SSS(s);
NB means that data availability is supported, but only when the NB is accessible;
G means that data availability  is supported only when shares are replicated;
IN and OUT stand for inner and outer code verification, respectively; 
B means that updates operate on data blocks instead of individual shares;
and I means that exact matches are run on indices.

\setlength{\rotFPtop}{0pt plus 1fil}
\begin{sidewaystable}
	\caption{Features of SSSs}
	\label{tab:Properties-of-each-scheme}
	\centering \resizebox{0.95\textheight}{!} {
    		\begin{tabular}{|c|c|c|c|c|c|c|c|c|c|c|c|c|c|c|}
    		
    			\hline 
				\multirow{4}{*}{\textbf{Type}} &
    			\multirow{4}{*}{\textbf{Group}} &
    			\multirow{4}{*}{\textbf{Scheme(s)}} &
    			\multicolumn{12}{c|}{\textbf{Features}} \\ \cline{4-15}
    			&  &  & & & & & &
    			\multicolumn{6}{c|}{\textbf{Data access}} & \\ \cline{9-14}
    			&  &  & \textbf{Data} & \textbf{Data} & \textbf{Key} & \textbf{Data} & \textbf{Key} &
    			\multicolumn{4}{c|}{\textbf{Updates}} &
    			\textbf{Exact} & \textbf{Aggregation} & \textbf{Other features} \\ \cline{9-12} 
    			&  &  & \textbf{privacy} & \textbf{availability} & \textbf{availability} &
    			\textbf{integrity} & \textbf{integrity} & \textbf{Add} & \textbf{Delete} & \textbf{Append} &
    			\textbf{Modify} & \textbf{match} &  &  \\ \hline\hline
    			    			
    			\multirow{5}{*}{SSS} &
    			\multirow{5}{*}{Group 1} &
    			\cite{Shamir-1979} & 
    			X & X & & & & X & X & X & & & SUM, AVG & \\ \cline{3-15}
    			
    			& & \cite{Blakley-1979} &
    			X & X & & & & X & X & X & & X & SUM, AVG & \\ \cline{3-15}
    			
    			& & \cite{Asmuth-Bloom-1983,Iftene-2007,Parakh-Kak-2009,Parakh-Kak-2011} &
    			X & X & & & & X & X & X & & & & \\ \cline{3-15}
    			
    			& & \cite{He-Dawson-1994} &
    			X & NB & & & & X & X & X & & & SUM, AVG & Strong $t$-consistency property \\ \cline{3-15}
    			
    			& & \cite{Harn-Lin-2010,Liu-et-al-2012}SSS &
    			X & X & & & & X & X & X & & & & Strong $t$-consistency property \\ \hline
    			
    			\multirow{6}{*}{MSSS} &
    			\multirow{2}{*}{Group 2} & 
    			\cite{Yang-et-al-2004} &
    			X & NB & X & & & & & & & & & \\ \cline{3-15}
    			 
    			& & \cite{Waseda-Soshi-2012} &
    			X & & X & & & B & B & & & & & $t$ and block size may vary \\ \cline{2-15}
    			
    			& \multirow{4}{*}{Group 3} &
    			\cite{Chan-Chang-2005} &
    			X & X & X & & & & & & & & & $t$ may vary \\ \cline{3-15}
    			
    			& & \cite{Runhua-et-al-2008} &
    			X & NB & NB & & & B & B & & & & & $t$ may vary \\ \cline{3-15}
    			
    			& & \cite{Liu-et-al-2012}MSSS &
    			X & X & & & & X & X & X & & & & Strong $t$-consistency property \\ \cline{3-15}
    			
    			& & \cite{Takahashi-Iwamura-2013} &
    			X & X & & & & X & X & X & & & & \\ \hline

    			\multirow{4}{*}{VSSS} &
    			\multirow{2}{*}{Group 4} & \cite{Pedersen-1991,Yue-Hong-2009} &
    			X & X & & OUT & & X & X & X & & & & \\ \cline{3-15}
    			
    			& &
    			\cite{Tang-Yao-2008} &
    			X & X & & IN & & X & X & X & & & & \\ \cline{2-15}
    			    			
    			& Group 5 &
    			\cite{Hwang-Chang-1998} &
    			X & G & X & & OUT & X & X & X & & & & $n$ may vary \\ \cline{2-15}
    			
    			& Group 6 & \cite{Zhao-et-al-2012} & 
    			X & NB & X & OUT & OUT & X & X & X & & & & \\ \hline
    			
    			\multirow{12}{*}{VMSSS} &
    			Group 7 &
    			\cite{Eslami-Ahmadabadi-2010} &
    			X & NB & & OUT & & & & & & & & \\ \cline{2-15}
    			
    			& \multirow{3}{*}{Group 8} &
    			\cite{Zhao-et-al-2007,Dehkordi-Mashhadi-2008a,Dehkordi-Mashhadi-2008b,Hu-et-al-2012,Eslami-Rad-2012} &
    			X & NB & X & & OUT & & & & & & & \\ \cline{3-15}
    			
    			& & \cite{Chen-et-al-2012} &
    			X & NB & X & & OUT & B & B & & & & &
    			$n$ and $t$ may vary \\ \cline{3-15}
    			 
    			& & \cite{Wang-et-al-2011,Li-et-al-2012} &
    			X & NB & X & & OUT & B & B & & & & & \\ \cline{2-15}
    			    			
    			& \multirow{4}{*}{Group 9} & 
    			\cite{Lin-We-1999,Chang-et-al-2005,Bu-Yang-2012} &
    			X & NB & X & OUT & OUT & X & X & X & & & & \\ \cline{3-15}
    			 
    			& & \cite{Shao-Cao-2005} & X & NB & X & OUT & OUT & & & & & & & \\ \cline{3-15}
    			
    			& & \cite{Wei-et-al-2007}  & X & NB & X & OUT & OUT & X & X & X & & & & $t$ may vary \\ \cline{3-15}

    			& & \cite{Das-Adhikari-2010} & X & & & IN & OUT & X & X & X & & X & & \\ \cline{2-15}
    			
    			& \multirow{3}{*}{Group 10} &
    			\cite{DB-Attasena-et-al-2014-J} &
    			X & X & & INT, OUT & & B & B & & & X & SUM, AVG & \\ \cline{3-15}
    			
    			& &
    			\multirow{2}{*}{\cite{DB-Attasena-et-al-2014-P}} &
    			\multirow{2}{*}{X} & 
    			\multirow{2}{*}{X} & 
    			\multirow{2}{*}{} & 
    			\multirow{2}{*}{IN, OUT} & 
    			\multirow{2}{*}{} & 
    			\multirow{2}{*}{X} & 
    			\multirow{2}{*}{X} & 
    			\multirow{2}{*}{X} & 
    			\multirow{2}{*}{} & 
    			\multirow{2}{*}{I} & 
    			\multirow{2}{*}{SUM, AVG} & 
    			Can share new data although \\ 
    			& & & & & & & & & & & & & & some PTs disappear \\ \cline{2-15}
    			
    			& Group 11 &
    			\cite{Shi-et-al-2007} &
    			X & X & X & OUT & OUT & B & B & & & & SUM, AVG & \\ \hline

    			
	
    		\end{tabular}
    	}
\end{sidewaystable}

\subsubsection{Data Privacy and Availability} 
\label{sec:dPdA}

Since all SSSs divide data into $n$ shares such that each individual share is meaningless, they enforce data privacy by design. Moreover, data availability is guaranteed as long as $t$ out of $n$ PTs are available, since $t$ PTs are enough to reconstruct secrets. However, a coalition of $t$ or more malicious PTs can break any secret. Thus, \cite{DB-Attasena-et-al-2014-P} provides further privacy by protecting data from PT group cheating, by having a number of shares at all PTs that is lower than $t$. Finally, since most (V)MSSSs type-I store all shares in the NB, they are vulnerable and can loose data access if the NB is compromised. 

The privacy level of all SSSs mainly depends on parameter $t$. Provided PTs independently enforce sound security measures, collecting at least $t$ shares, i.e., compromising at least $t$ PTs, is indeed harder and harder when $t$ increases. High data protection is thus achieved when $t$ is large \cite{Asmuth-Bloom-1983,JD-CR-2012}, but at the expense of computing overhead, especially when sharing and reconstructing data (Section~\ref{sec:cost-sss}). Moreover, some SSSs may be insecure for applications 
where $t$ is limited in practice. For instance, when $t$ is a number of CSPs or servers, budget constraints come into play. We discuss three frameworks for outsourcing data in the cloud that address this issue in Section~\ref{sec:Frameworks}.

The robustness of almost SSSs directly relies on the gap between the two parameters $n$ and $t$. The secret can be recovered although up to $n - t$ PTs disappear. Nevertheless, computing time and storage costs become prohibitive when $n \gg t$ (Section~\ref{sec:cost-sss}). Thus, $n$ should be only a little bigger than $t$ to achieve data availability with acceptable costs.

\subsubsection{Data Integrity}
\label{sec:dI}

The reconstruction process in SSSs always produces the correct result if secrets, shares and sharing and reconstruction functions are defined over a finite field \cite{Survey-SS}. However, if shares are altered, reconstructed secrets are mechanically incorrect. Thus, VSSSs and VMSSSs have been introduced to enforce data integrity. We categorize them into four classes: SSSs that verify keys, shares, secrets or both secrets and shares. 

First, all schemes in groups 5, 6, 8, 9 and 11 verify keys before reconstructing shares. Hence, they can detect PT cheating and prevent transferring any data back to the user when incorrect keys are detected. 

Second, most schemes in groups 4, 6, 7, 9, 10 and 11 verify the correctness of shares before reconstruction to reduce computation cost at the user's (no reconstruction occurs from incorrect shares). However, they require extra storage for signatures. 

Third, \cite{Tang-Yao-2008,Das-Adhikari-2010} verify the correctness of reconstructed secrets.  Their signature volumes are lower than that of the second class of VSSSs, since the number of shares is generally greater than that of secrets. However, incorrect secrets are detected only after they are already reconstructed. 

Fourth, \cite{DB-Attasena-et-al-2014-J,DB-Attasena-et-al-2014-P} verify the correctness of both secrets and shares with inner and outer code verification, respectively. Thus, no erroneous share is transferred to the user. Moreover, any PT cheating is detected.

Finally, although VSSSs and MVSSSs guarantee integrity, they consume more storage to handle signatures and more CPU power to verify keys, shares, and/or secrets. Moreover, to achieve the best possible verification performance, i.e., the lowest possible false positive rate, signatures must be big \cite{DB-Attasena-et-al-2014-P,DB-Attasena-et-al-2014-J}. A larger storage volume is thus required. We push the comparison of such costs in Section~\ref{sec:cost-sss}.

\subsubsection{Data Access}
\label{sec:dA}

SSSs manage data at two levels: data piece or data block. First, \cite{Takahashi-Iwamura-2013,DB-Attasena-et-al-2014-P} and most schemes in groups 1, 4, 5, 6, 9 share secrets independently. Hence, they can directly update data. For example, any secret can be deleted by removing its shares at all PTs'. Second, \cite{Waseda-Soshi-2012,Runhua-et-al-2008,Wang-et-al-2011,Chen-et-al-2012,Li-et-al-2012,Shi-et-al-2007,DB-Attasena-et-al-2014-J} share secrets as blocks and support the homomorphic property. Thus, they allow updating shared blocks without reconstruction. Moreover, they update data faster because several shares in the same data block can be updated at once. In contrast, the schemes that share all secrets at once cannot perform updates on shares. The whole database must indeed be reconstructed, updated and then shared again.
Thus, such schemes require longer execution times and use lots of memory when updating data. 

Some SSSs allow computing exact matches on shares. Since \cite{Shamir-1979,Blakley-1979,He-Dawson-1994,Shi-et-al-2007,DB-Attasena-et-al-2014-J,DB-Attasena-et-al-2014-P} use polynomial or linear equations to share data, they also allow sum and average operations on shares. Moreover, \cite{Blakley-1979,Das-Adhikari-2010,DB-Attasena-et-al-2014-J} allow exact matches on shares, because they use the same keys to share all secrets. In contrast, \cite{DB-Attasena-et-al-2014-P} uses indices to achieve exact match queries. Indices indeed help perform faster exact matches than operating directly on shares, although at the expense of extra storage volume. 
Thus, the tradeoff between security and query efficiency must be carefully considered before choosing an SSS. 
We further discuss this issue in Sections~\ref{sec:Frameworks} and \ref{sec:Applications}.

\subsubsection{Other Features}
\label{sec:dO}

More features are included in some schemes. 
\cite{He-Dawson-1994,Harn-Lin-2010,Liu-et-al-2012} verify a strong $t$-consistency property. Thus, they guarantee that any subset of $t$ shares or more always reconstruct the same data, but that any subset of $t$ shares or fewer cannot. 
\cite{Hwang-Chang-1998,Chen-et-al-2012} allow the user to add and remove PTs to/from the PT pool by updating the value of $n$. 
\cite{Waseda-Soshi-2012,Chan-Chang-2005,Runhua-et-al-2008,Chen-et-al-2012,Wei-et-al-2007} allow the user to assign different values of $t$ to different secrets, to enforce different security levels for each secret.
Eventually, \cite{DB-Attasena-et-al-2014-P} allows inserting new data even if some PTs disappear.

\subsection{Costs} 
\label{sec:cost-sss}

In the cloud pay-as-you-go paradigm, the cost of securing data must be balanced with the risk of data loss or pilfering, and thus the level of data security must be balanced with its cost. This is a particularly important issue with secret sharing, which basically multiplies secret data volume by $n$ in the worst case (provided individual share volume is not greater than secret data volume). We summarize the costs induced by SSSs in Table~\ref{tab:Costs-of-each-scheme}. 



\setlength{\rotFPtop}{0pt plus 1fil}
\begin{sidewaystable*}
	\caption{Costs induced by SSSs}
	\label{tab:Costs-of-each-scheme}
	\centering
		\resizebox{\textheight}{!} {
    		\begin{tabular}{|c|c|c|c|c|c|c|c|c|c|c|c|c|c|}					
			\hline 
			\multirow{3}{*}{\textbf{Type}} &
			\multirow{3}{*}{\textbf{Group}} &
			\multicolumn{1}{c|}{\multirow{3}{*}{\textbf{Scheme(s)}}} &
			\multicolumn{4}{c|}{\textbf{Average time complexity}} &
			\multicolumn{7}{c|}{\textbf{Storage volume}} \\ \cline{4-14} 
 			&  &  & 
 			\textbf{Sharing} &
 			\textbf{Reconstruction} &
 			\textbf{Key} &
 			\textbf{Data} &
 			\multicolumn{2}{c|}{\textbf{Shares}} &
 			\multicolumn{3}{c|}{\textbf{Keys}} &
 			\multicolumn{2}{c|}{\textbf{Signatures}} \\ \cline{8-14}
 			&  &  & 
 			\textbf{process} &
 			\textbf{process} &
 			\textbf{verification} &
 			\textbf{verification} &
 			\textbf{PTs} &
 			\textbf{NB} &
 			\textbf{PTs} &
 			\textbf{NB} &
 			\textbf{Client} &
 			\textbf{PTs} &
 			\textbf{NB} \\ \hline \hline 
 			
			\multirow{8}{*}{SSS} &
			\multirow{8}{*}{Group 1} &
			\cite{Shamir-1979} &
			$O\left(mnt\right)$ & $O\left(mt^{2}\right)$ & & & $mn\left\Vert d\right\Vert $ & & & & & &	\\ \cline{3-14}
			
 			&  & \cite{Blakley-1979} &
			$O\left(mnt\right)$& $O\left(mt^{2}\right)$ & & & $mn\left\Vert d\right\Vert $ & & & & $mt^{2}\left\Vert k\right\Vert $ & & \\ \cline{3-14}
			 
 			&  & \cite{Asmuth-Bloom-1983} &
 			$O\left(mn\right)$ & $O\left(mt\beta\right)$ & & & $mn\left\Vert d\right\Vert $ & & & & $\left(n+1\right)\left\Vert k\right\Vert $ & & \\ \cline{3-14} 
 			
 			&  & \cite{He-Dawson-1994} &
			$O\left(mnt\right)$ & $O\left(mt^{2}\right)$ & & & & $mn\left\Vert d\right\Vert $ & $2mn\left\Vert k\right\Vert $ & & & & \\ \cline{3-14}
 			 
 			&  & \cite{Iftene-2007} &
 			$O\left(mn\right)$ & $O\left(mt\beta\right)$ & & & $\frac{mn\left\Vert d\right\Vert }{t}$ & & & & $\left(n+1\right)\left\Vert k\right\Vert $ & & \\ \cline{3-14} 
 			 
 			&  & \cite{Parakh-Kak-2009} &
 			$O\left(mn^{2}t\right)$ & $O\left(mt^{3}\right)$ & & & $mn\left\Vert d\right\Vert $ & & $nt\left\Vert k\right\Vert $ & & & & \\ \cline{3-14	} 
 			 
 			&  & \cite{Harn-Lin-2010,Liu-et-al-2012}SSS &
 			$O\left(mn^{2}t\right)$ & $O\left(mt^{	3}\right)$ & & & $mn\left\Vert d\right\Vert $ & & & & & & \\ \cline{3-14} 
 			
 			&  & \cite{Parakh-Kak-2011} &
 			$O\left(mn^{2}t\right)$ & $O\left(mt^{3}\right)$ & & & $mn\left\Vert d\right\Vert/t$ & & & & & & \\ \hline 
 			
			\multirow{6}{*}{MSSS} &
			\multirow{2}{*}{Group 2} &
			\cite{Yang-et-al-2004} &
			$\begin{array}{ll} O\left(nt\right) & \text{if~} m\leq t\\ O\left(\left(n+m-t\right)t\right) & \text{ otherwise} \end{array}$ & 
			$O\left(\max\left(m^{3},mt^{2}\right)\right)$ &
			 & & &
			$\begin{array}{ll} n\left\Vert d\right\Vert  & \text{if~} m\leq t\\ \left(n+m-t\right)\left\Vert d\right\Vert  & \text{ otherwise} \end{array}$ &  $n\left\Vert k\right\Vert $ 
			& & & & \\ \cline{3-14}  
			
			&  & \cite{Waseda-Soshi-2012} &
			$O\left(\max\left(nt\beta,mt^{2}\right)\right)$ &
			$O\left(\max\left(t^{3},mt^{2}\right)\right)$ &
			 & & &
			$m\left\Vert d\right\Vert $ & $n\left\Vert k\right\Vert $ & $nt\left\Vert k\right\Vert $ 
			 & & & \\ \cline{2-14} 
			
			& \multirow{4}{*}{Group 3} & 
			\cite{Chan-Chang-2005} &
			$O\left(mnt\right)$ & $O\left(\max\left(mt^2,mt\beta\right)\right)$ & & & $mn\left\Vert d\right\Vert $ & & $n\left\Vert d\right\Vert $ & & $2m\left\Vert k\right\Vert $ & & \\ \cline{3-14} 
			 
			& &\cite{Runhua-et-al-2008} &
			$O\left(mn\right)$ or $O\left(bnt\right)$ & $O\left(mt^{2}\right)$ or $O\left(bnt^{3}\right)$ & & & $\geq mn\left\Vert d\right\Vert/t$ or $bn\left\Vert d\right\Vert $ & & & $\approx nt\left\Vert k\right\Vert $ & & & \\ \cline{3-14}
			 
			&  & \cite{Liu-et-al-2012}MSSS  &
			$O\left(mnt\right)$ & $O\left(mt^{2}\right)$ & & & $mn\left\Vert d\right\Vert $ & & & & & & \\ \cline{3-14}
			
			&  & \cite{Takahashi-Iwamura-2013} &
			$O\left(mnt\right)$ & $O\left(mt^{2}\right)$ & & & $\left(mn+m\gamma+\gamma\right)\left\Vert d\right\Vert $ & & $\beta\left\Vert k\right\Vert $ & & & & \\ \hline 
			
			\multirow{4}{*}{VSSS} & \multirow{2}{*}{Group 4} & \cite{Pedersen-1991} &
 			$O\left(mn^{2}t\right)$ & $O\left(mt^{3}\right)$ & & $O\left(mt^{2}\right)$ & $mn\left\Vert d\right\Vert $ & & & & & & $mt\left\Vert s\right\Vert $ \\ \cline{3-14}
 			& &
			\cite{Tang-Yao-2008} &
			$O\left(mnt\right)$ & $O\left(mt^{2}\right)$ & & $O\left(m\right)$ & $mn\left\Vert d\right\Vert $ & & & & $4\left\Vert k\right\Vert $ & & $m\left\Vert s\right\Vert $ \\ \cline{3-14} 
			
			& &
			\cite{Yue-Hong-2009} &
			$O\left(mnt\right)$ & $O\left(mt^{2}\right)$ & & $O\left(mt\right)$ & $mn\left\Vert d\right\Vert $ & &  $2n\left\Vert k\right\Vert$ & $n\left\Vert k\right\Vert$ & $4\left\Vert k\right\Vert$ &  & $mn\left\Vert s\right\Vert $ \\ \cline{2-14} 
			
			& Group 5 & 
			\cite{Hwang-Chang-1998} &
			$O\left(mt^{2}g\right)$ & $O\left(mt^{2}\right)$ & $O\left(t\right)$ & & & $g m\left\Vert d\right\Vert $ & $n\left\Vert k\right\Vert $ & $n\left\Vert k\right\Vert $ & & & $n\left\Vert s\right\Vert $ \\ \cline{2-14} 
			
			& Group 6 & 
			\cite{Zhao-et-al-2012} &
			$O\left(\max\left(\gamma t,mt^2\right)\right)$ & $O\left(\max\left(\gamma t,mt^2\right)\right)$ & $O\left(t\right)$ & $O\left(t^2\right)$ & $\gamma m\left\Vert d\right\Vert/t^2$ & $mn\gamma\left\Vert d\right\Vert/t^2$ & $n\left\Vert k\right\Vert $ & $\left(2\gamma t+2\right)\left\Vert k\right\Vert $ & & & $\left(mn\gamma+n+1\right)\left\Vert s\right\Vert $ \\ \hline
			
			\multirow{19}{*}{VMSSS} & Group 7 & 
			\cite{Eslami-Ahmadabadi-2010} &
			$O\left(\max\left(m,t^2\right)\right)$ & $O\left(\max\left(m,t^2\right)\right)$ & & $O\left(n\right)$ & $n\left\Vert d\right\Vert $ & $\left(m-t\right)\left\Vert d\right\Vert \text{if~} t>m$ & & & & & $n\left\Vert s\right\Vert $  \\ \cline{2-14} 
			
			& \multirow{9}{*}{Group 8} &
			\cite{Zhao-et-al-2007} &
			$\begin{array}{ll} O\left(nt\right) & \text{if~} m\leq t\\ O\left(\left(n+m-t\right)t\right) & \text{ otherwise} \end{array}$ &
			$O\left(\max\left(m^{3},mt^{2}\right)\right)$ &
			$O\left(t\right)$ & & &	$\begin{array}{ll} n\left\Vert d\right\Vert  & \text{if~} m\leq t\\ \left(n+m-t\right)\left\Vert d\right\Vert  & \text{ otherwise} \end{array}$ & 
			$n\left\Vert k\right\Vert $ & $\left\Vert k\right\Vert $ & $(n+2)\left\Vert k\right\Vert $ & & $(n+1)\left\Vert s\right\Vert $ \\ \cline{3-14} 
			 
			&  & \cite{Dehkordi-Mashhadi-2008a} &
			$\begin{array}{ll} O\left(nt\right) & \text{if~} m\leq t\\ O\left(\left(n+m-t\right)t\right) & \text{ otherwise} \end{array}$ &
			$O\left(\max\left(m^{3},mt^{2}\right)\right)$ &
			$O\left(t\right)$ & & &	$\begin{array}{ll} n\left\Vert d\right\Vert  & \text{if~} m\leq t\\ \left(n+m-t\right)\left\Vert d\right\Vert  & \text{ otherwise} \end{array}$ & 
			$n\left\Vert k\right\Vert $ & $4\left\Vert k\right\Vert $ & & & $n\left\Vert s\right\Vert $ \\ \cline{3-14} 
			 
			&  & \cite{Dehkordi-Mashhadi-2008b} &
			$O\left(m+n\right)$ & $O\left(\max\left(m,t^{2}\right)\right)$ & $O\left(t\right)$ & & & $\left(m+n\right)\left\Vert d\right\Vert $ & $n\left\Vert k\right\Vert $ & $4\left\Vert k\right\Vert $ & & & $n\left\Vert s\right\Vert $ \\ \cline{3-14}
			 
			&  & \cite{Wang-et-al-2011} &
			$O\left(mnt\right)$ & $O\left(mt^{2}\right)$ &  & $O\left(mt\right)$ & & $mn\left\Vert d\right\Vert/t$ & $n\left\Vert k\right\Vert $ & & & & $n\left\Vert s\right\Vert $  \\ \cline{3-14}
			 
			&  & \cite{Eslami-Rad-2012} &
			$O\left(\max\left(m^{2},t^{2}\right)\right)$ & $O\left(\max\left(m^{2},t^{2}\right)\right)$ & & $O\left(mt\right)$ & & $\left(n+m+t\right)\left\Vert d\right\Vert $ & $n\left\Vert k\right\Vert $ & & & & $\left(n+m\right)\left\Vert s\right\Vert $ \\ \cline{3-14} 
			 
			&  & \cite{Chen-et-al-2012} &
			$O\left(mnt^{2}\right)$ & $O\left(mnt^{2}\right)$ & $O\left(t\right)$ & & & $mn\left\Vert d\right\Vert/t$ & $n\left\Vert k\right\Vert $ & $2n\left\Vert k\right\Vert $ & & & $n\left\Vert s\right\Vert $ \\ \cline{3-14} 
			 
			&  & \cite{Hu-et-al-2012}-I &
			
			$\begin{array}{ll} O\left(nt\right) & \text{if~} m\leq t\\ O\left(\left(n+m-t\right)t\right) & \text{ otherwise} \end{array}$ &
			$O\left(\max\left(m^{3},mt^{2}\right)\right)$ &
			$O\left(t\right)$ & & &	$\begin{array}{ll} n\left\Vert d\right\Vert  & \text{if~} m\leq t\\ \left(n+m-t\right)\left\Vert d\right\Vert  & \text{ otherwise} \end{array}$ & 
			$n\left\Vert k\right\Vert $ & & & & $2n\left\Vert s\right\Vert $ \\ \cline{3-14}
			
			&  & \cite{Hu-et-al-2012}-II &
			$O\left(m+n\right)$ & $O\left(\max\left(m,t^{2}\right)\right)$ & $O\left(t\right)$ & & & $\left(m+n\right)\left\Vert d\right\Vert $ & $n\left\Vert k\right\Vert $ & $4\left\Vert k\right\Vert $ & & & $2n\left\Vert s\right\Vert $ \\ \cline{3-14}
			 
			&  & \cite{Li-et-al-2012} &
			$O\left(mnt\right)$ & $O\left(mt^{2}\right)$ & $O\left(t^{2}\right)$ & & & $m\left(n-1\right)\left(t^{2}-t\right)\left\Vert d\right\Vert/\left(2t\right)$ & $n\left\Vert k\right\Vert $ & & & & $n\left\Vert s\right\Vert $ \\ \cline{2-14} 
			
			& \multirow{6}{*}{Group 9} &  
			\cite{Lin-We-1999} &
			$O\left(mt\right)$ & $O\left(mt^{2}\right)$ & $O\left(t\right)$ & $O\left(mt\right)$ & & $4m\left\Vert d\right\Vert $ & $n\left\Vert k\right\Vert $ & $\left(t+3\right)\left\Vert k\right\Vert $ & $2\left\Vert k\right\Vert $ & & $n\left\Vert s\right\Vert $ \\ \cline{3-14} 
			 
			&  & \cite{Chang-et-al-2005} &
			$O\left(mt\right)$ & $O\left(mt^{2}\right)$ & $O\left(t\right)$ & $O\left(mt\right)$ & & $3m\left\Vert d\right\Vert $ & $n\left\Vert k\right\Vert $ & $\left(t+3\right)\left\Vert k\right\Vert $ & $2\left\Vert k\right\Vert $ & & $n\left\Vert s\right\Vert $ \\ \cline{3-14} 
			 
			&  & \cite{Shao-Cao-2005} &
			$\begin{array}{ll} O\left(nt\right) & \text{if~} m\leq t\\ O\left(\left(n+m-t\right)t\right) & \text{ otherwise} \end{array}$ &
			$O\left(\max\left(m^{3},mt^{2}\right)\right)$ &
			\multicolumn{2}{c|}{$O\left(\max\left(m^2,mt\right)\right)$} & & $\begin{array}{ll} n\left\Vert d\right\Vert  & \text{if~} m\leq t\\ \left(n+m-t\right)\left\Vert d\right\Vert  & \text{ otherwise} \end{array}$ & 
			$n\left\Vert k\right\Vert $ & & & & $\begin{array}{ll} n\left\Vert s\right\Vert  & \text{if~} m\leq t\\ \left(n+m-t\right)\left\Vert s\right\Vert  & \text{ otherwise} \end{array}$ \\ \cline{3-14}
			 
			&  & \cite{Wei-et-al-2007} &
			$O\left(mnt\right)$ & $O\left(mt^{2}\right)$ & \multicolumn{2}{c|}{$O\left(mt\right)$} & & $m\left(n+1\right)\left\Vert d\right\Vert $ & $n\left\Vert k\right\Vert $ & $2\left\Vert k\right\Vert $ & & & $\left(n+m\right)\left\Vert s\right\Vert $ \\ \cline{3-14} 
			&  & \cite{Das-Adhikari-2010} &
			$O\left(mt^{2}\right)$ & $O\left(mt^{2}\right)$ & $O\left(mnt\right)$ & $O\left(m\right)$ & & $mt\left\Vert d\right\Vert $ & $n\left\Vert k\right\Vert $ & $mt\left\Vert k\right\Vert $ & & & $\left(m+mnt\right)\left\Vert s\right\Vert $ \\ \cline{3-14}
			 
			&  & \cite{Bu-Yang-2012} &
			$O\left(\max\left(nt,m\right)\right)$ & $O\left(mt^{2}\right)$ & $O\left(t^{2}\right)$ & $O\left(mt\right)$ & & $3m\left\Vert D\right\Vert $ & $n\left\Vert k\right\Vert $ & $3\left\Vert k\right\Vert $ & $\left\Vert k\right\Vert $ & & $t\left\Vert s\right\Vert $\\ \cline{2-14}
			& \multirow{2}{*}{Group 10} &
			\cite{DB-Attasena-et-al-2014-J} &
			$O\left( mn \right)$ & $O\left( mt \right)$ & & $O\left( m \right)$ &
			$mn\left\Vert d\right\Vert / (t-1) $ &	 & 
			 & & $nt\left\Vert k\right\Vert$ &
			$mn\left\Vert s\right\Vert / (t-1) $ & \\ \cline{3-14} 
			
			& & \cite{DB-Attasena-et-al-2014-P} &
			$O\left( mnt \right)$ & $O\left( mt^2 \right)$ & & $O\left( mt \right)$ &
			$m(n-t+2)\left\Vert d\right\Vert$ &	 & 
			 & & $(2n+2)\left\Vert k\right\Vert$ &
			$\log m(n-t+2)\left\Vert s\right\Vert$ & \\ \cline{2-14} 
			 
			& Group 11 & \cite{Shi-et-al-2007} &
			$O\left(mt\right)$ & $O\left(mt^{2}\right)$ & \multicolumn{2}{c|}{$O\left(mt\right)$} & $\frac{mn\left\Vert d\right\Vert }{t}$ or $bn\left\Vert d\right\Vert $ & & $nt\left\Vert k\right\Vert $ & & & $m\left\Vert s\right\Vert $ & \\ \hline 

    		\end{tabular}
    	}
\end{sidewaystable*}

SSS time complexity and storage volume depend on a few parameters: $m$, $n$ and $t$. 
To determine time complexity and storage volume, we suppose that only $m$ is big. Other parameters $n$ and $t$ 
should remain quite small, because they relate to the number of PTs, i.e., the number of cloud service providers, which is limited in practice. Moreover, some SSSs such as \cite{Iftene-2007,Parakh-Kak-2011,Zhao-et-al-2012} cannot assign a big value to parameters $n$ and $t$ because neither can be greater than the size of a secret. 


\subsubsection{Time Complexity}
\label{sec:cT}


Data sharing and reconstruction complexity of most SSSs increases with $n$ and $t$.
In practice, $n$ is a little bigger than $t$ to guarantee data availability. Thus, the time complexity of sharing data is a little higher than that of reconstruction, e.g., $O(mnt)>O(mt^2)$ in \cite{Shamir-1979}. However, when availability is not enforced, data sharing and reconstruction complexity is the same.

In contrast, in most MSSSs type I, secret sharing time complexity is clearly lower than that of data reconstruction, e.g., $O\left(\left(n+m-t\right)t\right)<O(m^{3})$ in \cite{Yang-et-al-2004}, because they share several secrets at once but reconstruct each secret independently.

Overall, time complexities to share/reconstruct data by \cite{Eslami-Ahmadabadi-2010} are the lowest: $O(\max (m,t^2))$. Execution time actually depends only on $m$, because $m$ is large, while both $t$ and $n$ are small in the normal case ($m \gg n\geq t$).

Moreover, VSSSs and VMSSSs must verify the correctness of keys and/or data. Thus, extra computation time is required. The time complexity of data/key verification is generally lower than that of data sharing/reconstruction. Moreover, the time complexity of key verification is generally lower than that of data verification. Several schemes achieve the lowest key verification complexity: $O(t)$, but only \cite{Eslami-Ahmadabadi-2010} achieves the lowest data verification complexity: $O(n)$.

\subsubsection{Storage Volume}
\label{sec:cS}

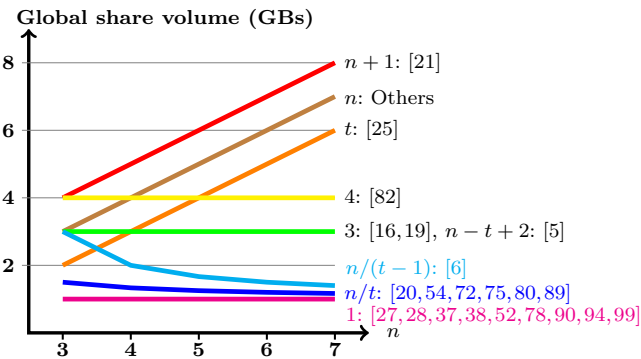
\begin{figure}[hbt]
	\centering 		
	\resizebox{0.5\textwidth}{!} {\begin{tikzpicture}
	
	\draw [<->,line width=0.5mm] (0.5,4.5) -- (0.5,0)--(5.5,0);
	\node[right] at (0.2,4.65) {\textbf{\small{Global share volume (GBs)}}};
	\node[right] at (5.65,0) {\textbf{\small{$n$}}};
	
	\draw [gray] (0.4,1)--(5,1); \node at (0.2,1) {\textbf{\small{2}}};
	\draw [gray] (0.4,2)--(5,2); \node at (0.2,2) {\textbf{\small{4}}};
	\draw [gray] (0.4,3)--(5,3); \node at (0.2,3) {\textbf{\small{6}}};
	\draw [gray] (0.4,4)--(5,4); \node at (0.2,4) {\textbf{\small{8}}};
	
	\draw [line width=0.3mm] (1,0)--(1,-0.1); \node at (1,-0.25) {\textbf{\small{3}}};
	\draw [line width=0.3mm] (2,0)--(2,-0.1); \node at (2,-0.25) {\textbf{\small{4}}};
	\draw [line width=0.3mm] (3,0)--(3,-0.1); \node at (3,-0.25) {\textbf{\small{5}}};
	\draw [line width=0.3mm] (4,0)--(4,-0.1); \node at (4,-0.25) {\textbf{\small{6}}};
	\draw [line width=0.3mm] (5,0)--(5,-0.1); \node at (5,-0.25) {\textbf{\small{7}}};
	
	\draw[color=red,line width=0.7mm,smooth,samples=10,domain=3:7,scale=1] plot (\x-2,\x/2+1/2) node[right,black] {\small{$n+1$: \cite{Wei-et-al-2007}}}; 
	\draw[color=brown,line width=0.7mm,smooth,samples=10,domain=3:7,scale=1] plot (\x-2,\x/2) node[right,black] {\small{$n$: Others}}; 
	\draw[color=orange,line width=0.7mm,smooth,samples=10,domain=3:7,scale=1] plot (\x-2,\x/2-1/2) node[right,black] {\small{$t$: \cite{Das-Adhikari-2010}}};
	
	\draw[color=yellow,line width=0.7mm,smooth,samples=10,domain=3:7,scale=1] plot (\x-2,4/2) node[right,black] {\small{$4$: \cite{Lin-We-1999}}};   
	\draw[color=green,line width=0.7mm,smooth,samples=10,domain=3:7,scale=1] plot (\x-2,3/2) node[right,black] {\small{$3$: \cite{Bu-Yang-2012,Chang-et-al-2005}, $n-t+2$: \cite{DB-Attasena-et-al-2014-P}}}; 
	 
	\draw[color=cyan,line width=0.7mm,smooth,samples=10,domain=3:7,scale=1] (1,3/1/2)--(2,4/2/2)--(3,5/3/2)--(4,6/4/2)--(5,7/5/2); \node[right,cyan] at (5.05,0.95) {\small{$n/(t-1)$: \cite{DB-Attasena-et-al-2014-J}}};
	 
	\draw[color=blue,line width=0.7mm,smooth,samples=10,domain=3:7,scale=1] (1,3/2/2)--(2,4/3/2)--(3,5/4/2)--(4,6/5/2)--(5,7/6/2) node[right] {\small{$n/t$: \cite{Chen-et-al-2012,Iftene-2007,Parakh-Kak-2011,Runhua-et-al-2008,Shi-et-al-2007,Wang-et-al-2011} }}; 
	
	\draw[color=magenta,line width=0.7mm,smooth,samples=10,domain=3:7,scale=1] plot (\x-2,1/2); \node[right,magenta] at (5.05,0.25) {\small{$1$: \cite{Dehkordi-Mashhadi-2008a,Dehkordi-Mashhadi-2008b,Eslami-Ahmadabadi-2010,Eslami-Rad-2012,Hu-et-al-2012,Shao-Cao-2005,Waseda-Soshi-2012,Yang-et-al-2004,Zhao-et-al-2007}}};  
	
	\end{tikzpicture} }

 	\caption{Global share volume comparison}
	\label{fig:GSV-comp}
\end{figure}

Figure~\ref{fig:GSV-comp} plots the estimated global share volume of all SSSs with respect to $n$, with $t=n-1$ and original data volume is 1~GB. \cite{Li-et-al-2012} is not plotted because global share volume grows very rapidly (about 252~GB when $n=7$).

Almost all SSSs require a volume about $n$ times that of secret data to store shares. Some SSSs propose solutions to minimize share volume. We categorize them into three classes. First, \cite{Iftene-2007,Parakh-Kak-2011,Zhao-et-al-2012} split data before sharing. Hence, share volume is only $n/t$ times that of secrets. However, since the size of shares decreases when $t$ increases, the value of $t$ cannot be bigger than the size of a secret. 

Second, global share volumes in \cite{Waseda-Soshi-2012} and \cite{Runhua-et-al-2008,Wang-et-al-2011,Chen-et-al-2012,Shi-et-al-2007,DB-Attasena-et-al-2014-J}  are only 1 and $n/t$ times that of secret data, respectively, because they construct $t$ and $n$ shares, respectively, per data block sizing $t$ secrets. 

Third, \cite{Eslami-Ahmadabadi-2010,Lin-We-1999,Chang-et-al-2005,Bu-Yang-2012,DB-Attasena-et-al-2014-P} share secrets independently, but they construct fewer than $n$ shares per secret (1, 4, 3, 3 and $n-t+2$ shares, respectively). Hence, share volumes are only 1, 4, 3, 3 and $n-t+2$ times that of secret data, respectively. 

Overall, \cite{Waseda-Soshi-2012,Eslami-Ahmadabadi-2010} require the lowest storage volume (the same as secret data volume) to store shares. However, \cite{Waseda-Soshi-2012} does not support data availability and  \cite{Eslami-Ahmadabadi-2010} supports data availability only when the NB is accessible.  Share volumes of \cite{Iftene-2007,Parakh-Kak-2011,Shi-et-al-2007,DB-Attasena-et-al-2014-J,DB-Attasena-et-al-2014-P} are a little higher than that of the lowest-share-volume approaches \cite{Waseda-Soshi-2012,Eslami-Ahmadabadi-2010} if $n$ is close to $t$, but they do support data availability. 

Some SSSs require extra storage to store keys. Most of them use only $n$ or $nt$ keys to share all secrets. Thus, they only consume a small storage volume. However, key volumes of \cite{Blakley-1979,He-Dawson-1994,Das-Adhikari-2010} are greater than the secret data volume (about $t^2$ \cite{Blakley-1979}, $2n$ \cite{He-Dawson-1994} and $t$ \cite{Das-Adhikari-2010} times data volume) because they use different key sets to share a secret. Hence, their overall storage volume (shares, keys and signatures) are greater than that of other SSSs, and thus incurs a higher storage cost.

Finally, all VSSSs and VMSSSs require extra storage to store signatures. The number of signatures is about the number of keys or shares, depending on the verified data type. Thus, overall signature volume is lower than share volume in all VSSSs and VMSSSs. However, if signatures are too small, verification accuracy becomes weak. 

Overall, \cite{Bu-Yang-2012} requires the lowest storage volume to store signatures. Hence, its overall storage volume is lower than $n$ times that of secret data. In contrast, \cite{Pedersen-1991,Yue-Hong-2009} require the greatest storage volume to store signatures. Hence, their overall storage volume turn to be greater than other SSSs, i.e., the same as \cite{Blakley-1979,He-Dawson-1994,Das-Adhikari-2010}, which construct a huge volume of keys.  


\section{Frameworks and Architectures for Sharing Secrets in the Cloud}
\label{sec:Frameworks}

\begin{figure}[hbt]
	\centering\includegraphics[width=0.45\textwidth]{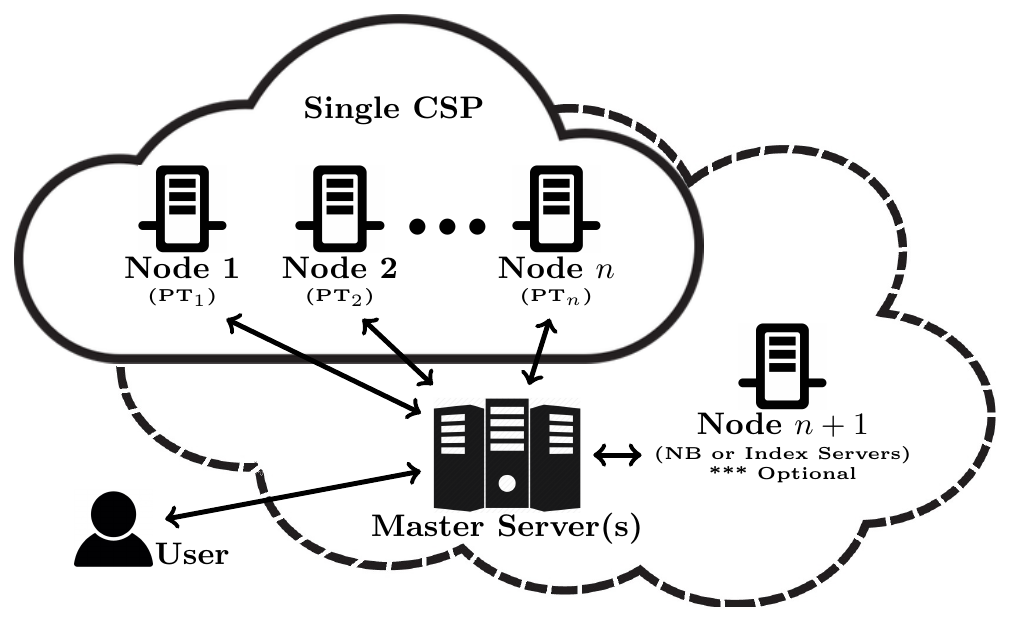}
 	\caption{Architecture from \cite{ST-KI-2013,DP-PK-JT-2015}}
	\label{fig:framework1}
\end{figure}

\begin{figure}[hbt]
	\centering \includegraphics[width=0.5\textwidth]{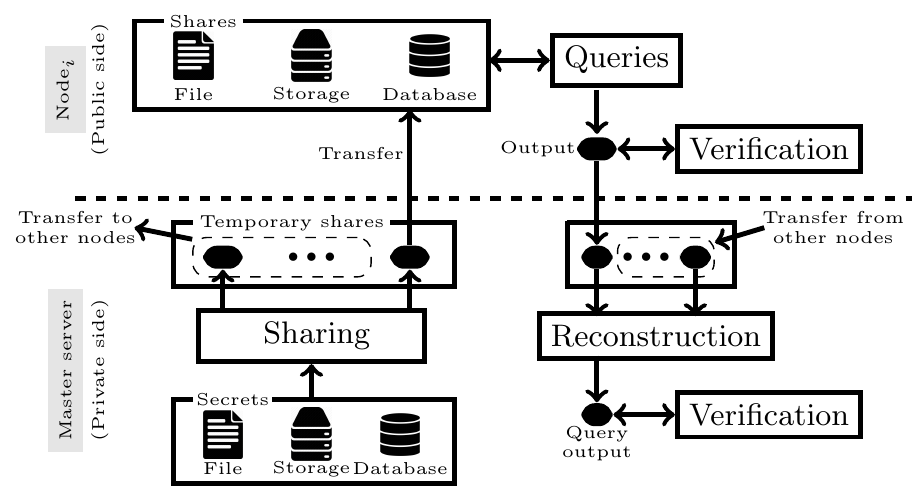}
 	\caption{Cloud SSS framework}
	\label{fig:framework-ex}
\end{figure}

Secret sharing-based cloud frameworks, such as the ones proposed by \cite{ST-KI-2013,DP-PK-JT-2015}, are similar to classical data distribution frameworks \cite{FZ-HC-2013,data-replication,DZ-DL-2012} in the cloud and distribute secrets over nodes at a single CSP's (Figure~\ref{fig:framework1}). They mostly differ in the SSSs they use. Unlike a classical data distribution framework, such frameworks guarantee data availability by default. Both secret sharing and data reconstruction processes run at a master server's (Figure~\ref{fig:framework-ex}). Although the master server may be a node in the cloud, to reduce privacy breaches in case of hacking, the master server usually stands at the user's side to hide all private parameters and keys from intruders collecting shares.

Two optional verification processes may be enforced by VSSSs. The first process helps verify the correctness of query results at PTs' so that no erroneous query results are transferred back to the master server. The second process runs at the master server's and verifies the correctness of reconstructed query results in case some PTs are not honest.

However, this framework bears a critical security weakness. Since all shares are stored at the same CSP’s, if the CSP is hacked, all data can be easily collected and reconstructed by the intruder.

 

\begin{figure}[hbt]
	\centering \includegraphics[width=0.45\textwidth]{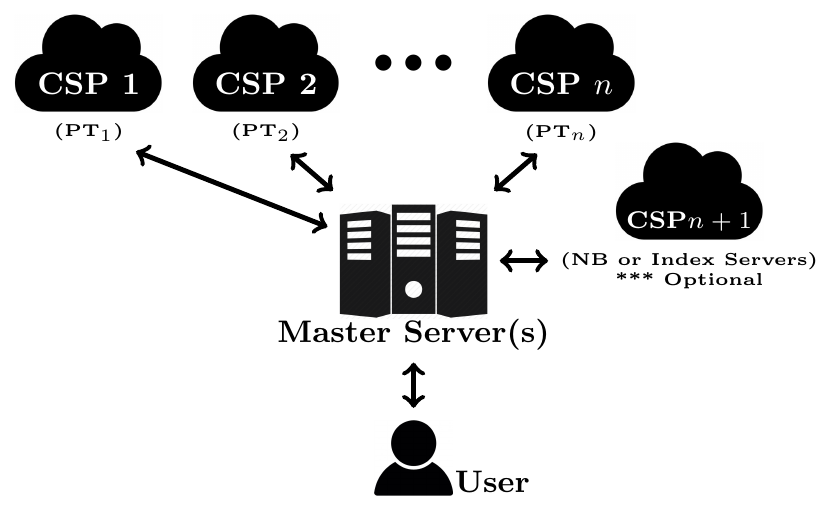}
 	\caption{Architecture from \cite{DB-Attasena-et-al-2014-J,DB-Attasena-et-al-2014-P,SD-YL-SS-2016,MM-UK-RK-MA-2016}}
	\label{fig:framework2}
\end{figure}

In contrast, the frameworks such as the ones proposed by \cite{DB-Attasena-et-al-2014-J,DB-Attasena-et-al-2014-P,SD-YL-SS-2016,MM-UK-RK-MA-2016} distribute secrets over multiple CSPs (Figure~\ref{fig:framework2}), thus providing better availability (it is unlikely that two or more CSPs all fail at the same time) and privacy (collecting all shares is more difficult than in the one-CSP case). 

As in the previous framework, storage and computation costs are still high. However, unlike global data volume, global storage monertary cost might not be $n$ times that of original data because storage cost differs from CSP to CSP. 
In contrast, data access time is bounded to the slowest CSP. Yet, this problem may be alleviated by both balancing data access time and providing the lowest possible costs \cite{DB-Attasena-et-al-2014-P,attasena-2015}.

\begin{figure}[hbt]
	\centering \includegraphics[width=0.45\textwidth]{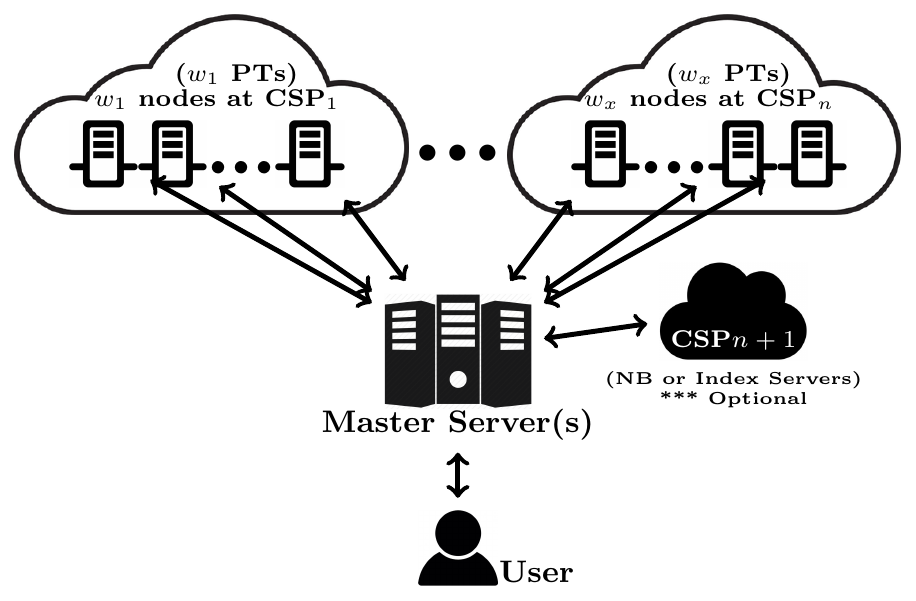}
 	\caption{SSSS-based framework \cite{SSSS-Nojoumian-et-al-2010,SSSS-Nojoumian-et-al-2012,SSSS-Zheng-et-al-2012,SSSS-Nojoumian-et-al-2012-2}}
	\label{fig:framework3}
\end{figure}

Finally, an SSSS-based framework \cite{SSSS-Nojoumian-et-al-2010,SSSS-Nojoumian-et-al-2012,SSSS-Zheng-et-al-2012,SSSS-Nojoumian-et-al-2012-2} generalizes the first two frameworks by distributing secrets over multiple nodes at multiple CSPs' (Figure~\ref{fig:framework3}). CSPs play the role of PTs and a number of nodes at CSPs' are the weight of PTs ($w_i$). Thus, security is not limited by the number of CSPs ($n$), but by the total number ($w=\sum _i w_i$) of nodes at all CSPs, which can be large. Moreover, shares stored in nodes at any CSP’s are not enough to reconstruct any secret since $w_i<t$.

There are some applications, e.g., secure data storage, secure databases and data warehouses, private information retrieval, and data management in the cloud, use the above frameworks. 

Eventually, let us briefly present query functionality in secure data storage solutions for public clouds that use or extend Shamir's SSS \cite{Shamir-1979}. Low-level data storage \cite{cStorage1,SD-YL-SS-2016} 
handle pattern search, equijoins and range queries on shares. 
Table~\ref{tab:DB-features} summarizes the querying features of secure cloud databases and data warehouses \cite{DB-Emekci-et-al-2005,DB-Emekci-et-al-2006,DB-Thompson-et-al-2009,DB-Agrawal-et-al-2009,DB-Hadavi-Jalili-2010,DB-Wang-et-al-2011,DB-Hadavi-et-al-2012,DB-Hadavi-et-al-2013,DB-Attasena-et-al-2014-J,DB-Attasena-et-al-2014-P,cDB2}.

\begin{table}
	\caption{Query types allowed by secret sharing-based cloud applications}
	\label{tab:DB-features}
		\resizebox{0.5\textwidth}{!} {
    	\begin{tabular}{ |l|c|c|c|c|c|c|c|}
    	\hline
    	 \textbf{Queries} & \textbf{\cite{DB-Agrawal-et-al-2009}} & \textbf{\cite{DB-Emekci-et-al-2006}} &
\textbf{\cite{DB-Emekci-et-al-2005,DB-Hadavi-et-al-2012,DB-Hadavi-Jalili-2010,DB-Hadavi-et-al-2013}} &
\textbf{\cite{DB-Thompson-et-al-2009}} & \textbf{\cite{DB-Wang-et-al-2011}} & \textbf{\cite{DB-Attasena-et-al-2014-J,DB-Attasena-et-al-2014-P,cDB2}} \\ \hline
    	 Update			& N & N & Y & Y & Y & Y \\ \hline
    	 Exact match	& N & Y & Y & N & Y & Y \\ \hline
    	 Range			& N & Y & Y & N & Y & Y \\ \hline
    	 Aggregate		& Y & Y & Y & Y & N & Y \\ \hline
    	 Grouping		& N & N & N & N & N & Y \\ \hline 
    	\end{tabular}
    }
\end{table}

\section{Conclusion}
\label{sec:Conclusions}

In this final section, we first draw a critical overview of all SSSs surveyed in this paper, including current challenges when using SSSs and in a cloud computing context. Finally, we present some sample applications that can benefit from SSSs.

\subsection{Secret Sharing Schemes}

Classic SSSs handle data security and availability with high sharing/reconstruction time and storage costs. MSSSs share data at once and reduce both costs. In addition, MSSSs type~I support data availability by using a NB, but are vulnerable if the NB is attacked. Hence, to share data with MSSSs type~I in the cloud, the NB should be located at a PT's that guarantees high security and availability. In contrast, PSSSs and \cite{DB-Attasena-et-al-2014-P} enhance data privacy by periodically refreshing shares and protecting data from CSP group cheating, respectively.

In addition, VSSSs and VMSSSs can verify the correctness of either or both of data and keys, but these operations induce additional time overhead and require to store signatures in addition to shares. Outer code verification still necessitates to trust PTs, because it is done at PTs'.   
Moreover, since almost all VMSSSs are also MSSSs type~I, their total storage volume (keys, shares and signatures) is still lower than $n$ times that of secret data.

Only \cite{Shi-et-al-2007} verifies the correctness of both data and keys. Although it is an MSSS type~II, its total storage volume is only about twice that of secret data. Moreover, its data sharing complexity is also reasonable, i.e., $O(mt)$, while most SSSs have a cubic sharing complexity. 

Eventually, only \cite{DB-Attasena-et-al-2014-J,DB-Attasena-et-al-2014-P} verify the correctness of both data and shares. They also minimize global share volume to lower than $n$ times that of secret data. Moreover, \cite{DB-Attasena-et-al-2014-P} can insert new data even though some PTs disappear, i.e., even though some CSPs fail due to technical or economic reasons.

Moreover, PSSSs refresh shares and verify their correctness to improve data privacy. However, computation (to renew shares) and storage (to store signatures) costs induce extra overhead in the refreshment process. Communications to synchronous shares from PTs to PTs are also numerous, thus provoking network bottlenecks.

Some SSSs support features such as updates, search operations, aggregation operations, etc. These features help minimize computation cost at the user's side and reduce communication overhead. Only three SSSs \cite{Blakley-1979,DB-Attasena-et-al-2014-J,DB-Attasena-et-al-2014-P} support all three operation types: update, exact match and aggregation. However, none can handle composite operations on shares, e.g., simultaneous exact match and aggregation. Performing composite operations on shares remains a challenge in SSSs as of today. Among SSSs that support search and aggregation operations, again only \cite{Shi-et-al-2007,DB-Attasena-et-al-2014-J,DB-Attasena-et-al-2014-P} minimize storage cost. \cite{Shi-et-al-2007} also optimizes data sharing time.

Finally, \cite{Hwang-Chang-1998,Chen-et-al-2012} allow the user to add and remove PTs to/from the PT pool. In the cloud, users can thus add and remove CSPs on demand. However, estimating monetary storage cost and detecting attacks or CSP failures is difficult. Thus, taking (or worse, automating) a decision regarding the CSP pool under CSP pricing or privacy constraints is still an open issue.


\subsection{Secure Applications in the Cloud}
\label{sec:Applications}

SSSs addressed various issues over time (Section~\ref{sec:Secret-sharing-schemes}). Let us describe below some applications that can benefit from secret sharing for data security.

Textual documents such as emails could be shared with \cite{Blakley-1979,Das-Adhikari-2010,DB-Attasena-et-al-2014-J}, since these SSSs optimize cost and update and search performance by allowing updates and exact matches directly over shares. Moreover, \cite{Das-Adhikari-2010,DB-Attasena-et-al-2014-J} also guarantee data integrity with inner and both inner and outer code verification, respectively. Finally, only \cite{DB-Attasena-et-al-2014-J} optimizes both storage volume and data sharing and reconstruction time.

In databases and data warehouses, update, exact match and aggregation operators are casually used. To optimize query response time, such SSSs as \cite{Das-Adhikari-2010,DB-Attasena-et-al-2014-J,DB-Attasena-et-al-2014-P} can be used to leverage cloud databases and warehouses. All these SSSs indeed guarantee data integrity. Moreover, \cite{DB-Attasena-et-al-2014-J,DB-Attasena-et-al-2014-P} also optimize storage cost and \cite{DB-Attasena-et-al-2014-P} allows inserting new data although some CSPs fail. Several secret sharing-based database or warehousing approaches \cite{DB-Emekci-et-al-2005,DB-Emekci-et-al-2006,DB-Thompson-et-al-2009,DB-Agrawal-et-al-2009,DB-Hadavi-Jalili-2010,DB-Wang-et-al-2011,DB-Hadavi-et-al-2012,DB-Hadavi-et-al-2013,DB-Attasena-et-al-2014-J,DB-Attasena-et-al-2014-P,cDB2} exploit the above-mentioned SSSs.


To handle data streams, SSSs such as \cite{Iftene-2007,Lin-We-1999,Chang-et-al-2005,Bu-Yang-2012} can be used, because they optimize sharing time and share secrets independently. Moreover, they require an overall storage volume that is lower than $n$ times that of secret data. \cite{Iftene-2007}'s storage volume is even close to the secret's volume if $n$ and $t$ are big and $n$ is close to $t$. However, only \cite{Lin-We-1999,Chang-et-al-2005,Bu-Yang-2012} guarantee data integrity.

Since memory is still limited in practice, SSSs that share data at once \cite{Yang-et-al-2004,Chan-Chang-2005,Eslami-Ahmadabadi-2010,Zhao-et-al-2007,Dehkordi-Mashhadi-2008a,Dehkordi-Mashhadi-2008b,Hu-et-al-2012,Eslami-Rad-2012,Shao-Cao-2005} cannot handle \emph{big data} volumes. However, SSSs that share individual secrets \cite{Shamir-1979,Blakley-1979,Asmuth-Bloom-1983,Iftene-2007,Parakh-Kak-2009,Parakh-Kak-2011,He-Dawson-1994,Harn-Lin-2010,Liu-et-al-2012,Takahashi-Iwamura-2013,Pedersen-1991,Tang-Yao-2008,Yue-Hong-2009,Hwang-Chang-1998,Zhao-et-al-2012,Lin-We-1999,Chang-et-al-2005,Wei-et-al-2007,Das-Adhikari-2010,Bu-Yang-2012,DB-Attasena-et-al-2014-P} or data blocks \cite{Waseda-Soshi-2012,Runhua-et-al-2008,Chen-et-al-2012,Wang-et-al-2011,Li-et-al-2012,Shi-et-al-2007,DB-Attasena-et-al-2014-J} allow the execution of the sharing process in main memory or even its parallelization, and thus can share huge data volumes efficiently.

Finally, potential users of database outsourcing should be aware that even frameworks based on normally highly secure SSSs might still be insecure because of inadequate architectural choices or a strong tradeoff in favor of query power \cite{JD-CR-2012}. To circumvent this problem, (V)MSSSs that primarily protect multiple secrets are a better choice than (V)SSSs for cloud applications. In any case, users should carefully evaluate the limitation of target SSSs before using them in any applicative context. We hope this survey will help them for this sake.


\bibliographystyle{spmpsci}      

\bibliography{cloud-security}

\end{document}